\documentclass[aps,pre,twocolumn,groupedaddress,amssymb,amsmath,nofootinbib,showpacs]{revtex4-1}

\usepackage{hyperref}

\usepackage{wasysym}
\usepackage{amssymb,amsmath}
\usepackage{graphicx}
\usepackage{bbm,color}

\graphicspath{{./figures/},{./}}
\renewcommand{\log}{\ln}
\newcommand{\beq}{\begin{equation}}
\newcommand{\eeq}{\end{equation}}
\newcommand{\be}{\begin{equation}}
\newcommand{\ee}{\end{equation}}
\newcommand{\bea}{\begin{eqnarray}}
\newcommand{\eea}{\end{eqnarray}}
\newcommand{\rmd}{\mathrm{d}}

\newcommand{\nn}{\nonumber}
\newcommand{\du}{\dot{u}}

\newcommand{\tu}{\tilde{u}}

\newcommand{\hu}{\hat{u}}

\newcommand{\dw}{\dot{w}}

\newcommand{\tih}{\tilde{h}}
\newcommand{\hh}{\hat{h}}
\newcommand{\hQ}{\hat{Q}}
\newcommand{\hP}{\hat{P}}
\newcommand{\hsh}{\hat{\mathfrak{s}}}

\newcommand{\rme}{{\rm e}}
\newcommand{\ti}{{t_{\rm i}}}
\newcommand{\tf}{{t_{\rm f}}}

\usepackage[USenglish]{babel}
\begin{document}

\title{Statistics of Avalanches with Relaxation, and  Barkhausen Noise: A Solvable Model}

\author{Alexander Dobrinevski}
\email[]{Corresponding author: Alexander.Dobrinevski@lpt.ens.fr}
\author{Pierre Le Doussal}
\author{Kay J\"org Wiese}
\affiliation{CNRS-Laboratoire de Physique Th\'eorique de l'Ecole
Normale Sup\'erieure, 24 rue Lhomond, 75005 Paris, France
\thanks{LPTENS is a Unit\'e Propre du C.N.R.S.
associ\'ee \`a l'Ecole Normale Sup\'erieure et \`a l'Universit\'e Paris Sud}
}

\date{\today\ }

\begin{abstract}
We study a generalization of the Alessandro-Beatrice-Bertotti-Montorsi (ABBM) model of a particle in a Brownian force landscape, including retardation effects.
We show that under monotonous driving the particle moves forward at all times, as it does in absence of retardation (Middleton's theorem). This remarkable property allows us to develop an analytical treatment. The model with an exponentially decaying memory kernel is realized in Barkhausen experiments with eddy-current relaxation, and has previously been shown numerically to account for the experimentally observed asymmetry of Barkhausen-pulse shapes. We elucidate another qualitatively new feature: the breakup of each avalanche of the standard ABBM model into a cluster of sub-avalanches, sharply delimited for slow relaxation under quasi-static driving. These conditions are typical for earthquake dynamics. With relaxation and aftershock clustering, the present model includes important ingredients for an effective description of earthquakes.
We analyze quantitatively the limits of slow and fast relaxation for stationary driving with velocity $v>0$. The $v$-dependent power-law exponent for small velocities, and the critical driving velocity at which the particle velocity never vanishes, are modified. We also analyze non-stationary avalanches following a step in the driving magnetic field. Analytically, we obtain the mean avalanche shape at fixed size, the duration distribution of the first sub-avalanche, and the time dependence of the mean velocity. We propose to study these observables in experiments, allowing to directly measure the shape of the memory kernel, and to trace eddy current relaxation in Barkhausen noise. 
\end{abstract}


\maketitle

\section{Introduction and model\label{sec:RetABBM}}
\subsection{Barkhausen noise}
The Barkhausen noise \cite{Barkhausen1919} is a characteristic magnetic signal emitted when a soft magnet is slowly magnetized. It can be measured and made audible as \textit{crackling} through an induction coil: periods of quiescence followed by pulses, or \textit{avalanches}, of random strength and duration. The statistics of the emitted signal depends on material properties and its state. By analyzing the Barkhausen signal, one can deduce for example residual stresses \cite{GauthierKrauseAtherton1998,StewartStevensKaiser2004} or grain sizes \cite{RanjanJilesBuckThompson1987,YamauraFuruyaWatanabe2001} in metallic materials. Understanding how particular details of the Barkhausen noise statistics depend on microscopic material properties is important for such applications. 

On the other hand, Barkhausen noise pulses are just one example for avalanches in disordered media. Such avalanches also occur in the propagation of cracks during fracture \cite{HerrmanRoux1990,ChakrabartiBenguigui1997,MaloySantucciSchmittbuhlToussaint2006}, in the motion of fluid contact lines on a rough surface \cite{RolleyGuthmannGombrowiczRepain1998,MoulinetGuthmannRolley2002,LeDoussalWiese2010,LeDoussalWieseMoulinetRolley2009}, and as earthquakes driven by motion of tectonic plates \cite{BenZionRice1993,MehtaDahmenBenZion2006,FisherDahmenRamanathanBenZion1997,BenZionRice1997}. Some features of the avalanche statistics, like size and duration distributions \cite{LeDoussalWiese2011b,LeDoussalWiese2012a}, are universal for many of these phenomena \cite{SethnaDahmenMyers2001}. Barkhausen noise is easily measurable experimentally, and provides a good way to study aspects of avalanche dynamics common to all these systems.

A first advance in the theoretical description of Barkhausen noise was the stochastic model postulated by Alessandro, Beatrice, Bertotti and Montorsi \cite{AlessandroBeatriceBertottiMontorsi1990,AlessandroBeatriceBertottiMontorsi1990b} (\textit{ABBM model}). They proposed modeling the domain-wall position $u(t)$ through the stochastic differential equation (SDE)
\beq
\label{eq:EOMABBM}
\Gamma \du(t) = 2  I_s\big[H(t) - k u(t) + F(u(t))\big].
\eeq
We follow here the conventions of \cite{ZapperiCastellanoColaioriDurin2005} and \cite{ColaioriDurinZapperi2007}.
$I_s$ is the saturation magnetization, and $H(t)$ the external field which drives the domain-wall motion. A typical choice is a constant ramp rate $ c$, $ H(t)=c\, t = k v\, t$, which leads to a constant average domain-wall velocity $v  = c/k$ \cite{AlessandroBeatriceBertottiMontorsi1990}. $k$ is the demagnetizing factor characterizing the strength of the demagnetizing field $-k u$ generated by effective free magnetic charges on the sample boundary \cite{AlessandroBeatriceBertottiMontorsi1990,Colaiori2008}. The domain-wall motion induces a voltage proportional to its velocity $\du(t)$, which is the measured Barkhausen noise signal. Here $F(u(t))$ is a random local pinning force. It is assumed to be a Brownian motion, i.e.\ Gaussian with correlations
\beq \label{gauss1}
\nn
\overline{\left[F(u)-F(u')\right]^2} = 2\sigma |u-u'|.
\eeq
This choice may seem unnatural, since the physical disorder does not exhibit such long-range correlations. It is only recently that it has been shown \cite{LeDoussalWiese2011,LeDoussalWiese2011b,LeDoussalWiese2012a} that the ``ABBM guess'' emerges as an effective disorder to describe the avalanche motion of the center-of-mass of the interface, denoted $u(t)$, in the mean-field limit of the field theory of an elastic interface with $d$ internal dimensions. This correspondence holds both for interfaces driven quasi-statically \cite{LeDoussalWiese2011,LeDoussalWiese2012a}, and for static interfaces at zero temperature \cite{LeDoussalWiese2011b}. The mean-field description is accurate above a certain critical internal dimension $d_{\rm c}$. For $d < d_{\rm c}$, a systematic expansion in $\epsilon = d_{\rm c} - d$ using the functional renormalization group yields universal corrections to the scaling exponents \cite{NattermannStepanowTangLeschhorn1992,ChauveLeDoussalWiese2000a,LeDoussalWieseChauve2002} and avalanche size \cite{LeDoussalWiese2011b,LeDoussalWiese2012a} and duration \cite{LeDoussalWiese2011,LeDoussalWiese2008c,LeDoussalWiese2012a} distributions.

For the particular case of magnetic domain walls, the predictions of the ABBM model are well verified experimentally in certain ferromagnetic materials, for example FeSi alloys \cite{AlessandroBeatriceBertottiMontorsi1990b,OBrienWeissman1994,DurinZapperi2000}. These are characterized by long-range dipolar forces decaying as $1/r^3$ between parts of the domain wall a distance $r$ apart. This leads \cite{CizeauZapperiDurinStanley1997} to a critical dimension $d_{\rm c}=2$ coinciding with the physical dimension of the domain wall. In this kind of systems, as expected, the mean-field approximation is reasonably well satisfied. Measurements on other types of ferromagnets, for example FeCoB alloys \cite{DurinZapperi2000} indicate a universality class different from the mean-field ABBM model. This may be explained by short-range elasticity, and a critical dimension $d_{\rm c} > 2$. To describe even the center of mass mode in this class of domain walls, one needs to take into account the  spatial structure of the domain wall. Predictions for roughness exponents \cite{ChauveLeDoussalWiese2000a,LeDoussalWieseChauve2002} and avalanche statistics \cite{LeDoussalWiese2011,LeDoussalWiese2008c,LeDoussalWiese2011b,LeDoussalWiese2012a} for this non-mean-field universality class have been obtained using the functional renormalization group.

On the other hand, even for magnets in the mean-field universality class, a careful measurement of Barkhausen pulse shapes \cite{SpasojevicBukvicMilosevicStanley1996,KuntzSethna2000,SethnaDahmenMyers2001,ZapperiCastellanoColaioriDurin2005} shows that they differ from the simple symmetric shape predicted by the ABBM model \cite{PapanikolaouBohnSommerDurinZapperiSethna2011,LeDoussalWiese2011}. 
This hints at a more complicated equation of motion than the first-order overdamped dynamics usually considered for elastic interfaces in disordered media. 

In a physical interface, there may be additional degrees of freedom. One example was studied in \cite{LecomteBarnesEckmannGiamarchi2009,BarnesEckmannGiamarchiLecomte2012}. Other examples include deformations of a plastic medium, 
or eddy currents arising during the motion of a magnetic domain wall. 
For viscoelastic media, these can be modeled by a memory term which is non-local in time \cite{MarchettiMiddletonPrellberg2000,Marchetti2005}. At mean-field level, this is equivalent to a model with dynamical stress overshoots \cite{SchwarzFisher2001}.
Such memory terms may lead to interesting new phenomena, like coexistence of pinned and moving states \cite{MarchettiMiddletonPrellberg2000,Marchetti2005,LeDoussalMarchettiWiese2008}. 
A similar memory term, non-local in time, is argued in \cite{ZapperiCastellanoColaioriDurin2005} to describe the dissipation of eddy currents in magnetic domain-wall dynamics,
\beq
\label{eq:EOMZapperi}
 \frac{1}{\sqrt{2\pi}} \int_{-\infty}^t \rmd s\, \mathfrak{f}(t-s) \, \du(s) =  2 I_s\big[H(t) - k u(t) + F(u(t))\big].
\eeq
The response function $\mathfrak{f}$, derived by solving the Maxwell equations in a rectangular sample \cite{Bishop1980,ZapperiCastellanoColaioriDurin2005,ColaioriDurinZapperi2007,Colaiori2008}, is
\beq
\label{eq:EOMZapperi1}
\mathfrak{f}(t) = \sqrt{2\pi}\frac{64 I_s^2}{a b^2 \sigma \mu^2} \sum_{n,m=0}^\infty \frac{e^{-t/\tau_{m,n}}}{(2n+1)^2\omega_b}.
\eeq
$\tau_{m,n}$ are relaxation times for the individual eddy current modes,
\begin{align*}
& \tau_{m,n}^{-1} = (2m+1)^2 \omega_a + (2n+1)^2 \omega_b \\
& \omega_a = \frac{\pi^2}{\sigma \mu a^2},\quad\quad \omega_b = \frac{\pi^2}{\sigma \mu b^2}.
\end{align*}
They depend on the sample width $a$, thickness $b$, permeability $\mu$ and conductivity $\sigma$.  
\eqref{eq:EOMZapperi} and \eqref{eq:EOMZapperi1} correspond to Eqs.~(13), (17) and (21) in \cite{ColaioriDurinZapperi2007}; we refer the reader there for  details of the derivation.

Zapperi et al.\ \cite{ZapperiCastellanoColaioriDurin2005} showed numerically that avalanche shapes in the model \eqref{eq:EOMZapperi} are asymmetric. They concluded that eddy-current relaxation may be one way of explaining the experimentally observed skewness of Barkhausen noise pulses. They also argue that similar relaxation effects may be relevant for other physical situations where asymmetric pulse shapes are observed\footnote{For example, \cite{ZapperiCastellanoColaioriDurin2005} mentions slip velocity profiles during earthquakes \cite{HoustonBenzVidale1998,MehtaDahmenBenZion2006}. However, it is not clear if there are physical reasons to expect a relaxation of the form \eqref{eq:EOMZapperi}.}.

A simplification of Eq.~\eqref{eq:EOMZapperi1} occurs when considering only the leading contributions for small and large relaxation times\footnote{We have $f(0) \propto \sum_{m,n}\frac{1}{(2n+1)^2} = \infty$ and $\int_0^{\infty} f(t) \rmd t \propto \sum_{m,n}\frac{1}{(2n+1)^2\left[(2n+1)^2+(2m+1)^2\right]} = \rm const$. Thus, for small times, $f(t)$ is well approximated by $\Gamma \delta(t)$ for some constant $\Gamma$. On the other hand, for long times, only the mode that relaxes slowest remains. Hence, for long times one can set $f(t) \approx \frac{\Gamma_0}{\tau_{0,0}}e^{-t/\tau_{0,0}}$. }. Then one obtains \cite{ZapperiCastellanoColaioriDurin2005} a natural generalization of the ABBM equation \eqref{eq:EOMABBM}:
\begin{align}
\nn
&\Gamma \du(t) + \frac{\Gamma_0}{\tau}\int_{-\infty}^t \rmd s\, e^{-(t-s)/\tau}\du(s) \\
\label{eq:EOMZapperi2}
&\qquad =  2 I_s\big[H(t) - k u(t) + F(u(t))\big].
\end{align}
Here, $\tau$ is the longest relaxation time of the eddy-current modes, $\tau=\tau_{0,0}=\frac{\mu\sigma}{\pi^2}\left(\frac{1}{a^2}+\frac{1}{b^2}\right)^{-1}$. $\Gamma$ and $\Gamma_0$ are damping coefficients given in \cite{ZapperiCastellanoColaioriDurin2005}.

\subsection{The ABBM model with retardation}

For the remainder of this work, we adopt the conventions used in the study of elastic interfaces. Let us introduce a more 
general model than \eqref{eq:EOMZapperi2},
\beq
\label{eq:EOMU}
\eta \du(t) + a \int_{-\infty}^t \rmd s\,f(t-s) \du(s) = F\big(u(t)\big) + m^2 \big[w(t) - u(t)\big].
\eeq
which describes a particle driven in a force landscape $F(u)$, {\it with retardation}. At this stage $F(u)$ is arbitrary. Here
$f(t)$ is a general memory kernel with the following properties:
\begin{enumerate}
        \item $f(0) = 1$ (without loss of generality, since a constant may be absorbed into the parameter $a$).
        \item $f(x) \to 0$ as $x \to \infty$.
        \item $f'(x) \leq 0$ for all $x$, i.e.~memory of the past trajectory always decays with time.
\end{enumerate}
This model possesses a remarkable property for any such kernel $f(t)$ and any landscape $F(u)$. 
It has monotonicity, i.e.\ it satisfies the Middleton theorem: For non-negative driving $\dw \geq 0$, after an initial transient period one has $\du \geq 0$ at all times. A more precise statement and a proof are given in Appendix \ref{sec:Monot}. It has very important consequences, both in the driven regime, and in the
limit of quasi-static driving, i.e.\ small $\dot w \to 0^+$. In that limit it converges to the quasi-static process $u(t) \to u(w(t))$, where $u(w)$ is the
(forward) Middleton metastable state, defined as the smallest (leftmost) root of
\be  \label{soluuw}
m^2 u - F(u) = m^2 w \quad \Leftrightarrow \quad u=u(w) \ .
\ee 
It is {\it independent} of the precise form of the kernel $f(t)$. Hence the domain-wall position $u(t)$ is uniquely determined by the value of the driving field $w(t)$, due to the monotonicity property \cite{Middleton1992}. This process $u(w)$ exhibits jumps at a set of individual points, the avalanche locations $w_i$, and the quasi-static avalanche sizes \be 
S_i = u(w_i^+)-u(w_i^-)
\ee  
are thus independent of the retardation kernel. What depends on the kernel is the  dynamics  within these avalanches,
and that is studied here. The quasi-static avalanche sizes $S_i$ have a well-defined distribution $P(S)$ which has been computed for a particle in various force landscapes \cite{LeDoussalWiese2009,DobrinevskiLeDoussalWiese2012} and for the non-trivial case of a $d$-dimensional elastic interface using functional RG methods \cite{LeDoussalWiese2008c,LeDoussalMiddletonWiese2009,RossoLeDoussalWiese2009}. As long as the dynamics obeys the Middleton theorem, the {\em avalanche-size } distribution remains {\em independent} of the details of the dynamics \cite{LeDoussalWiese2012a}.

While monotonicity holds for any $F(u)$, in this article we focus on the case of the Brownian force landscape
which can be solved analytically.
As in the standard ABBM model, we choose the effective random pinning force $F(u)$ to be a random walk, i.e.~Gaussian with correlator 
given by (\ref{gauss1}). \footnote{It can be realized as a stationary landscape, $\overline{F(u)F(u')}=\Delta_0-\sigma |u-u'|$ with a cutoff at  scale $u \sim \Delta_0/\sigma$, or by $\overline{F(u)F(u')}=2\sigma \min( u,u')$ (non-stationary landscape with $F(0)=0$).
In both cases $F'(u)$ is a white noise, and that is the important feature. }
We call this the \textit{ABBM model with retardation}. In view of the application to  Barkhausen noise, the parameter
$a>0$ describes the overall strength of the force exerted by eddy currents on the domain wall. 
For $a=0$, \eqref{eq:EOMU} reduces to the equation of the standard ABBM model in the conventions 
of \cite{LeDoussalWiese2009,DobrinevskiLeDoussalWiese2012}.

The retarded ABBM model is particularly interesting in view of the monotonicity property. Other ways of generalizing the ABBM model to include inertia, e.g.\ by a second-order derivative \cite{LeDoussalPetkovicWiese2012}, do not inherit this property from the standard ABBM model. This makes the ABBM model with retardation very special, and it will be important for its solution in section \ref{sec:RetSol}. 

When considering the particularly interesting case of exponential relaxation motivated in \cite{ZapperiCastellanoColaioriDurin2005}, we set
\beq \label{tau}
f(t) =  e^{-t/\tau},
\eeq
$\tau$ is the longest time scale of eddy-current relaxation, as discussed above. 
In this approximation, \eqref{eq:EOMU} can be re-written as two coupled, \textit{local} equations for the domain-wall velocity $\du(t)$, and the eddy-current pressure $h(t)$, 
\begin{align} h(t) &=  \frac{1}{\tau}\int_{-\infty}^t \rmd s\,e^{-(t-s)/\tau} \du(s)
\\
\eta \du(t)+ a \tau h(t) &= F\big(u(t)\big) + m^2 \big[w(t) - u(t)\big]\\
 \tau \partial_t h(t) &= \du(t) - h(t) .
\label{eq:EOMExp}
\end{align}
Although most of our quantitative results will be derived for this special case only, most 
qualitative features carry over to more general kernels with sufficiently fast decay. 

By rescaling $u$, $w$ and $t$ in Eq.~\eqref{eq:EOMU} (for details, see section \ref{sec:Dim}), one finds 
the characteristic time scale $\tau_m = \eta/m^2$ and length scale
$S_m = \sigma/m^4$ of the standard ABBM model ($a=0$). They set the
scales for the durations and sizes of the largest avalanches. There are 
of course avalanches of  smaller size (up to some microscopic cutoff
if one defines it). The velocity scale is  $v_m =\sigma/(\eta m^2)$ and one
can define a renewal time for the large avalanches as $\tau_v = S_m/v$, 
the limit of quasi-static driving being $\tau_m \ll \tau_v$, equivalent to $v/v_m \ll 1$. 
In the retarded ABBM model (\ref{tau}) one introduces an additional memory time scale $\tau$
and various regimes will emerge depending on how $\tau$ compares with the
other time scales (whose meaning will be changed). 

Eq.\ (\ref{eq:EOMExp}) then describes a depinning model with relaxation, i.e.\ one can think of the
disorder landscape as relaxing via the additional degree of freedom $h(t)$.
This is a feature of interest for earthquake models as discussed below.
In this context one considers the limit of well separated time scales, 
$\tau_m \ll \tau \ll \tau_v$. 

Other
 features of Barkhausen noise predicted for the ABBM model with retardation are quite different from those of the standard ABBM model.  Zapperi et al.\ \cite{ZapperiCastellanoColaioriDurin2005} already realized that the inclusion of eddy currents leads to a skewness in the avalanche shape. In this article, we go further and discuss changes in the avalanche statistics. The relaxation of eddy currents introduces an additional slow time scale into the model. This leads to avalanches which stretch further in time. In particular, avalanches following a kick (or, more generally, stopped driving) never terminate, by contrast with the standard ABBM model. This is because of the exponentially decaying retardation kernel, which never vanishes \footnote{For a model such that $f(t)=0$ for $t>t_0$,  avalanches would remain of finite duration.}. Avalanche sizes however, are not changed by retardation in the limit of quasi-static driving, as discussed above. In that limit, retardation leads to a break-up of avalanches into sub-avalanches, which can also be called aftershocks. Avalanches at continuous driving overlap stronger, and the velocity threshold for the infinite avalanche (i.e.\ the velocity $\dot u$ no longer vanishes) is decreased. We now describe these effects in detail and formulate more precise statements.
 
\subsection{Protocols\label{sec:Protocols}}

Let us first review qualitatively the main situations that we will study, and define the terminology. 

(i) stationary driving: The driving velocity is constant, $w(t) = v t$, and the distributions of the domain-wall velocity $\du$ and of the eddy-current pressure $h$ reach a steady state,
which we study. If $v$ is large enough the velocity will never vanish and one has a single infinite avalanche, also called ``continuous motion". At smaller $v>0$ the
velocity will sometimes vanish. That defines steady state avalanches. These are more properly called sub-avalanches of the infinite avalanche since
at finite $v>0$ they immediately restart. Only in the limit $v=0^+$ they become well separated in time and can then be called steady state avalanches. 

(ii) Avalanches following a kick: We consider an initial condition at $t=0$ prepared to lie in the ``Middleton attractor'' at $u=u(w(t<0))$, as discussed above. 
It can be obtained by driving the system monotonously in the far past with $\dot w >0$, until memory of the initial condition is erased; then let it relax for a long time with $\dot w=0$
until time $t=0$. Hence the initial condition is $\dot u(t=0)=h(t=0)=0$. At $t=0,$ one changes the external magnetic field instantaneously by $w_{0}$, i.e.\ sets $\dw(t) = w_0 \delta(t)$. For $t>0$, the external field does not change anymore, thus a kick in the driving velocity corresponds to a step in the applied force. At $t=\infty$ the system has settled again into the Middleton attractor 
at $u(t=\infty)=u(w+w_0),$ because of the properties discussed above. One can thus consider the total motion to define a single avalanche following a kick, which is thus
unambiguously defined. The total size $S=\int_0^{\infty} \dot u(t)\,\rmd t$ is the same as in the absence of retardation. We will ask about the
total duration (which becomes infinite) and whether the velocity has vanished at intermediate times, i.e.\ whether the avalanche has broken
into sub-avalanches.

Avalanches following a kick are called {\em non-stationary avalanches} (since driving is non-stationary). However, in the limit of
 $w_0\to 0^+$ they become identical to the steady-state avalanches obtained by stationary driving
discussed above (conditioned to start at $t=0$).

\begin{figure*}
\begin{minipage}{0.48\textwidth} 
\includegraphics[width=0.9\columnwidth]{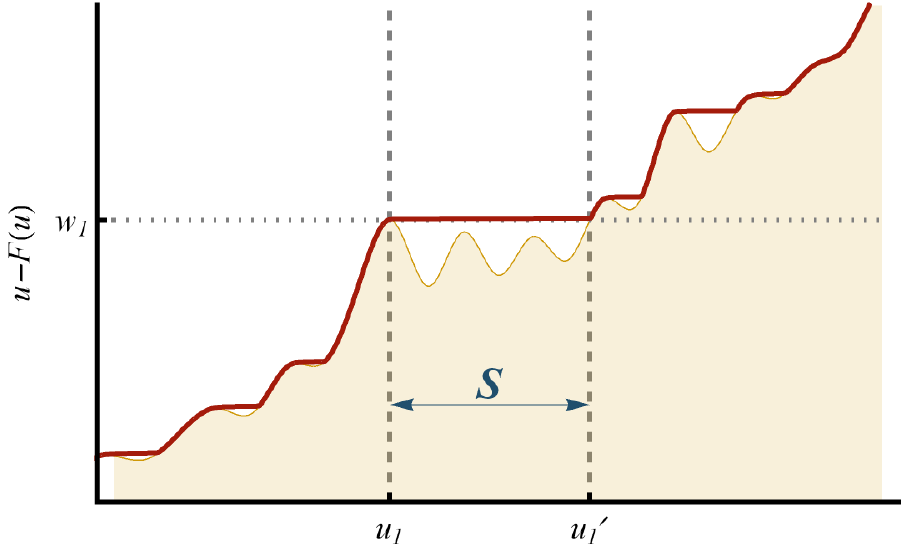}\\
{\small (a) Standard ABBM model.}
\end{minipage}~
\begin{minipage}{0.48\textwidth} 
\includegraphics[width=0.9\textwidth]{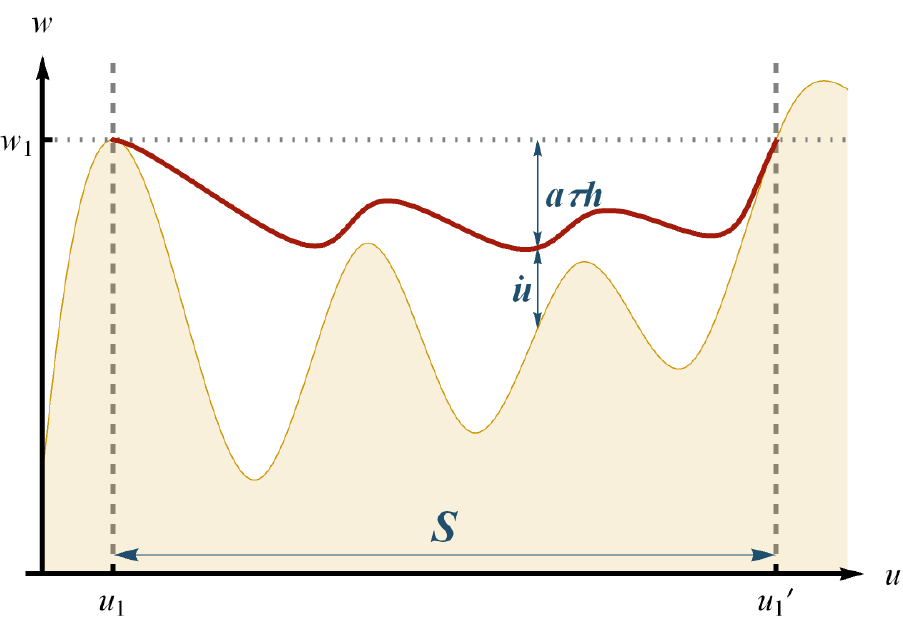}\\
{\small (b) ABBM model with retardation, $\tau = 1$.}
\end{minipage}
\begin{minipage}{0.48\textwidth} 
\includegraphics[width=0.9\columnwidth]{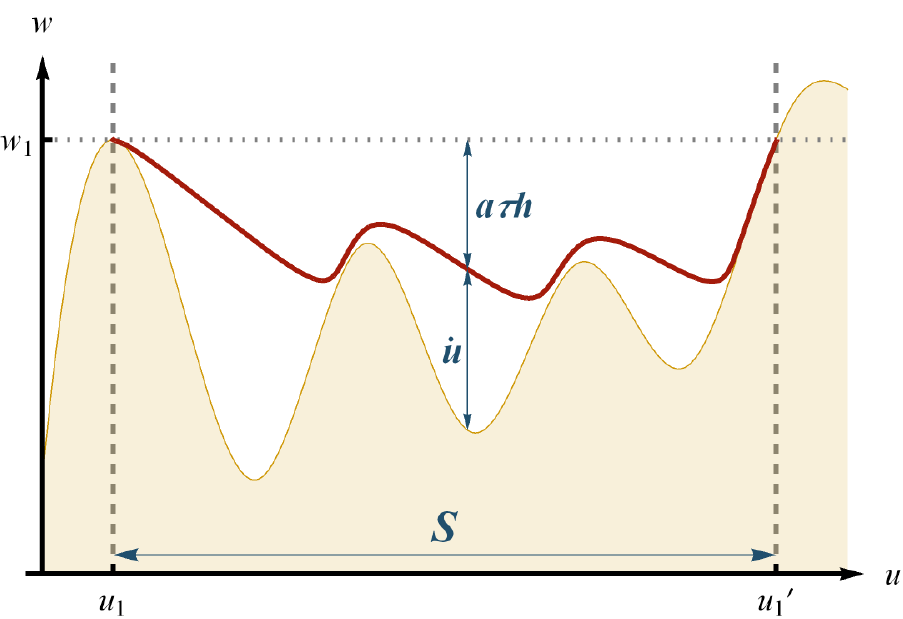}\\
{\small (c) ABBM model with retardation, $\tau = 3$.}
\end{minipage}~
\begin{minipage}{0.48\textwidth} 
\includegraphics[width=0.9\columnwidth]{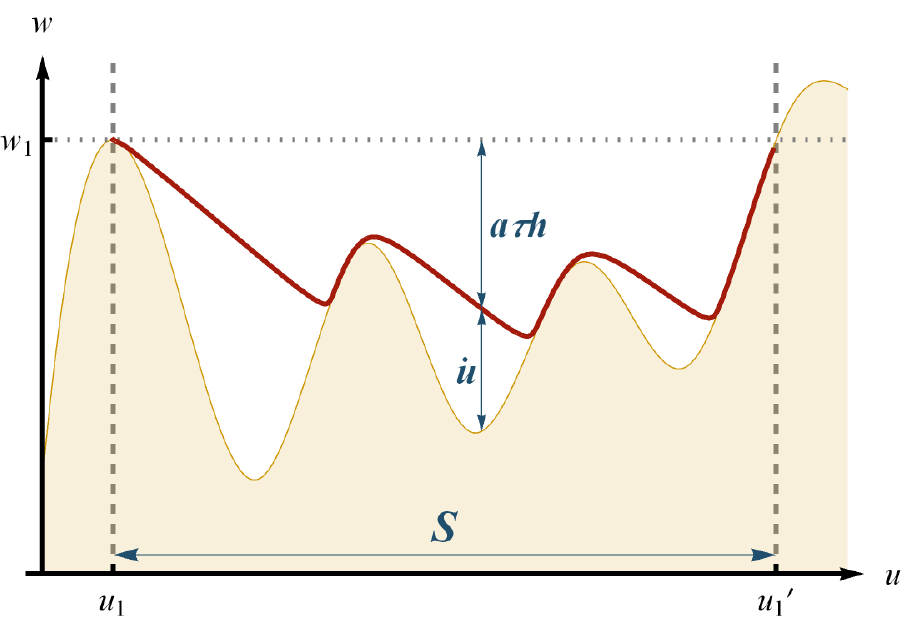}\\
{\small (d) ABBM model with retardation, $\tau = 10$.}
\end{minipage}
\caption{Splitting of an avalanche into sub-avalanches through the retardation mechanism. We have set $m^2=1$ and $a=1$ and we vary the relaxation time $\tau$.}\label{fig:Splitting}\end{figure*}
\subsection{Organization of this article}

The remainder of this article is structured as follows: 
In section \ref{sec:Physics}, we discuss in more detail the phenomenology and the qualitative physics of the ABBM model with retardation. We discuss the splitting of a quasi-static avalanche into {\em\ sub-avalanches}, and the effects of retardation on the stationary and the non-stationary dynamics. 

In section \ref{sec:RetSol} we explain how the probability distribution of observables linear in the domain-wall velocity can be computed by solving a non-linear, non-local ``instanton'' equation. By this, the stochastic model is mapped onto a purely deterministic problem of non-linear dynamics. This is a generalization of the method developed in \cite{LeDoussalWiese2011,DobrinevskiLeDoussalWiese2012} for the standard ABBM model with arbitrary driving. 

Section \ref{sec:MeanVel} discusses how the explicit form of the memory kernel $f(t)$ can be extracted in an experiment from the response to a kick. 

Section \ref{sec:QSLim} is devoted to an analysis of the instanton equations in the limit $\frac{\eta}{m^2} \ll \tau $. This means that eddy currents relax much more slowly than the domain wall moves. In this limit, we obtain the stationary distributions of the eddy-current pressure and domain-wall velocity, as well as their behavior following an instantaneous kick in the driving field. The instanton solution reflects the two time scales in the problem: A short time scale, on which eddy currents build up but do not affect the dynamics, and a long time scale, on which they relax quasi-statically. We prove that, even after the driving has stopped, the velocity never becomes zero permanently.

In section \ref{sec:FastRel} we discuss the fast-relaxation limit $\frac{\eta}{m^2} \gg \tau$.  In this limit, eddy currents relax much faster than the domain wall moves. The instanton solutions again exhibit two time scales, but now eddy currents are irrelevant for the long-time asymptotics. Qualitative results (like the fact that the domain-wall motion never stops entirely) are in agreement with those for the slow-relaxation limit, considered in section \ref{sec:QSLim}.

In section \ref{sec:AvStatSize} we discuss non-stationary avalanches following an instantaneous kick in the driving. In particular, we compute their average shape at fixed size.

In section \ref{sec:AvStatDur}, we show how to include an absorbing boundary in the instanton solution of section \ref{sec:RetSol}. This is required for treating avalanches during stationary driving. We then derive the distribution of avalanche durations in the standard ABBM model at finite driving velocity, $v>0$, and the leading corrections for weak relaxation and $\tau=\tau_m$. We also show numerical results for more general situations, and give some conjectures on the modification of size and duration exponents by retardation effects.

Last, in section \ref{sec:Conclusion}, we summarize our results. We discuss how they can be used to learn more about the dependence of Barkhausen noise on eddy current dissipation.

\section{Physics of the model and summary of the results\label{sec:Physics}}

\subsection{Quasi-static driving: Sub-avalanches and aftershocks\label{sec:PhysicsQS}} 

Consider the system either under stationary driving at $v=0^+$, or following an infinitesimal kick $w_0 = 0^+$ as discussed above, and
call $t=0$ the starting time of the avalanche. The main physics can be understood from  figure \ref{fig:Splitting}
and keeping in mind the equations (\ref{eq:EOMExp}). 

In Fig.\ \ref{fig:Splitting}a we represent the usual construction for $u(w)$ in the standard ABBM as the left-most solution of 
the equation (\ref{soluuw}) (in the figure we set $m=1$). Assuming $\tau_m \ll \tau_v$ this construction indicates the position of 
the domain wall as a function of $w=w(t)$ on time scales of order $\tau_v$. At $w_1$ the solution jumps from $u_1=u(w_1^-)$ to $u_1'=u(w_1^+)=u_1+S$
corresponding to an avalanche of size $S$; the latter occurs on the much faster time scale $\tau_m$. 
During the avalanche the velocity $\dot u(u)$ (setting $\eta=1$) is given
by the difference in height between the line $m^2 w= m^2 w_1$ and the landscape $ m^2 u-F(u)$, providing a graphic representation of the
motion. The velocity $\dot u(u)$ vanishes at $u=u_1$ and $u=u_1'$. For illustration we have represented a force landscape which is ABBM like at
large scales but smooth at small scales. For the continuous ABBM model the construction is repeated at all scales and
one has avalanches of all smaller sizes.

Let us now add retardation, setting $a>0$, and varying the memory time $\tau$. The graphical construction corresponding to Eq.\ (\ref{eq:EOMExp})
is represented in Fig.\ \ref{fig:Splitting}b to \ref{fig:Splitting}d. The difference in height is now the sum of $\dot u$ and $a \tau h$ (in the Figure we chose $a=1$), which evolves according to
the second equation in Eq.\ (\ref{eq:EOMExp}). It can be rewritten as
\be 
\tau \partial_u h = 1 - \frac{h}{\dot u(u)} \ .
\ee  
Hence $h$ increases from $h=0$, initially as $h \approx (u-u_1) /\tau$ (since $\dot u \sim \sqrt{u-u_1}$). Thus
the curve $w - a\tau h$ versus $u$ starts with a negative slope $ -a$. 

Another way to see this is to note that for $t\ll \tau$, the second equation of \eqref{eq:EOMExp} gives
\beq
\tau h(t) = \int_0^t \du(t) + \mathcal{O}(t/\tau) = u(t)-u_1 + \mathcal{O}(t/\tau).
\eeq
Inserting this into the first equation of \eqref{eq:EOMExp}, we obtain
\beq
\eta \du(t) = F\big(u(t)\big) + m^2 \left[w(t) - u(t)\right] - a \left[u(t)-u_1\right]+....
\eeq
Effectively, for short times the mass is modified from $m^2 \to m^2+a$.
Thus, while $w$ is fixed, the end of the first sub-avalanche is determined not by the roots of $m^2 w=m^2 u - F(u)$, but by the roots of $m^2 w=(m^2+a)u-au_1 - F(u)$. Equivalently, in the landscape $m^2 u - F(u)$, instead of looking at intersections with the horizontal curve  $m^2 w$, we should look at intersections with $m^2 w-a(u-u_1)$, a line with slope $-a$.

At the point
where this curve intersects first the landscape $m^2  u-F(u)$ we get a point $u_{s1}<u'_1$ where $\dot u$ first vanishes.
This defines the size $S_1=u_{s1}-u_1$ of the first {\it sub-avalanche}. If $\tau$ is small this usually occurs near 
the end, but if $\tau$ is larger the original avalanche (called main avalanche) is divided -- in size -- in a sequence of sub-avalanches
$S= \sum_\alpha S_\alpha$. The number of sub-avalanches in the main avalanche is finite for a smooth landscape, and infinite for
the continuous Brownian landscape. The total size $S=u'_1-u_1$ is however the same as for $a=0$, due to 
the Middleton theorem. For instance in the landscape of figure \ref{fig:Splitting}d, the main avalanche is divided into three large sub-avalanches, and
for the continuous Brownian landscape the intermediate segments are also divided into smaller sub-avalanches, at infinitum. 
 Figure \ref{fig:Splitting}  illustrates the correlation between the sub-avalanche structure (in $u$) and the 
realization of the random landscape, where larger hills favor the breakup into sub-avalanches. Note also 
that in intermediate regions where $\dot u$ is very small, $\tau h$ starts decreasing again (it decreases whenever
$\dot u< h$). The effective driving seen by the particle then becomes $m^2 w - a  \tau h$ and increases. 
This mechanism triggers a new sub-avalanche, and so on. 

To obtain the dynamics one must solve the equations (\ref{eq:EOMExp}), which we do below. 
For the standard ABBM model \cite{DobrinevskiLeDoussalWiese2012}, and in the mean-field theory 
of the elastic interface \cite{LeDoussalWiese2011,LeDoussalWiese2012a}, it was seen that an avalanche terminates with probability 1, i.e.\ $\du(t) = 0$ for $t>T$. This allowed defining and computing the distribution of avalanche durations \cite{LeDoussalWiese2011,DobrinevskiLeDoussalWiese2012}, and their average shape \cite{PapanikolaouBohnSommerDurinZapperiSethna2011,LeDoussalWiese2011,DobrinevskiLeDoussalWiese2012}.

In presence of retardation, and for an exponential kernel, the avalanche duration defined in the same way becomes infinite. 
Inside one avalanche, the velocity $\du(t)$ becomes zero infinitely often, but is then pushed forward again by the relaxation of the eddy-current pressure. 
Thus, an avalanche in the ABBM model with retardation splits into an infinite number of sub-avalanches, delimited by zeroes of $\du$. 
Each sub-avalanche has a finite size $S_i$ and duration $T_i$ with $S = \sum_i S_i$ (the same size as in the standard ABBM model), but $\sum_i T_i = \infty$.

Below we study in detail two limits:

In the slow-relaxation limit $\tau \gg \tau_m$  the duration of the largest sub-avalanches remains of order
$\tau_m$, while the total duration is of order $\tau$.  This leads to the estimate that 
the main avalanche breaks into $\sim \tau/\tau_m$ significant (i.e.\ non-microscopic) sub-avalanches.

For the fast-relaxation limit $\tau \ll \tau_m$ ($=1$ here) $h \approx \dot u$. The correction to the domain-wall velocity $\du(u)$ is small in this limit, and vanishes as $\tau/\tau_m \to 0$ (in contrast to the limit $\tau/\tau_m \to \infty$ discussed above). In fact, the correction due to retardation amounts to a rescaling of the velocity as $\du \to (1+a\tau)\du$. 

Of course, in presence of driving, the total duration is not strictly infinite since at some time-scale the driving will
kick in again, and lead to another main avalanche, itself again divided in sub-avalanches and so on. 
We can call that scale again $\tau_v$ but its precise value may differ from
the estimate for the case $a=0$. 

Thus one main property of the retarded ABBM model is that it leads to {\it aftershocks}, a feature not contained in
the standard ABBM model. The main avalanche is divided into a series of aftershocks (the sub-avalanches) 
which can be unambiguously defined and attributed to a main avalanche (which basically contains all of them) in the limit of small driving. This sequence of sub-avalanches is also called an avalanche cluster. The aftershocks are
triggered by the relaxation of the additional degree of freedom $h$. That in turn changes the force acting on the
elastic system. Relaxation and aftershock clustering have been recognized as important ingredients of an effective description of earthquakes; the present model is a solvable case in this class.  In some earthquake models considered previously,
relaxation was implemented in the disorder landscape itself \cite{Jagla2010,JaglaKolton2010,Jagla2011}. 
Here the relaxation mechanism is simpler, which makes it amenable to an analytic treatment. Note of course that at this stage it
is still rudimentary. First it is not clear how to identify the``main shock" among the
sequence of sub-avalanches; while there is indeed some tendency, see e.g.\ Fig.\ \ref{fig:Splitting}d, that the earliest sub-avalanche is the largest, this is not necessarily true.
Second, to account for features such as the decay of activity in time as a power-law   (Omori law  \cite{Scholz1998}) one
needs to go beyond the exponential kernel, to a power-law one. Finally, more ingredients
are needed if one wants to account for other features 
of realistic earthquakes, such as quasi-periodicity. 

\subsection{Stationary motion}
In the case where the driving velocity is constant, $w(t) = v t$, the distributions of the domain-wall velocity $\du$ and of the eddy current pressure $h$ become stationary. 
 The distribution of $\du$ for small $\du$ has a power-law form with an exponent depending on $v$,
\beq
\label{eq:PofDuStat}
P(\du) \sim \du^{-1+\frac{v}{v_c}}.
\eeq
There is no contribution $\sim \delta(\du)$. $v_c$ is a critical driving velocity, which separates several different regimes:

\begin{enumerate}
        \item For $ v > v_c$, the velocity $\du$ never becomes zero. 
        It is not possible to identify individual avalanches, one can say that there is a single infinite avalanche. 
        \item For $0 < v < v_c$, the velocity $\du$ vanishes infinitely often. The times $\{t_i | \du ({t_i}) = 0\}$ delimit individual (sub-)avalanches\footnote{Note that there are no finite-time intervals where the velocity $\du$ is identically zero, since else the probability distribution \eqref{eq:PofDuStat} would have a $\delta(\du)$ part. Thus the times $t_i$ are single points, which may, however, be spaced arbitrarily close. Scaling arguments suggest that the set of points \(\{ t_i\}\) has a  fractal dimension of $\frac{v}{v_c}$.}. Their durations $T_i := t_{i+1}-t_i$ and sizes $S_i = \int_{t_i}^{t_{i+1}}\rmd t' \du({t'})$ have distributions $P_v(T)$ and $P_v(S)$ depending on the driving velocity $v$. In section \ref{sec:AvStatDur} we compute $P_v(T)$ for the standard ABBM model and for a special case of the ABBM model with retardation. 
For sub-avalanches, starting at $\du_{\rm i}=0,$ and a fixed value of the eddy-current pressure $h_{\rm i}$, in the limit of small $a$ and $\tau=\tau_m$, we show that
\beq
\nn
P_v(T) \sim T^{-2+v+a h_{\rm i}}\quad\quad \text{for } \, T\to 0\,.
\eeq
In particular, the pure ABBM power-law exponent $P_v(T)\sim T^{-2+v}$ is not modified for the first sub-avalanche, starting at $h_{\rm i}=0$. Since the typical $h_{\rm i}$ goes to zero as $v\to 0$, we conjecture that the quasi-static exponents are still given by the mean-field values $P(S)\sim S^{-3/2}$, $P(T)\sim T^{-2}$.
\end{enumerate}
In sections \ref{sec:VelStat} and \ref{sec:FastRel}, we compute $v_c$ in several limiting regimes. For $ \tau \gg  \tau_m$, i.e.\ eddy-current relaxation  slow with respect to the domain-wall motion, we obtain in section \ref{sec:VelStat}
\beq
\nn
v_c =  \frac{\sigma}{\eta(m^2+a)} + \mathcal{O}\left( \tau_m/\tau \right).
\eeq
This means that slow eddy-current relaxation decreases the critical velocity. The stronger the eddy-current pressure $a$, the smaller $v_c$ becomes.
On the other hand, for $ \tau \ll \tau_m$, i.e.\ fast eddy-current relaxation, we obtain in section \ref{sec:FastRel}
\beq
\nn
v_c =  \frac{\sigma}{\eta m^2}\left[ 1 - a\frac{\tau}{\eta} + \mathcal{O}\left({ \tau/\tau_m}\right)^2 \right].
\eeq
Hence, fast eddy-current relaxation also decreases the critical velocity. However, the correction in this case is small and vanishes, as the time-scale separation  between $\tau_m$ and $\tau$ becomes stronger. 

The above regimes 1 and 2  do not change qualitatively compared to the standard ABBM model. This means that features like the power-law behavior of $P(\du)$ around $\du=0$ are robust towards changes in the dynamics, as long as it remains monotonous.

\begin{figure*}
\begin{minipage}{0.48\textwidth} 
\includegraphics[width=\textwidth]{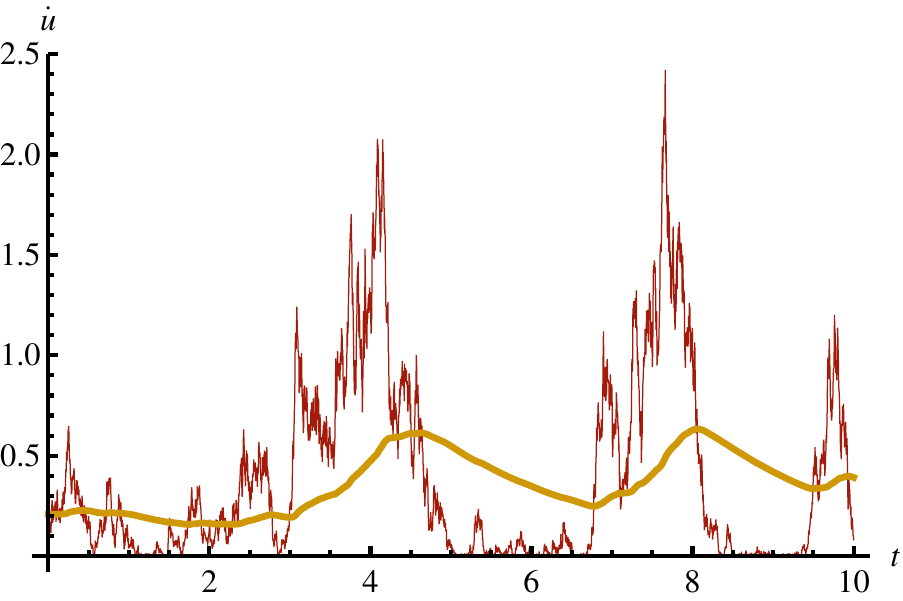}
{\small (a1) $\dw(t)=0.5$, $a=0$, $\tau=2$, $\eta=1$, $m=1$.}
\end{minipage}~
\begin{minipage}[l]{0.48\textwidth}\includegraphics[width=\textwidth]{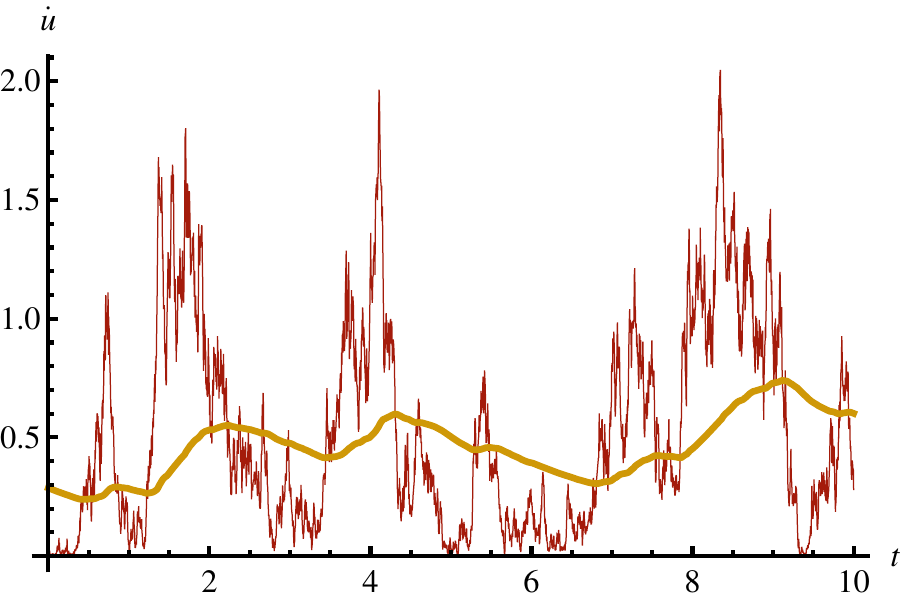}\\{\small (a2) $\dw(t)=0.5$, $a=1$, $\tau=2$, $\eta=1$, $m=1$.}
\end{minipage}
\begin{minipage}{0.48\textwidth} 
\includegraphics[width=\textwidth]{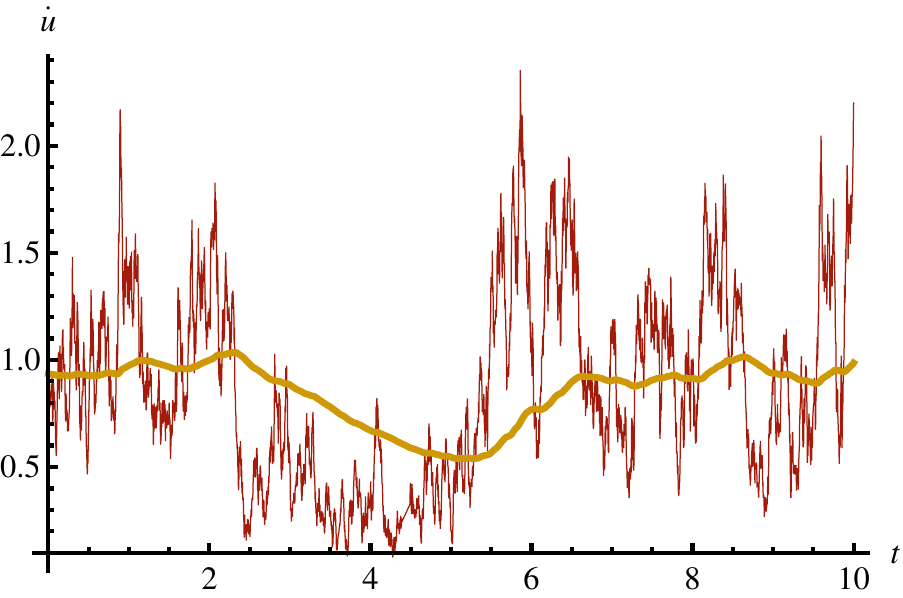}
{\small (b1) $\dw(t)=1.5$, $a=0$, $\tau=2$, $\eta=1$, $m=1$.}
\end{minipage}~
\begin{minipage}[l]{0.48\textwidth}
\includegraphics[width=\textwidth]{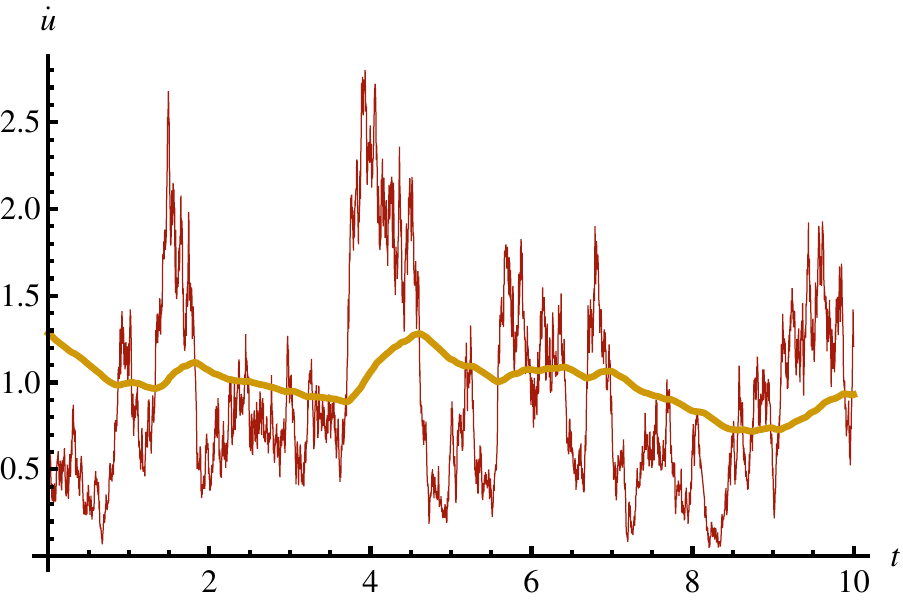}\\
{\small (b2) $\dw(t)=1.5$, $a=1$, $\tau=2$, $\eta=1$, $m=1$.}
\end{minipage}
\begin{minipage}{0.48\textwidth} 
\includegraphics[width=\textwidth]{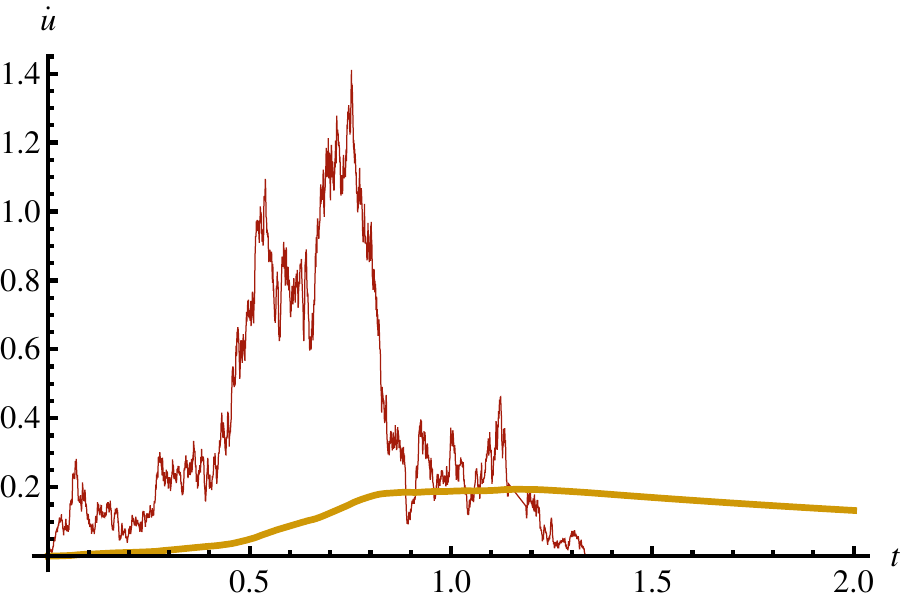}
{\small (c1) $\dw(t)=\delta(t)$, $a=0$, $\tau=2$, $\eta=1$, $m=1$.}
\end{minipage}~
\begin{minipage}[l]{0.48\textwidth}
\includegraphics[width=\textwidth]{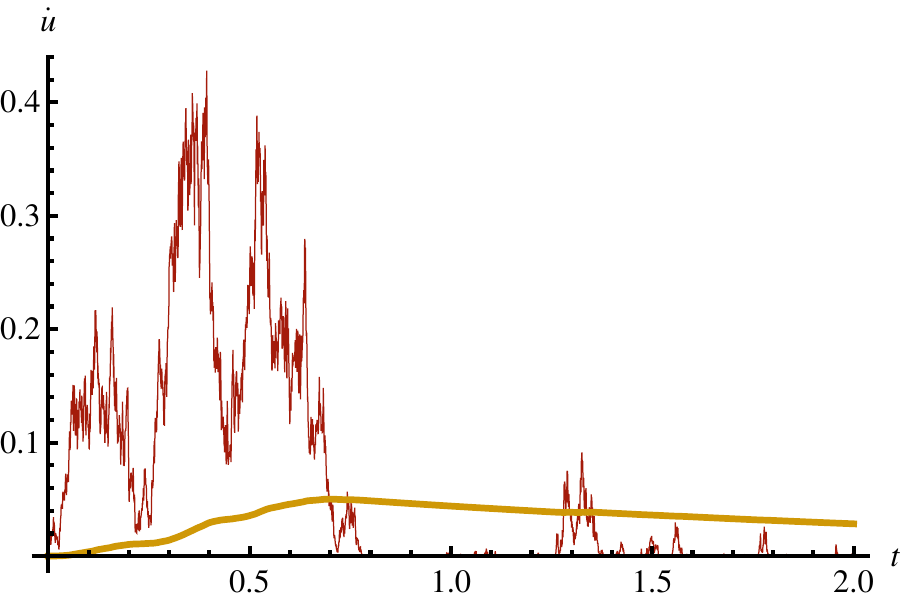}\\
{\small (c2) $\dw(t) = \delta(t)$, $a=1$, $\tau=2$, $\eta=1$, $m=1$.}
\end{minipage}
\caption{(Color online) Example trajectories for $\du(t)$ (thin, red) and $h(t)$ (thick, yellow), for various parameter values. The left column (a1),(b1),(c1) corresponds to the standard ABBM model ($a=0$), the right column to the model with retardation (here $a=1$). Figures (a), (b) correspond to stationary driving with a constant velocity, whereas the driving  in (c) has a kick at $t=0$. Observe that after a kick, $\du(t)$ in the standard ABBM model becomes zero permanently after a certain time, see figure (c1) , whereas in the ABBM model with retardation (c2) sub-avalanches restart infinitely often.} 
\label{fig:VelocitySims}
\end{figure*}

\subsection{Non-stationary driving: Response to a finite kick}

Instead of continuous driving, let us now perform a kick as defined in section~\ref{sec:Protocols}. 
In the standard ABBM model, like for the quasi-static driving discussed above, this leads to an avalanche on a time scale of order $\tau_m$, which terminates with probability 1. At some time $T$, the domain-wall velocity $\du$ becomes zero. The domain wall then stops completely, so that $\du(t)=0$ for all $t\geq T$. This gives an unambiguous definition for the size and duration of the non-stationary avalanche following a kick \cite{DobrinevskiLeDoussalWiese2012}. 
Formally, this behavior is seen by computing the probability $p_{\du(t) =0}$. It turns out that $p_{\du(t) =0}> 0$ for any $t>0$, and tends to $1$ as $t\to\infty$. The distribution $P(\du_t)$ (with $\du_{t}\equiv \du(t)$) for $t>0$ has a continuous part and a $\delta$-function part: $P(\du_t) = p_{\du_t =0} \delta(\du_t) + \mathcal{P}(\du_t)$ \cite{DobrinevskiLeDoussalWiese2012}.

In the ABBM model with retardation, the situation is different. We show in section \ref{sec:VelKick} that $p_{\du_t =0} = 0$ following a kick, so that the dynamics never terminates completely. If one defines the avalanche duration $T$ as $T=\min\{t|\du_s=0\,\text{for}\, s\geq t\}$, $T$ is infinite. This is also seen from the example trajectories in figure \ref{fig:VelocitySims}b.
However, the velocity intermittently becomes zero  an infinite number of times. Thus, the avalanche following a kick is split into an infinite number of {\em sub-avalanches}, just like a quasi-static avalanche discussed above.

On the other hand, the sub-avalanches become smaller and smaller with time. In section \ref{sec:AvSizes}, we show that the total avalanche size $S:= \int_0^\infty \rmd t \, \du_t$ following a kick of size $w_0$ is finite and distributed according to the same law as in the standard ABBM model \cite{LeDoussalWiese2009},
\beq
P_{w_0}(S) = \frac{w_0 }{2
   \sqrt{\pi \sigma} S^{\frac{3}{2}}}e^{\textstyle -\frac{(w_0-m^2 S)^2}{4 \sigma S}}.
\eeq
This result holds independently of the memory kernel $f$.
For infinitesimal kicks, $w_0 \to 0$, $P_{w_0}(S)$ becomes the distribution of quasi-static avalanche sizes discussed above.

The disorder-averaged velocity $\overline{\du_t}$ following the kick decays smoothly. In the standard ABBM model, the decay is exponential \cite{DobrinevskiLeDoussalWiese2012}. With retardation, we show in section \ref{sec:MeanVel} that 
 the dependence of $\overline{\du_t}$ on $t$ 
is directly related to the form of the memory kernel $f$.

Another interesting observable is the mean avalanche shape. Conventionally, it is defined at stationary driving for a \textit{sub-avalanche}: One takes two neighboring zeroes $\du(0)=0$ and $\du(T)=0$ which delimit a (\textit{sub}-)avalanche of duration $T$. The mean  avalanche shape is then the average of the domain-wall velocity $\overline{\du(t)}$ as a function of time, in the ensemble of all such (\textit{sub}-)avalanches of duration $T$. It has been realized \cite{ZapperiCastellanoColaioriDurin2005} that the skewness of this shape provides information on the relaxation of eddy currents.

However, this definition is hard to treat analytically. Instead of considering the mean (sub-)avalanche shape at a constant duration, we discuss the mean shape of a complete avalanche (consisting of infinitely many sub-avalanches, with infinite total duration) of a fixed size $S$, triggered by a step in the force at $t=0$. In section \ref{sec:AvShapeFixedSize} we give an explicit expression for this shape at fixed size, for exponential eddy-current relaxation. We show how it reflects the time scale of eddy-current relaxation.

The phenomenology discussed here is expected to be similar if instead of a kick at $t=0$, one takes some arbitrary driving $w_t$ for $t<0$, which stops at $t=0$ so that $\dw_{t>0} = 0$.

We see that the non-stationary relaxation properties of the retarded ABBM model differ qualitatively from those of the standard ABBM model. They provide a more sensitive way of distinguishing experimentally the effect of eddy currents  than stationary observables at finite velocity, and allow one to identify the form of the memory kernel $f$. In the following sections, we provide quantitative details underlying this picture.

\section{Solution of the retarded ABBM model\label{sec:RetSol}}

In this section, we apply the methods developed in \cite{LeDoussalWiese2011,DobrinevskiLeDoussalWiese2012,LeDoussalWiese2012a,LeDoussalPetkovicWiese2012} to obtain the following exact formula for the generating functional of domain-wall velocities,
\beq
\label{eq:RetSolution}
\overline{e^{\int_t \lambda_t \du_t\, \rmd t}} = e^{ m^2 \int_t \dw_t \tu_t \, \rmd t}.
\eeq
It is valid for an arbitrary monotonous driving $\dw_t \geq 0$, where $\tu_t$ is the solution of the following
nonlocal instanton equation,
\begin{align}
&\eta \partial_t \tu(t) - (m^2 + a) \tu(t) + \sigma \tu(t)^2 - a\! \int^{\infty}_t \rmd s\, f'(s-t) \tu(s) \nn\\&~= - \lambda(t),
\label{eq:RetInstanton}
\end{align}
with boundary condition $\tu(\infty)=0$. The important observation that allows such an exact formula is that for monotonous driving, 
the motion in the ABBM model with retardation is still monotonous, as in the standard ABBM model (see appendix \ref{sec:Monot})
as discussed above.

To prove \eqref{eq:RetSolution} we apply the same series of arguments as in the absence of retardation\cite{LeDoussalWiese2011,DobrinevskiLeDoussalWiese2012,LeDoussalWiese2012a}.
Taking one derivative of Eq.~\eqref{eq:EOMU} gives a closed equation of motion for $\du(t)$, instead of $u(t)$:
\begin{align}
\nn
&\eta \partial_t \du(t) + a \du(t) + a \int_{-\infty}^t \rmd s f'(t-s) \dot u(s) \\
\label{eq:EOMUd}
&= \sqrt{\du(t)}\xi(t) + m^2 \left[\dw(t) - \du(t)\right].
\end{align}
$\xi(t)$ is a Gaussian white noise, with $\overline{\xi(t) \xi({t'})} = 2\sigma \delta(t-t')$.  The term $\sqrt{\du(t)}$ comes from rewriting the {\em position-dependent} white noise in terms of a {\em time-dependent} white noise,
\begin{align}
&&\overline{\xi(u(t))\xi(u(t'))} &= 2\sigma \delta(u(t)-u(t')) = \frac{2\sigma}{\du(t)}\delta(t-t') \nn \\
&\Rightarrow & \xi(u(t)) &= \frac{1}{\sqrt{\du(t)}}\xi(t) \nn \\
&\Rightarrow & \partial_t F(u(t)) &= \du(t) \xi(u(t)) = \sqrt{\du(t)}\xi(t). \label{noise1}
\end{align}
This uses crucially the monotonicity of each trajectory.

Using the Martin-Siggia-Rose method, we express the generating functional for solutions of \eqref{eq:EOMUd} as a path integral,
\begin{align}
\nn
\overline{e^{\int_t \lambda_t \du_t}} = & \int \mathcal{D}[\du,\tu] e^{-S[\du,\tu] + \int_t \lambda_t \du_t} \\
\nn
S[\du,\tu] = \int_t \tu_t &  \Big[\eta \partial_t \du_{t} + a \du_{t} + a \int_{-\infty}^t \rmd s\, f'(t-s) \dot u_{s}  \\
&  - m^2 \big(\dw_t - \du_{t}\big)  \Big] - \sigma \int_t \tu_t^2 \du_t.
\label{eq:MSRAction}
\end{align}
For compactness, we have noted time arguments via subscripts. We will use this notation from now on when convenient. 

As in the standard ABBM model, the action \eqref{eq:MSRAction} is linear in $\du$. Thus, the path integral over $\du$ can be evaluated exactly. It gives a $\delta$-functional enforcing the instanton equation \eqref{eq:RetInstanton}. The only term not involving $\du$ in the action is $m^2\int_t \tu_t \dw_t$, which yields the result \eqref{eq:RetSolution} for the generating functional. For more details, see section II in \cite{DobrinevskiLeDoussalWiese2012} and sections II B-E in \cite{LeDoussalWiese2012a}.

Similarly to the discussion in \cite{DobrinevskiLeDoussalWiese2012,LeDoussalWiese2012a} the solution \eqref{eq:RetSolution} generalizes 
to an elastic interface with $d$ internal dimensions in a Brownian force landscape (i.e.\ elastically coupled ABBM models). There is indeed
a simple way to introduce retardation in that model to satisfy the monotonicity property. We will not study this extension here. 

For the case of exponential relaxation, $ f(x)=e^{-x/\tau}$, Eq.~\eqref{eq:EOMUd} can be simplified to  a set of two local Langevin equations for the velocity $\du$ and the eddy-current pressure $h$:
\begin{align}
\eta \partial_t \du_{t} &= \sqrt{\du_t}\xi_t + m^2 [\dw_{t}- \du_{t}] -a\left( \du_{t}
- h_{t}\right) \\
\tau \partial_t h_{t} &= \du_{t} -h_{t}.
\label{eq:EOMUdExp}
\end{align}
The action for this coupled system of equations is
\begin{align*}
S[\du,\tu] =& \int_t \Big\{\tu_t \Big[ \eta \partial_t \du_{t}+ a\big( \du_{t} -h_{t}\big) + m^2 \big(\du_{t} - \dw_{t}\big)  \Big] \\
&~~~~~ - \sigma  \tu_t^2 \du_t + \tilde h_t \big( \tau\partial_t h_t+h_t-\du_t \big) \Big\}\ .
\end{align*}
This action is linear in $\dot u_t$ and $h_t$. Thus, integrating over these fields gives $\delta$-functionals enforcing a set of two local instanton equations for $\tu_t$ and $\tih_t$,
\begin{align}
\label{eq:RetInstantonExpTu}
\eta \partial_t \tu_{t} - \left(m^2 + a\right) \tu_{t} + \sigma \tu_{t}^2 + \tih_{t} &=
- \lambda_{t}, \\
\label{eq:RetInstantonExpTh}
\tau\partial_t \tih_{t} -\tih_{t} + a \tu_{t} &=  - \mu_{t}.
\end{align}
 We then obtain the generating functional for the joint distribution of velocity $\du$ and
eddy-current pressure $h$,
\beq
\label{eq:RetSolutionExp}
\overline{e^{\int_t \left(\lambda_t \du_t + \mu_t h_t\right)\, \rmd t}} =
e^{m^2 \int_t \dw_t \tu_t \, \rmd t}\ ,
\eeq
in terms of the solution to these two instanton equations. It 
 reduces to \eqref{eq:RetSolution} for $\mu_t=0$.

Now, the remaining difficulty for  arbitrary observables is to obtain sufficient information on the solutions of \eqref{eq:RetInstantonExpTu}
and \eqref{eq:RetInstantonExpTh} with the corresponding source terms. We shall see that this is more difficult than in the standard ABBM model, but can be done for certain observables and certain parameter values.

\subsection{Dimensions and scaling\label{sec:Dim}}
Before we proceed to compute observables, let us discuss the scaling behaviour of our model, and determine the number of free parameters. The mass $m$ can be eliminated by dividing both sides of \eqref{eq:EOMUd} by $m^2$,
\begin{align}
\nn
&\frac{\eta}{m^2} \partial_t \du(t) + \frac{a}{m^2} \du(t) + \frac{a}{m^2} \int_{-\infty}^t \rmd s \partial_t f(t-s) \du(s) \\
\label{eq:EOMUdS1}
&= \frac{1}{m^2}\sqrt{\du(t)}\xi(t) + \dw(t) - \du(t)
\end{align}
 The time derivative $\frac{\eta}{m^2} \partial_t \du(t)$ shows that there is a natural time scale $\tau_m=\eta/m^2$ so that  $t = t' \tau_m$, where $t'$ is dimensionless. The nonlinear term $\frac{1}{m^2}\sqrt{\du(t)}\xi(t)$ shows that there is a natural length scale $S_m=\frac{\sigma}{m^4}$, so that $u = S_m u'$, where $u'$ is dimensionless. We thus rescale velocities as
\beq
\du(t) = \frac{\sigma}{\eta m^2} \du'(t')\ , \quad \dw(t) = \frac{\sigma}{\eta m^2} \dw'(t'),
\eeq
using the natural unit of velocity $v_m = S_m/\tau_m$.
Multiplying with $m^2 \eta/\sigma$, we get the equation
\begin{align}
\label{eq:EOMUdS2}
&\partial_{t'} \du'(t') + \frac{a}{m^2} \du'(t') + \frac{a}{m^2}\int_{-\infty}^{t'} \rmd s'\, \partial_{t'}f(t'-s') \du'(s') \nn\\
&= \sqrt{\du'(t')}\xi'(t') + \dw'(t') - \du'(t'),
\end{align}
where the noise is now $\left<\xi'(t_1) \xi'(t_2)\right>=2 \delta(t_1-t_2) $. Effectively, for the dynamics in terms of the primed variables we have $m=\sigma=\eta=1$
(i.e.\ we have fixed the units of time and space so that $\tau_m=S_m=1$). 

For the standard ABBM model, $a=0$, and Eq.~\eqref{eq:EOMUdS2} is a dimensionless equation without any free parameters. To describe a signal $\du(t)$ produced by the standard ABBM model, it thus suffices to fix the velocity (amplitude) scale $v_m = \frac{\sigma}{\eta m^2}$, and the time scale $\tau_m = \frac{\eta}{m^2}$.

For the ABBM model with retardation, we have an additional time scale $\tau$, on which the memory kernel $f(t-s)$ in \eqref{eq:EOMUd} changes. The ratio of $\tau$ to the time scale of domain-wall motion  $\tau_m=\frac{\eta}{m^2}$ is a dimensionless parameter $\tau' := \tau /\tau_m$. Eq.~\eqref{eq:EOMUdS2} also contains a second dimensionless parameter $a':=\frac{a}{m^2}$, which gives the strength of the eddy-current pressure, as compared to the driving $\dw$ by the external magnetic field. We thus remain with \textit{two} dimensionless parameters $\tau'$ and $a'$, which cannot be scaled away.

From now on, we will use the rescaled (primed) variables only. To simplify the notation, we drop all primes; we thus remain with the dimensionless equation of motion
\begin{align}
\label{eq:EOMUdS3}
&\partial_{t} \du(t) + a \du(t) + a\int_{-\infty}^{t} \rmd s\, \partial_{t}f(t-s) \du(s) \\
\nn
&= \sqrt{\du(t)}\xi(t) + \dw(t) - \du(t).
\end{align}
This amounts to setting $m=\sigma=\eta=1$ in the original equation of motion, i.e.\ to working in the
natural units for the ABBM model without retardation.

\section{Measuring the memory kernel $f$}\label{sec:MeanVel}
First, we discuss how the function $f$ in equation \eqref{eq:EOMU} can be measured in an experiment or in a simulation. This allows verifying the validity of the exponential approximation \eqref{eq:EOMZapperi2}.
We consider the mean velocity $\overline{\du(t)}$ at $t>0$ following a kick by the driving field $w(t)$ at $t=0$, i.e.\ $\dw(t) = w_0 \delta(t)$. Our claim is that its Fourier transform and the Fourier transform of the memory kernel $f$ are related via 
\beq
\label{eq:MeanVel}
u_\omega :=\int_0^\infty \rmd t \, e^{-i\omega t}\overline{\du(t)} = \frac{w_0}{m^2 +i \omega\left[ \eta + a f(\omega)\right]},
\eeq
where $f(\omega) := \int_0^\infty \rmd t\,e^{-i\omega t}f(t)$.

To show this, we apply \eqref{eq:RetSolution} to express the mean velocity at time $t_0>0$ as
\begin{align}
\nn
\overline{\du(t_0)} &= \partial_{\lambda}\big|_{\lambda = 0} \overline{e^{\lambda \du(t_0)}} = \partial_{\lambda}\big|_{\lambda = 0} e^{\int_t \rmd t \,\tu(t) \dw(t)} \\
\label{eq:MeanVel1}
& = \partial_{\lambda}\big|_{\lambda = 0} e^{w_0 \tu(t=0;t_0)}\ .
\end{align}
The function $\tu(t)$ is the solution of \eqref{eq:RetInstanton} with $\lambda(t) = \lambda \delta(t-t_0)$. Since above we only need  the term of order $\lambda$, and $\tu(t;t_0)$ is of order $\lambda$ itself, the nonlinear term in \eqref{eq:RetInstanton} can be neglected. In other words, the disorder does not influence the mean velocity $\overline{\du(t_0)}$, and to obtain $\tu(t;t_0)$, it suffices to solve the linear equation
\begin{align}
\nn
&\eta \partial_t \tu(t;t_0) - (m^2 + a) \tu(t;t_0) - a \int^{\infty}_t \rmd s\, f'(s-t) \tu(s;t_0) \\
& = - \lambda\delta(t-t_0).
\label{eq:MeanVel2}
\end{align}
Its solution is a function of the time difference $t-t_0$ only, $\tu(t;t_0) = \tu(t-t_0)$, which can be obtained  by taking the Fourier transform $\tu(\omega) := \int_{-\infty}^\infty \rmd \tau\,e^{-i\omega \tau} \tu(\tau)$ as
\begin{align}
\nn
&(i\omega\eta-m^2-a)\tu(\omega) -a\tu(\omega)\left[-i\omega f(-\omega)-1\right] = -\lambda \\
 &\Rightarrow ~~~ \tu(\omega) = \frac{\lambda }{-i\omega \eta + m^2 - a i\omega f(-\omega)}.
\end{align}
Here $f(\omega) = \int_0^\infty \rmd t \, e^{-i\omega t}f(t)$ is the Fourier transform of the memory kernel.
Inserting this relation into \eqref{eq:MeanVel1}, the Fourier transform of the mean velocity after a kick is
\begin{eqnarray}
\nn
\int_0^\infty \rmd t_0\, e^{-i\omega t_0}\overline{\du(t_0)} &=& w_0 \int_{-\infty}^\infty \rmd t_0\,e^{-i\omega t_0}\tu(0-t_0)\nn\\& =& w_0 \tu(-\omega),
\end{eqnarray}
which then gives \eqref{eq:MeanVel}, as claimed. In fact, it is easy
to see from (\ref{eq:MeanVel1}) that a more general relation holds for a kick of
arbitrary shape,
\be  \label{gen}
\overline{\du_\omega} := \int_0^\infty \rmd t\, e^{-i\omega t}\overline{\du_t} = \frac{w_0(\omega)}{m^2 +i \omega\left[ \eta + a f(\omega)\right]},
\ee 
where $w_0(\omega) := \int_0^\infty \rmd t\,e^{-i\omega t} \dot w(t)$.

This relation allows one to obtain, at least in principle, the memory kernel $f(t)$ by measuring $\overline{\du(t)}$ following a kick. 
This permits to verify the validity of the exponential approximation \eqref{eq:EOMExp} experimentally.
It also allows to test the validity of the ABBM model. Indeed, while (\ref{gen}) at small $w_0(\omega)$ is
simply a linear response, the fact that it holds for a kick of arbitrary amplitude is a very distinctive property
of the ABBM model. Alternatively, it may allow to determine the frequency range in which the model provides
a good description of the experiment.

\section{ The slow-relaxation limit $\frac{\eta}{m^2} \ll \tau$\label{sec:QSLim}}
In order to go beyond the mean velocity and see the influence of disorder, one needs to solve the instanton equation \eqref{eq:RetInstanton} including the nonlinear term. 
Even in the special case of exponential relaxation, where \eqref{eq:RetInstanton} reduces to the local  equations \eqref{eq:RetInstantonExpTu} and \eqref{eq:RetInstantonExpTh}, their solution is complicated.
However, we can analyze the latter in the slow-relaxation limit $\tau_m = {\eta/m^2} \ll {\tau}$. In this limit,  the relaxation of the domain wall to the next (zero force) metastable state, occurring on a time scale  $\tau_m$, is much faster than the relaxation of eddy currents (occurring on a time scale  $\tau$). Using the expressions for the relaxation times derived in \cite{ZapperiCastellanoColaioriDurin2005}, one sees that this is the case for very thick or very permeable samples\footnote{In the pure ABBM model, the small-dissipation limit $\eta \to 0$ is equivalent -- up to a choice of time scale -- to the limit of quasi-static driving $v\to 0^+$. However, these two limits are different for the retarded ABBM model which we discuss here.}.
To simplify the expressions, we rescale $\du$ as discussed in section \ref{sec:Dim}. This amounts to setting $m=\sigma=\eta=1$. Thus, the time scale of domain-wall motion becomes $\tau \gg 1$.

In the following sections, we will compute stationary distributions of the eddy-current pressure $h_t$ and domain-wall velocity $\du_t$ at constant driving $w_t = vt$, as well as their behaviour following a kick. A similar calculation for position differences at constant driving velocity is relegated to appendix \ref{sec:QSPositions}.

\subsection{Stationary distribution of eddy-current pressure\label{sec:QSEddyStat}}

Using \eqref{eq:RetSolutionExp}, the generating functional for the eddy-current pressure $h=h(t=0)$,  at constant driving $w_t = v t$ is
\beq
\label{eq:QSStatGenFct} 
\overline{e^{\mu h}} = e^{v\int_t \tu(t)}.
\eeq
$\tu(t)$ is obtained from the instanton equations \eqref{eq:RetInstantonExpTu}, \eqref{eq:RetInstantonExpTh} with the sources $\mu(t) = \mu \delta(t)$, and $\lambda(t) = 0$.  From \eqref{eq:RetInstantonExpTh}, one sees that $\tih(t)$ evolves on a time scale $s = t/\tau$. On this scale, both $\tih(t)$ and $\tu(t)$ have a finite limit for $\frac{\eta}{m^2\tau} \to 0$. In this limit they are related via
\begin{align}
\tih(s) &= -\tu(s)^2 + (1+a)\tu(s), \label{32}\\
\tu(s) &= \frac{1}{2} \left(a+1-\sqrt{(a+1)^2-4 \tih(s)}\right).
\end{align}
The equation \eqref{eq:RetInstantonExpTh} for $\partial_s \tih(s)$ reads 
\beq
\partial_s \tih(s) =  \tih(s) - a \tu(s)\ .
\eeq Replacing $\tih(s)$ on both sides of this equation using Eq.~\eqref{32} yields a  closed equation for $\tu(s)$,
\beq
\label{eq:QSInstanton}
[1+a-2\tu(s)]\partial_s \tu(s) = \tu(s)-\tu(s)^2.
\eeq
The boundary condition at $s=0$ is fixed by the source, $\mu( s) = \mu \delta(t)  = \frac{\mu}{\tau} \delta(s) =: \mu_r \delta(s)$ (note $\tu(s>0)=\tih(s>0)=0$ by causality):
\beq
\label{eq:QSStatIC}
\tih(0) = \mu_r \Rightarrow \tu(0) = \frac{1}{2} \left(a+1-\sqrt{(a+1)^2-4 \mu_r}\right).
\eeq
Using Eq.\ \eqref{eq:QSInstanton}, we can now compute the generating functional \eqref{eq:QSStatGenFct},
\begin{align}
\nonumber
\int_{-\infty}^0 \tu(t) \, \rmd t &=  \tau\int_{-\infty}^0 \tu(s) \, \rmd s=\tau \int_0^{\tu(0)} \frac{\tu \,\rmd \tu }{\partial_s \tu(\tu)} \\
\nn
&=  \tau\int_0^{\tu(0)} \frac{ (1+a-2\tu(s)) \rmd \tu}{1-\tu}\\
\label{eq:QSStatInt}
&=  \tau \left[2\tu(0) +(1-a) \ln \left(1-\tu(0)\right)\right].
\end{align}
Inserting this result into Eq.~\eqref{eq:QSStatGenFct} we get
\beq
\label{eq:QSStatLT}
\overline{e^{\mu h_0}} = e^{2 v_r\tu_0(\mu_r)}\left[1-\tu_0(\mu_r)\right]^{(1-a)v_r},
\eeq
 where $\tu_0(\mu_r)\equiv \tu(0)$ is given by \eqref{eq:QSStatIC} and we have defined a rescaled velocity $v_r:=v\tau$
(i.e.\ the driving length during the relaxation time)

The stationary distribution of $h_r := \tau h$, obtained by inverting the Laplace transform, is
           \begin{align}
\nonumber
P(h_r) =& \frac{v_r}{\sqrt{\pi
   }} 2^{\frac{1}{2} (a-1) v_r-1} h_r^{\frac{1}{2} \left[(a-1) v_r-3\right]} \times \\
\nonumber
& \times \exp
   \left\{h_r - \frac{\left[2v_r - (3+a)h_r\right]^2}{8h}\right\} \times \\
\nonumber
        & \times \left\{(a-1) \sqrt{2h_r} D_{(1-a)v_r-1}\left[\frac{(1-a) h_r+2 v_r}{\sqrt{2h_r}}\right] \right. \\
\label{eq:QSStatH}
        & \quad \left.+2 D_{(1-a)v_r}\left[\frac{(1-a) h_r+2 v_r}{\sqrt{2h_r}}\right]\right\}.
\end{align}
$D$ is a parabolic cylinder function \cite{AbramowitzStegun1965}.

For small $h_r$, the distribution \eqref{eq:QSStatH} behaves as
\bea
\nn
P(h_r)&=&\frac{1}{\sqrt{\pi}} e^{-\frac{((1+a)h_r-2 v_r)^2}{4 h_r}} v_r\,h_r^{-\frac{3}{2}}\left(\frac{h_r}{v_r}\right)^{(a-1)v_r}\nn\\
&& ~~~~~\times \left(1+\mathcal{O}(\sqrt{h_r})\right).
\eea
Thus, there is a small-$h$ cutoff (due to the exponential term) and a power-law regime with a non-trivial exponent, $h_r^{-\frac{3}{2}+v_r(a-1)}$.
Note that the above results hold in the double limit $v \to 0$ and $\tau \to \infty$ with $v_r = v \tau$ fixed. Restoring units
this is $v_r = v \tau/S_m = \tau/\tau_v$ fixed, with both $\tau,\tau_v \gg \tau_m$, hence $v_r$ compares the two longest time scales, the driving time scale
and the eddy-relaxation time scale.

In the limit where the driving is slow  compared to the eddy-relaxation time scale,
$v_r \to 0^+$, the stationary distribution \eqref{eq:QSStatH} takes the form
of a limiting (un-normalized) density $\rho(h_r)$ proportional to
\begin{align}
\nn
&\partial_{v_r} \big|_{v_r=0} P(h_r) =   \\
\nn
&   \frac{e^{-a h_r}}{2 h_r}  (a-1)
   \left\{\text{erf}\left[\frac{1}{2} (a-1)
   \sqrt{h_r}\right]+1\right\}+\frac{e^{-\frac{1}{4} (a+1)^2 h_r}}{\sqrt{\pi } h_r^{3/2}}.
\end{align}
In this limit, the small-$h_r$ behaviour is a pure power law,
\beq \label{aga} 
\partial_v \big|_{v=0} P(h_r) = \frac{1}{\sqrt{\pi} h_r^\frac{3}{2}} + \frac{a-1}{2h_r} + \mathcal{O}(h_r^{-\frac{1}{2}})\ .
\eeq
We note the resemblance of the tail of the distribution of $h$ and the one of the size $S$ in the usual ABBM model
with the $3/2$ exponent in both cases. If we assume that during avalanches (sub-avalanches) $\dot u$ varies much
faster than the relaxation time $\tau$ (i.e.\ on scales $\tau_m \ll \tau$) we can rewrite $h(t) \sim \sum_{\alpha, t_\alpha<t} S_\alpha e^{- (t-t_\alpha)/\tau}$ 
where sub-avalanche $\alpha$ occurs at $t_\alpha$. Schematically $h(t)$ integrates avalanche sizes occurring in a time window  of order $\tau$,
which could account for the similarity.

\subsection{Eddy-current pressure following a kick\label{sec:QSEddyKick}}

Still in the limit $\frac{\eta}{\tau m^2}\to 0$, let us now discuss a non-stationary situation: The dynamics following a kick of size $w_0$ at $t=t_0<0$, $\dw_t = w_0 \delta(t-t_0)$. Using \eqref{eq:RetSolutionExp}, the generating function for the eddy-current pressure at time $0$ is given by
\beq
\label{eq:QSGenFctKick}
\overline{e^{\mu h_0}} = e^{w_0 \tu_{t_0}},
\eeq
where $\tu_t$ is the solution of \eqref{eq:RetInstantonExpTu}, \eqref{eq:RetInstantonExpTh} with the sources $\lambda(t)=0$, $\mu(t) = \mu \delta(t)$, as in the previous section. Now, we need its time-dependence and not just the total integral. An implicit solution for $\tu(t)$ at $t<0$ is obtained from Eq.~\eqref{eq:QSInstanton}:
\begin{align}
\nn
\frac{t}{\tau} &= \int_{\tu_0}^{\tu_t} \frac{\rmd \tu}{\tau\partial _t \tu(\tu)} \\
\label{eq:QSKickTu}
&= (1+a)\ln\frac{\tu_t}{\tu_0} + (1-a)\ln\frac{1-\tu_t}{1-\tu_0}.
\end{align}
$\tu_0$ is fixed by \eqref{eq:QSStatIC}.  As in the previous section, we define a rescaled time $s := \frac{t}{\tau}$. There is no expression in closed form  for $\tu_s$ for general $a$, but for specific values one  obtains simple expressions (see table \ref{tbl:QSDecayUt}).
\begin{table}
\begin{tabular}{|l|l|}
\hline
$~a = 0~$ & $\rule[-3mm]{0mm}{8mm}\displaystyle~\tu_s = \frac{1}{2}\left(1-\sqrt{1-4e^{s}\tu_0(1-\tu_0)}\right)$ \\ \hline
$~a = 1$ & $\rule[-3mm]{0mm}{8mm}\displaystyle~\tu_s = e^{\frac{s}{2}}\tu_0$ \\ \hline
$~a = 3$ & ~$\rule[-4mm]{0mm}{10mm}\displaystyle \tu_s =\displaystyle \frac{\tu_0^2 e^{s/2}-\tu_0 e^{s/4} \sqrt{\tu_0^2 e^{s/2}-4 \tu_0+4}}{2 (\tu_0-1)}$~ \\ \hline
\end{tabular}
\caption{Some particular solutions of the implicit equation \eqref{eq:QSKickTu} describing the eddy-current pressure following a kick, in the limit $\frac{\eta}{m^2 \tau}=0$.}
\label{tbl:QSDecayUt}
\end{table}
In the case $a=1$, the solution is particularly simple. Equation \eqref{eq:QSGenFctKick} gives
\beq
\overline{e^{\mu h(0)}} = e^{w_0 e^{t_0/(2\tau)}\left(1-\sqrt{1-\mu_r}\right)}.
\eeq
Taking the inverse Laplace transform, one obtains the distribution of eddy-current pressure $h_r := \tau h(0)$ after a kick of size \(w_{0}\) at time $t_0<0$,
\beq
P( h_r)=\frac{w_0}{2 \sqrt{\pi } h_r^{3/2}}\exp\bigg({\frac{t_0}{2\tau}-\frac{[e^{t_0/(2\tau)}w_0-2
   h_r]^2}{4 h_r}}\bigg).
\eeq
The average pressure $\overline{h_r}=e^{t_0/(2 \tau)} w_0/2$ decays exponentially with time. Note that
the limit $t_0=0^-$ leads to a non-trivial $P(h_r)$ which should hold within the
entire matching region $\tau_m \ll |t_0| \ll \tau$.

For $a \neq 1$, we did not obtain an exact solution. 
However, for any $a$, taking the limit $\mu \to -\infty$, or equivalently $\tu_0 \to -\infty$, Eq.~\eqref{eq:QSKickTu} shows that $\tu_t \to -\infty$. This implies that the probability to find zero pressure, $p_{h=0} = \lim_{\lambda \to -\infty} e^{w_0 \tu_{t_0}} = 0$. Thus, after a kick at $t=t_0$, there is no time $T > 0$ such that $h_t=0$ for {\em all} $t>T$; the eddy current pressure never stops. A similar discussion for the domain-wall velocity follows in section \ref{sec:VelKick}.

Another interesting statement can be made regarding the time integral of the eddy-current pressure following a kick. 
From Eq.~\eqref{eq:EOMExp} it must equal the total avalanche size (integrating this equation and using that for a kick $h(0)=h(\infty)=0$), i.e.\ $\int_0^{\infty} h(t) \rmd t = S$.

\subsection{Distribution of instantaneous velocities\label{sec:VelStat}}
The distribution of instantaneous velocities $P(\du)$ at stationary driving is one of the simplest observables that can be determined from an experimental Barkhausen signal. For the standard ABBM model, it has been obtained in \cite{AlessandroBeatriceBertottiMontorsi1990}. For a $d$-dimensional elastic interface driven quasi-statically through short-range correlated disorder, this form is modified by universal corrections below the critical dimension $d_{\rm c}$. These corrections have been computed using the functional renormalization group to one loop in $\epsilon = d_{\rm c}-d$ in \cite{LeDoussalWiese2011,LeDoussalWiese2012a}.
Using \eqref{eq:RetSolutionExp}, the generating function of the instantaneous velocity, for constant driving $w_t = v t$,
is\beq
\label{eq:QSVelGenFct}
\overline{e^{\lambda\du}} = e^{v \int_t \tu_t}.
\eeq
Now $\tu_t$ is the solution of the instanton equations \eqref{eq:RetInstantonExpTu}, \eqref{eq:RetInstantonExpTh} with the sources $\lambda(t) = \lambda \delta(t)$, $\mu(t)=0$. 

To obtain the leading-order velocity distribution for $\tau \gg 1$, we need to solve \eqref{eq:RetInstantonExpTu} to order $\tau^{-1}$. The solution $\tu_t$, $\tih_t$ has two time scales, which are well-separated in the $\tau \to \infty$ limit: $t \propto 1$ and $t \propto  \tau$. We thus introduce 
\beq 
s := t/\tau \label{46},
\eeq 
and assume the scaling
\begin{align}
\label{eq:QSVelInstB}
\tu(t) &=: \tu^{(b)}\left(t\right) + \tau^{-1} \tu^{(b)}_1(t) + \mathcal{O}(\tau^{-2}),\quad |t| \propto 1 \\
\tih(t) &=: \tau^{-1}\tih^{(b)}\left(t\right) + \mathcal{O}(\tau^{-2}),\quad |t| \propto 1 \\
\label{eq:QSVelInstA}
\tu(t) &=: \tau^{-1} \tu^{(a)}\left(s\right) + \mathcal{O}(\tau^{-2}),\quad |t| \propto \tau \\
\tih(t) &=: \tau^{-1} \tih^{(a)}\left(s\right) + \mathcal{O}(\tau^{-2}),\quad |t| \propto \tau.
\end{align}
Physically, the first regime $|t|\propto 1$ is the regime where the eddy currents have not yet built up ($\tih \ll \tu$) and are negligible. Hence the instanton is, up to a parameter change, identical to that of the standard ABBM model. The second regime is the regime of quasi-static relaxation of the eddy currents built up during the first stage. In that regime the instanton will be related to the instanton for the eddy-current relaxation discussed in the previous section.

The source terms enforce the boundary conditions $\tu^{(b)}(0) = - \lambda$, $\tih^{(b)}(0) = 0$.
 We  now construct $\tu^{(a,b)}$ and $\tih^{(a,b)}$ in turn.

\subsubsection{Boundary layer: $|t|\propto 1$\label{sec:VelBdry}}
Let us first compute the leading term $\tu^{(b)}$, which is of order $ 1$.
For $-\tau \ll t < 0$, inserting Eq.~\eqref{eq:QSVelInstB} into Eq.~\eqref{eq:RetInstantonExpTh}, the term $\tih_t$ is subdominant compared to $\tau\partial_t \tih_t$ and $\tu_t$. We  therefore obtain
\beq
\label{eq:QSVelInstShortTimeH}
\tih^{(b)}(t) = a\int^{0}_t \rmd t'\, \tu^{(b)}(t') + \mathcal{O}(\tau^{-1})\ .
\eeq
Thus, the term $\tih(t)$ in Eq.~\eqref{eq:RetInstantonExpTu} is of order $\tau^{-1}$, and negligible in this regime. This is consistent with the interpretation of the boundary layer as the regime where the eddy currents have not yet built up. Eq.~\eqref{eq:RetInstantonExpTu} reduces to
\beq
\partial_t \tu^{(b)}_{t} - \left(1 + a\right) \tu^{(b)}_{t} + \sigma (\tu^{(b)}_{t})^2 =
- \lambda_{t}.
\eeq
This is just the instanton equation (Eq.~(13) of Ref.~\cite{LeDoussalWiese2011}) of the standard ABBM model, but with a modified mass, $m^2 =1 \to 1+a$. We obtain the known solution \cite{LeDoussalWiese2011,DobrinevskiLeDoussalWiese2012}
\beq
\label{eq:QSVelInstShortTime}
\tu^{(b)}(t) = \frac{(a+1) \lambda  e^{(a+1) t}}{a+1 +\lambda 
   \left(e^{(a+1) t}-1\right)}
.\eeq
Consequently, for $t \to -\infty$, $\tih^{(b)}(t)$ is given by 
\beq
\tih^{(b)}(-\infty) = a\int^{0}_\infty \rmd t\, \tu^{(b)}(t) = - a\log \left(1-\frac{\lambda}{a  +1}\right).
\eeq  
To compute the correction $\tu^{(b)}_1$ of order $\tau^{-1}$, we need to expand \eqref{eq:RetInstantonExpTu} to the next order. We get the linear equation
\beq
\partial_t \tu^{(b)}_1(t) -(1+a) \tu^{(b)}_1(t) + 2 \tu^{(b)}(t) \tu^{(b)}_1(t) + \tih^{(b)}(t) = 0.
\eeq
 Using the expressions \eqref{eq:QSVelInstShortTimeH}, \eqref{eq:QSVelInstShortTime} for $\tih^{(b)}(t)$, its solution is given by
\begin{align}
&\!\!\!\tu^{(b)}_1(t) = \frac{a}{(1+a)\left[1+a+(1-e^{-(1+a)t})\lambda\right]^2} \times  \nn \\
& \times \bigg\{-\lambda  e^{(a+1) t} \bigg[2 (1+a -\lambda )\times  \nn \\
&
   ~~~~~~~~\times\left(\text{Li}_2\Big(-\frac{e^{(a+1) t} \lambda }{1+a -\lambda}\Big)
        -\text{Li}_2\Big(-\frac{\lambda }{1+a -\lambda }\Big)\right) \nn \\
&  ~~~~~~~~~-(1+a) t (1+a -\lambda )+\lambda  \left(e^{(a+1)
   t}-1\right)\bigg] \nn \\
&
~~~~~-\left[(1+a -\lambda )^2-\lambda ^2 e^{2 (a+1) t}\right]
   \times \nn \\ &~~~~~~~~\times\log\! \left(1+\frac{\lambda  \left(e^{(a+1) t}-1\right)}{a+1 }\right) \nn \\
&\left. ~~~~+2
   (a+1) \lambda  t e^{(a+1) t} (1+a -\lambda ) \log \left(1-\frac{\lambda
   }{a+1 }\right)\right\}. \label{secondu}
\end{align}
For $t\to -\infty$, this tends to a constant,
\beq
\label{eq:QSVelInstBLAs}
\lim_{t\to-\infty} \tu^{(b)}_1(t) = -\frac{a}{1+a} \log \left(1-\frac{\lambda}{1+a}\right).
\eeq
Since $\tu^{(b)}(-\infty)=0$, see Eq.~(\ref{eq:QSVelInstShortTime}), this is  the dominant contribution of the boundary-layer solution for $\tu$ in the limit $t \to -\infty$.

\subsubsection{Long-time regime: $|t| \propto \tau$}
\begin{figure}
\includegraphics[width=\columnwidth]{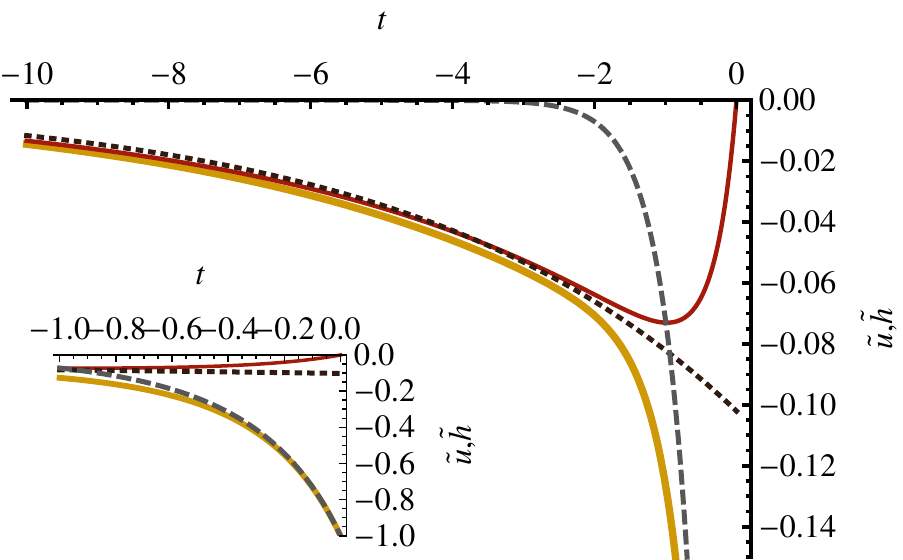}
\caption{(Color online). Instanton solution $\tu_t$, $\tih_t$ of \eqref{eq:RetInstantonExpTu}, \eqref{eq:RetInstantonExpTh} with sources $\lambda(t) = -\delta(t)$, $\mu(t)=0$ in the slow-relaxation limit $\tau \gg 1$. Parameters are $a=1.3$, $\tau=2$. Yellow (thick) curve: $\tu_t$, red (thin) curve: $\frac{1}{1+a}\tih_t$, black dotted curve: long-time asymptotics \eqref{eq:QSVelInstLongTime}, grey dashed curve: short-time asymptotics \eqref{eq:QSVelInstShortTime}. The inset shows details of the boundary layer $|t|\propto 1$.}
\label{fig:StatVelInstanton}
\end{figure}

Now, let us consider the regime $t \lesssim -\tau$.  Inserting the rescaled time $s := t/\tau$ into Eq.~\eqref{eq:RetInstantonExpTu}, we see that the term $\partial_t \tu(t) = \tau^{-1} \partial_s \tu(s)$ is subdominant in $\tau^{-1}$. The instanton in this regime is thus a special case of the instanton discussed in section \ref{sec:QSEddyStat}. Applying Eq.~\eqref{eq:QSVelInstA} we see that $\tu(t) \propto \tau^{-1}$ is small. Thus, we can also neglect the non-linear term $\tu(t)^2$ in \eqref{eq:RetInstantonExpTu}. This gives the simple relation
\beq
\nn
\tu^{(a)}(s) = \frac{1}{1+a}\tih^{(a)}(s).
\eeq
Consequently Eq.~\eqref{eq:RetInstantonExpTh} reduces to
\beq
\nn
\partial_s \tih^{(a)}(s) = \frac{1}{1+a} \tih^{(a)}(s).
\eeq
The boundary condition at $s=0$ is now non-trivial, and given not by the sources, but by the asymptotics of the boundary layer as $t \to -\infty$, 
\begin{align}
\tih^{(a)}(0) &= \tih^{(b)}(-\infty) = - a\log \left(1-\frac{\lambda}{a  +1}\right) \ .
\end{align}
The resulting solution of Eq.~\eqref{eq:RetInstantonExpTu} is
\begin{align}
\label{eq:QSVelInstLongTime}
&\tih^{(a)}(s) = \tih^{(a)}(0)e^{\frac{s}{1+a}} = - a\log \left(1-\frac{\lambda}{a  +1}\right)e^{\frac{s}{1+a}}  \\
&\Longrightarrow~~ \tu^{(a)}(s) = - \frac{a}{1+a}e^{\frac{s}{1+a}}\log \left(1-\frac{\lambda}{a  +1}\right).\label{eq:QSVelInstLongTime2}
\end{align}
A non-trivial consistency check is  that this expression matches the $\mathcal{O}(\tau^{-1})$ term of the $t\to -\infty$ asymptotics of the boundary layer given in Eq.~\eqref{eq:QSVelInstBLAs},
\beq
\tu^{(a)}(0) = \tu^{(b)}_1(-\infty).
\eeq
The  boundary-layer solution \eqref{eq:QSVelInstShortTime} for $t\propto 1$, and the long-time asymptotics \eqref{eq:QSVelInstLongTime} for $t\propto \tau$ compare  well to a direct numerical solution of \eqref{eq:RetInstantonExpTu}, \eqref{eq:RetInstantonExpTh}  in the corresponding regimes, see figure \ref{fig:StatVelInstanton}.

\subsubsection{The velocity distribution}
From the combined knowledge  of  the previous sections, we can extract the generating function for the velocity distribution \eqref{eq:QSVelGenFct}. We have
\begin{align}
\int_t \tu_t &= \left[\int_{-\infty}^0 \rmd t\, \tu^{(b)}(t) + \int_{-\infty}^0 \rmd s\, \tu^{(a)}(s)\right] \nn\\
& = \left[- \log \left(1-\frac{\lambda}{a  +1}\right) - a\log \left(1-\frac{\lambda}{a  +1}\right) \right] \nn\\
& = -\left(1+a\right)\log \left(1-\frac{\lambda}{a  +1}\right)\ .
\end{align}
Thus, the generating function for the velocity distribution \eqref{eq:QSVelGenFct} is
\beq
\overline{ e^{\lambda \du}} = \left(1-\frac{\lambda}{a  +1}\right)^{-\left(1+a\right)v}\left[1+ \mathcal{O(\tau)}^{-1}\right]
\ .\eeq
The Laplace inversion is easy to do, giving the distribution of instantaneous velocities to leading order in $\tau^{-1}$ (but without any approximation in $v$). Restoring units, this is 
\beq
P(\du) = \frac{e^{-\frac{\eta}{\sigma}\left(m^2+a\right)\du}}{\Gamma\left[\frac{\eta}{\sigma}\left(m^2+a\right)v\right]} \frac{1}{\du}\left[\du \frac{\eta}{\sigma}\left(m^2+a\right)\right]^{\frac{\eta}{\sigma}\left(m^2+a\right) v}.
\label{64}
\eeq
We can compare this to the  distribution of the  standard ABBM model,
\beq
P(\du) = \frac{e^{- \frac{\eta}{\sigma}m^2\du}}{\Gamma\left(\frac{\eta}{\sigma}m^2 v\right)} \frac{1}{\du}\left( \du \frac{\eta}{\sigma}m^2\right)^{\frac{\eta}{\sigma}m^2 v }. \label{65}
\eeq
We  see that the effect of eddy currents on the instantaneous velocity distribution, in the limit of $\tau \to \infty$, is  the same as if the mass in the standard ABBM model were increased from $m^2$ to $m^2 + a$. In particular, this means that the transition between intermittent avalanches and continuous motion happens for  driving velocities $v$ reduced by a factor of \((1+a/m^2)\).
In dimensionful units
\beq
v_c =\frac{\sigma}{\eta (m^2+a)}\ .
\label{63}\eeq

\subsection{Velocity following a kick\label{sec:VelKick}}
Let us now assume that the driving velocity undergoes a kick at  time $t<0$, i.e.\ $\dw(t') = w_0 \delta(t'-t)$.
As discussed in  section~\ref{sec:Protocols}, we consider an initial condition at $t=0$ prepared in the ``Middleton attractor'' $u(w)$
with $\dot u(0)=h(0)=0$. The kick gives deterministically $\du(t^+) = m^2 w_0 =: \du_{\rm i}$, so that the distribution of velocities $\du_{\rm f} := \du_0$ is the propagator $\mathcal{P}(\du_{\rm f},0 | \du_{\rm i}, t)$ at zero driving velocity\footnote{In this section, we still keep $\eta=1$ but restore units of $m^2$ in some places for clarity.}. 
Applying \eqref{eq:RetSolution}, the generating function for the velocity $\du_0$ at time $t=0$ is given by\beq
\overline{e^{\lambda \du_0}} = e^{w_0 \tu_t}.  \label{vel0}
\eeq
One can obtain the probability distribution of $\du_0$ in the slow-relaxation limit $\tau \gg 1$
by inverse Laplace transformation as follows. There are two time regimes:

(i) for $|t| \sim 1$ we need to insert $\tu_t = \tu^{(b)}\left(t\right) + \tau^{-1} \tu^{(b)}_1(t)$ in (\ref{vel0}) as given by
(\ref{eq:QSVelInstShortTime}) and (\ref{secondu}). For small $t$ we need to use only (\ref{eq:QSVelInstShortTime})
and we recover the velocity distribution and propagator of the standard ABBM model at zero driving velocity, but with modified parameters. The propagator of the standard ABBM model has been discussed, among others, in \cite{Colaiori2008,LeDoussalWiese2011,DobrinevskiLeDoussalWiese2012}. Here we use the result from Eq.~(24) in Ref.~\cite{LeDoussalWiese2011}, or equivalently (19) in Ref.~\cite{DobrinevskiLeDoussalWiese2012}, with $v=0$ and $m^2 \to m^2+a$:
\begin{align}
\nn
P(\du) =& P(\du;0|m^2 w_0;t) = \\
\nn
= &\frac{m^2 \sqrt{w_0/\du}}{2 \sinh \left[\frac{1}{2} t \left(a+m^2\right)\right]} \times \\
\nn 
& \times \exp \left\{-\frac{\left(a+m^2\right) \left[\du +w_0 e^{\left(a+m^2\right)t}\right]}{1-e^{\left(a+m^2\right)t}}\right\}\times \\
 & \times I_1\left[\frac{\left(m^2+a\right)
   \sqrt{\du w_0}}{\sinh\left(\frac{1}{2}
   \left(m^2+a\right) t\right)}\right]
\end{align}
This distribution is not normalized; formally, there is an additional $\delta$-function term as in Eq.~(24) in Ref.~\cite{LeDoussalWiese2011}. This indicates that this formula is valid for $\du \sim 1$ only; there is another regime $\du \sim 1/\tau \ll 1$ which requires a more careful treatment but will not be considered here. 

At later times there is a complicated crossover to regime (ii), which requires to
keep the $1/\tau$ correction, which we will not detail here. 

(ii) for long times $|t| \sim \tau$, $\tu_t$ is given by \eqref{eq:QSVelInstA} and \eqref{eq:QSVelInstLongTime2}. 
This gives
\beq
\overline{e^{\lambda \du_0}} =\left(1-\frac{\lambda
   }{a+1}\right)^{-\frac{a w_0
   }{(a+1) \tau} \exp(\frac{t}{\tau(a+1)})}.
\eeq
Inverting the Laplace transform, one obtains the velocity distribution and  propagator
\beq
P(\du) \simeq  \frac{({a+1})^{
   \epsilon_ t } e^{-\du (1+a)}}{\Gamma (\epsilon_t )\du^{1-\epsilon_t}}\ , \qquad
  \epsilon_t:=\frac{a w_0 }{(a+1) \tau} e^{\frac{t}{\tau(a+1)}}\ .
  \label{67}
\eeq
It approaches rather quickly a (normalized and regularized) power-law distribution proportional to $1/\dot u$. 
Note that for an infinitesimal kick we recover the limiting density obtained from the stationary motion
$\rho(\du) = \partial_{w_0}\big|_{w_0=0} P(\du) = \tau_{\rm typ}^{-1} \partial_v\big|_{v=0} P^{\rm stat}(\du)  \sim \dot u^{-1} e^{-(1+a) \dot u}$ 
where $P^{\rm stat}(\dot u)$ was obtained in (\ref{64}) and $\tau_{\rm typ}$ is a typical time scale.

 In general, the velocity distribution $P(\du)$ following a kick can be decomposed as
\beq
\label{eq:PAfterKick}
P(\du) = p_{\du=0} \delta(\du) + \mathcal{P}_{\rm reg}(\du)\ ,
\eeq
where $p_{\du=0}$ is the probability that the domain-wall has come to a complete halt. From \eqref{67} one sees that in the ABBM model with retardation, $p_{\du=0} =0$ and the domain-wall motion following a kick never stops completely. This is in contrast to the standard ABBM model, where one has (cf. \cite{DobrinevskiLeDoussalWiese2012}, Eq.~(28))
\beq
\nn
p_{\du=0} = \exp\left(-\frac{w_0}{e^{-t}-1}\right).
\eeq
This can also be seen directly from the instanton solution $\tu$: The decomposition \eqref{eq:PAfterKick} implies 
\beq
\overline{e^{\lambda \dot u}} = \int_{\dot u} e^{\lambda \du} P(\du) ~\stackrel{\lambda\to -\infty}{-\!\!\!-\!\!\!-\!\!\!-\!\!\!\longrightarrow}~ p_{\du=0}
\ .\eeq
Since $\overline{e^{\lambda \dot u}} =\mathrm{e}^{w_{0} \tilde  u_{t}}$, we can conclude that  $p_{\du=0}$ is zero, if and only if $\lim_{{\lambda\to -\infty}}\tilde u_{t} = -\infty $.
 This is the case in the retarded ABBM model, where \eqref{eq:QSVelInstLongTime2} shows that $\tu_t \propto -\ln\left(1-\frac{\lambda}{1+a}\right)\stackrel{\lambda\to -\infty}{-\!\!\!-\!\!\!-\!\!\!-\!\!\!\longrightarrow}-\infty$. However, it is not the case in the standard ABBM model, where $\lim_{\lambda \to -\infty} \tu_t = \frac{1}{1-e^{-t}}$ is finite (cf. \cite{DobrinevskiLeDoussalWiese2012}, Eq.~(14)).

We conclude that in the ABBM model with retardation, the velocity following a kick never becomes zero permanently, even though its mean decays exponentially in time
over time scales of order $\tau$, with for $t \lesssim -\tau$,
\beq
\left<\du_0 \right>=\frac{a w_0 }{(a+1)^2 \tau} e^{-\frac{|t|}{(a+1)\tau}}\ . 
\eeq Although the calculation above was done at leading order in $\tau^{-1}$, we expect the phenomenology to be similar for arbitrary $\tau$. To make this explicit, we  now  consider the opposite limit of fast relaxation, $\tau \ll \frac{\eta}{m^2}$, in which analytical progress is also possible.

\section{The fast-relaxation limit $\tau \ll \frac \eta{m^2}$ \label{sec:FastRel}}
We can also consider the limit $\tau \ll \tau_m = {\eta/m^2}$, where the eddy currents relax much faster than the domain-wall motion.
Experimentally, this limit is even more relevant than the slow-relaxation limit discussed in section \ref{sec:QSLim}: As a function of sample thickness \(b\), the eddy-current relaxation time $\tau\propto b^2$, whereas the domain-wall motion occurs on a time scale $\frac{\eta}{m^2} \propto a b$.  For typical experimental setups \cite{ZapperiCastellanoColaioriDurin2005,DurinZapperi2006b} $a \gg b$ and hence $\tau \ll \eta/m^2$.

We now discuss the stationary velocity distribution, and the velocity following a kick in the driving velocity, in the fast-relaxation limit. As in section \ref{sec:VelStat}, we need to construct the instanton $\tu$, $\tih$ solving Eqs.\ \eqref{eq:RetInstantonExpTu}, \eqref{eq:RetInstantonExpTh} with sources $\lambda(t)=\lambda \delta(t),\,\mu(t)=0$. Now, however, $\tau$ and not $\tau^{-1}$ is a small parameter. We expect a two-scale solution: A boundary layer for $|t| \propto \tau$ around $t=0$, and an asymptotic regime for $|t| \propto 1$. 
We thus introduce the rescaled time $s := t/\tau$ and make the ansatz 
\begin{align}
\label{eq:SRVelInstB}
\tu(t) &=: \tu^{(b)}_0\left(t\right) + \tau\tu^{(b)}_1(t) + \mathcal{O}(\tau)^2,\quad |t| \propto 1 \\
\tih(t) &=: \tih^{(b)}_0\left(t\right) + \mathcal{O}(\tau),\quad |t| \propto 1 \\
\label{eq:SRVelInstA}
\tu(t) &=: \tu^{(a)}_0\left(s\right) + \tau\tu^{(a)}_1(s) + \mathcal{O}(\tau)^2,\quad |t| \propto \tau \\
\tih(t) &=: \tih^{(a)}_0\left(s\right) + \mathcal{O}(\tau),\quad |t| \propto \tau.
\end{align}

\subsection{Leading order}
At order $\tau^0$, the instanton equations \eqref{eq:RetInstantonExpTh} and \eqref{eq:RetInstantonExpTu} reduce in the asymptotic regime  to
\begin{align*}
\partial_t \tu_0^{(b)}(t) - (1+a)\tu_0^{(b)}(t) + \left[\tu_0^{(b)}(t)\right]^2 + \tih_0^{(b)}(t) &= 0 \\
-\tih_0^{(b)}(t) + a\tu_0^{(b)}(t) &=0,
\end{align*}
with boundary condition $\tu_0(0) = \lambda$. This gives the leading-order solution
\begin{align}
\label{eq:SRVelLO}
\tu_0^{(b)}(t) &= \frac{\lambda}{\lambda + (1-\lambda)e^{-t}} \\
\nn
\tih_0^{(b)}(t) &= a \tu_t^{(b)}.
\end{align}
In the boundary layer, the corresponding solution is $\tu_0^{(a)}(s) = \tu_0^{(b)}(0) = \lambda$, and Eq.~\eqref{eq:RetInstantonExpTu} gives
\begin{align*}
&\partial_s \tih_0^{(a)}(s) -\tih_0^{(a)}(s) + a\lambda =0, \\
&\Rightarrow   ~~\tih_0^{(a)}(s) = a \lambda\left(1-e^{s}\right).
\end{align*}

\subsection{Next-to-leading order}
We obtained  in Eq.~\eqref{eq:SRVelLO}  the leading-order solution $\tu_0(t)$, valid in both regimes. Expanding around it, \ setting $\tu(t) = \tu_0(t) + \tau \tu_1(t) + \mathcal{O}(\tau)^2$, we get an equation for \(\tu_1(t)\) 
\beq
\nn
\partial_t \tu_1(t) - (1+a)\tu_1(t) + 2 \tu_0(t) \tu_1(t) + \frac{1}{\tau}\left[\tih(t) - a \tu_0(t) \right] = 0.
\eeq
In the boundary layer, $\partial_t \tu_1(t) = \frac{1}{\tau} \partial_s \tu_1^{(a)}(s),$ and 
\beq
\nn
\tih_0(s) - a \tu_0(s) = a \lambda\left(1-e^{s}\right) - a\lambda = - a \lambda e^s.
\eeq
Hence, the next-to-leading-order contribution $\tu_1^{(a)}(s)$ in the boundary layer satisfies
\begin{align}
\nn
&&\partial_s \tu_1^{(a)}(s) &= a \lambda e^s \\
&\Rightarrow & \tu_1^{(a)}(s) & = -a\lambda(1-e^s)~~~~
\label{eq:SRVelNLOBL}
\end{align}
On the other hand, Eq.~\eqref{eq:RetInstantonExpTh} gives in the asymptotic regime 
\beq
\partial_t \tih^{(b)}_0(t) = \tih_1^{(b)}(t) - a \tu_1^{(b)}(t).
\eeq
Inserting this relation into Eq.~\eqref{eq:RetInstantonExpTu} gives
\begin{align}
\nn
&\partial_t \tu_1^{(b)}(t)-\tu_1^{(b)}(t) + 2 \tu^{(b)}_0(s) \tu_1^{(b)}(t) + a \partial_t \tu^{(b)}(t) = 0 \\
&\Rightarrow  ~\tu_1^{(b)}(t) = \frac{a \lambda e^t (\lambda t-t-1)}{(\lambda e^t-\lambda+1)^2} \ .
\label{eq:SRVelNLOAs}
\end{align}
Here we used the matching condition $\tu_1^{(b)}(0) = \tu_1^{(a)}(-\infty) = -a\lambda$, as given by Eq.~\eqref{eq:SRVelNLOBL}.

\begin{figure}
\includegraphics[width=\columnwidth]{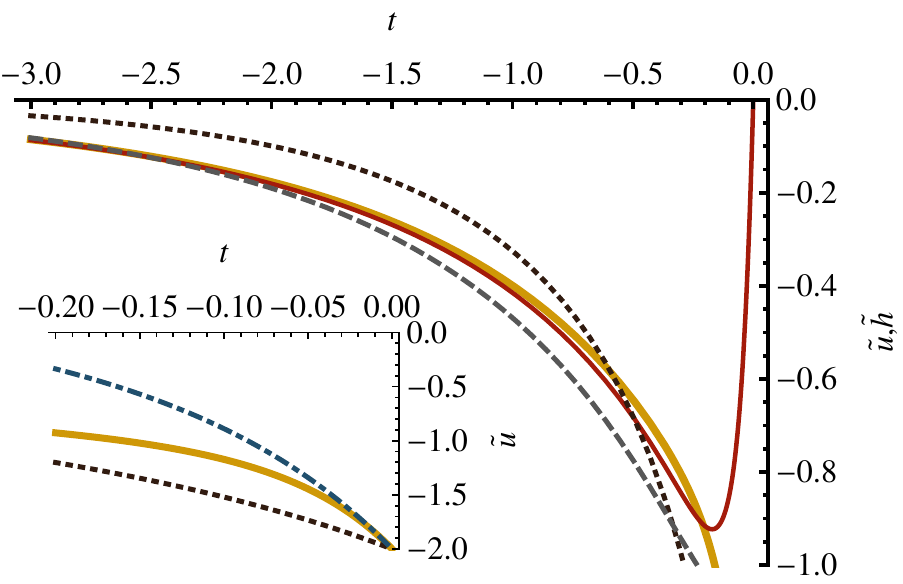}
\caption{(Color online) Instanton solution $\tu$, $\tih$ of Eqs.~\eqref{eq:RetInstantonExpTu}, \eqref{eq:RetInstantonExpTh} with sources $\lambda(t) = -\delta(t)$, $\mu(t)=0$ for fast eddy-current relaxation, $\tau \ll \eta/m^2$. Parameters are $a=5$, $\tau=0.1$. Yellow (thick) curve: $\tu(t)$, red (thin) curve: $\frac{1}{a}\tih(t)$. Black dotted curve: Leading-order result \eqref{eq:SRVelLO}, corresponding to the standard ABBM model. Grey dashed curve: Solution \eqref{eq:SRVelLO} plus next-to-leading order correction \eqref{eq:SRVelNLOAs} in the long-time regime $|t| \propto 1$. 
 The inset shows details of the boundary layer $|t| \propto \tau$. Blue dot-dashed curve (inset only): Solution \eqref{eq:SRVelLO} plus next-to-leading order correction \eqref{eq:SRVelNLOBL} in the boundary layer.}
\label{fig:StatVelInstantonFastRel}
\end{figure}

The next-to-leading order corrections \eqref{eq:SRVelNLOBL} and \eqref{eq:SRVelNLOAs} compare  well to a direct numerical solution of Eqs.~\eqref{eq:RetInstantonExpTu} and \eqref{eq:RetInstantonExpTh}, see figure \ref{fig:StatVelInstantonFastRel}.

\subsection{Stationary velocity distribution}
With the above analysis, we can obtain some results on the velocity distribution. The integral over the instanton solution gives
\begin{align}
\nn
\int_t \tu(t) \rmd t = & \int_{-\infty}^0 \rmd t\,\tu^{(b)}_0(t) + \tau\int_{-\infty}^0 \rmd t\,\tu_1^{(b)}(t)  + \mathcal{O}(\tau)^2 \\
\nn
= &  -\ln(1-\lambda) + a\tau\left[\frac{\lambda}{\lambda-1} - \ln(1-\lambda)\right] \\ &+ \mathcal{O}(\tau)^2 
\end{align}
Inserting  this result into Eq.~\eqref{eq:QSVelGenFct} for the generating functional of instantaneous velocities gives
\beq
\overline{e^{\lambda\du}} = \left(1-\lambda\right)^{-v\left(1+a \tau\right)}e^{a v \tau\frac{\lambda}{\lambda-1} + \mathcal{O}(\tau)^2 }.
\eeq
To the same order in $\tau$, this can also be rewritten as 
\begin{align}
\int_t \tu(t) \rmd t = & -(1+a\tau)\ln\left[1-\frac{\lambda}{1+a\tau}\right] + \mathcal{O}(\tau)^2  \\
\nn
\Rightarrow
\overline{e^{\lambda\du}} = & \left[1-\frac{\lambda}{(1+a \tau)}\right]^{-v\left(1+a \tau\right)} + \mathcal{O}(\tau)^2 .
\end{align}
This makes it clearer that, to leading order, the form of the velocity distribution is not modified, and only the parameters are rescaled.

For small $\du$, it indicates that, in the fast-relaxation limit, the instantaneous velocity distribution $P(\du)$ has a power-law behaviour
\beq
P(\du) \sim \du^{-1+v\left(1+a \tau+\mathcal{O}(\tau)^2\right)}.
\eeq
Putting back the units, the power-law exponent becomes $-1+\frac{v}{v_c}$, where
\bea
v_c &=& \frac{\sigma}{\eta m^2}\left[ 1 - a\frac{\tau}{\eta} + \mathcal{O}\left(\frac{\tau}{\eta/m^2}\right)^2 \right] \nn \\ &=& \frac{\sigma}{ m^2(\eta +a \tau  )} + \mathcal{O}\left(\frac{\tau}{\eta/m^2}\right)^2\ .
\eea
We see that fast eddy-current relaxation decreases $v_c$, just as  in Eq.~\eqref{64} for slow eddy-current relaxation. Both formulas have the form $v_c=\frac{\sigma}{m^2(\eta + a \tau_{\rm f})} $, where $\tau_{\rm f}$ is the fastest time scale in the problem, $\tau_{\rm f}=\eta/m^2$ for the slow-relaxation limit, and $\tau_{\rm f}=\tau$ for the fast-relaxation limit. 
In contrast to Eq.~\eqref{64} however, the correction we obtain here is perturbative: 
It vanishes as $\tau \to 0$. In the limit $\tau\to0$ we recover the standard ABBM model.

\subsection{Velocity following a kick}
\begin{figure}\includegraphics[width=\columnwidth]{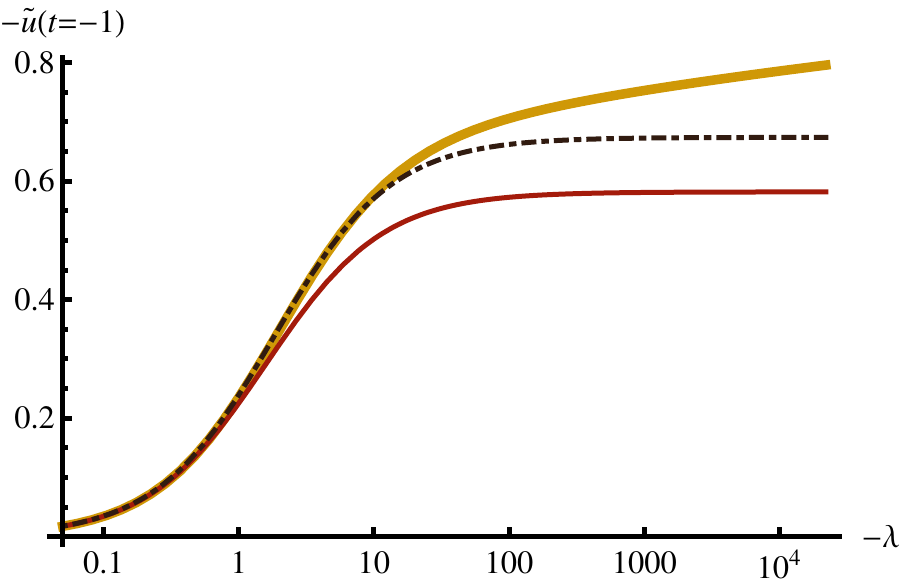}\caption{(Color online) Instanton $\tu(t)$ at a fixed time $t=-1$, as a function of $\lambda$, for $\tau=0.1$ and $a=1$. Thick yellow line: Numerical solution of \eqref{eq:RetInstantonExpTu}, \eqref{eq:RetInstantonExpTh}. Thin red line: Leading-order, ABBM solution \eqref{eq:SRVelLO}, corresponding to $a=0$. Dot-dashed black line: Next-to-leading order correction \eqref{eq:SRVelNLOAs}.}\label{fig:KickInstantonFastRel}\end{figure}

The generating functional for the distribution of velocities $\du(0)$ following a kick of size $w$ at $t < 0$ can also be expressed in terms of the instanton solution \eqref{eq:SRVelInstB},
\beq
\label{eq:SRGenFctKick}
\overline{e^{\lambda\du_0}} = e^{w\tu(t)} = \exp\left[w \tu^{(b)}_0(t) +  w \tau \tu^{(b)}_1(t) + \mathcal{O}(\tau)^{2}\right].
\eeq
$\tu_0^{(b)}$ and $\tu^{(b)}_1$ are given by Eqs.~\eqref{eq:SRVelLO} and \eqref{eq:SRVelNLOAs} above. They have a finite limit as $\lambda \to -\infty$:
\begin{align}
\lim_{\lambda \to -\infty} \tu^{(b)} &= \frac{1}{1-e^{-t}}\ , \\
\lim_{\lambda \to -\infty} \tu^{(b)}_1 &= \frac{a t e^{-t}}{(1-e^{-t})^2}\ .
\end{align}
This would suggest, that the velocity distribution  contains a term $\sim \delta(\dot u)$. However, for large negative $\lambda$, the expansion above breaks down, and higher orders in $\tau$ become non-negligible. By solving the complete instanton equations numerically one obtains figure \ref{fig:KickInstantonFastRel}.
One observes that the leading order (standard ABBM) instanton \eqref{eq:SRVelLO} goes to a fixed value for $\lambda \to -\infty$. The next-to-leading order correction \eqref{eq:SRVelNLOAs} coincides better with the numerical solution, but still breaks down around $\lambda \approx -10$, and goes to a fixed value, too. However, the true (numerically obtained) solution  of the instanton equations goes to $-\infty$ as $\lambda\to-\infty$. Hence, $\lim_{\lambda\to -\infty} \overline{e^{\lambda\du_0}} = \lim_{\lambda\to -\infty} e^{w\tu(t)} = 0$, and the distribution $P(\du_0)$ does not have a $\delta(\du_0)$ piece, consistent with the results obtained above in section \ref{sec:VelKick} in the $\tau \to \infty$ limit.

From the instanton expansion \eqref{eq:SRVelLO}, \eqref{eq:SRVelNLOAs}, valid for $\lambda \propto 1$, we can obtain the velocity distribution $P(\du_0)$ following a kick at $t<0$, in the regime $\du_0 \propto 1$. Using Eq.~\eqref{eq:SRGenFctKick}, we write its generating function to order $\tau$ as 
\beq
\overline{e^{\lambda \du_0}} = \exp\left(w A - \frac{w B}{\lambda-q} + \frac{w C}{(\lambda-q)^2}\right),
\eeq
with 
\begin{align*}
q = &  \frac{1}{1-e^{t}} \\
A = & \frac{e^t}{e^t-1} + \tau\frac{a e^{t}t}{(e^{t}-1)^2} \\
B = &  \frac{e^{t}}{(e^{t}-1)^2}+\tau\frac{a e^{t}\left[e^t(t+1)+t-1\right]}{(e^t-1)^3}  \\
C = & \tau \frac{a e^t\left[e^t(t+1)-1\right]}{(e^t-1)^4} .
\end{align*}
The inverse Laplace transform of $\overline{e^{\lambda \du_0}}$ can be written as
\bea
P(\du) &=&e^{wA-q\dot u} \\
&& \times \int\limits_0^{2\pi} \frac{\rmd \phi}{2\pi} \,re^{i\phi} \exp\Big(w\Big[ \frac{C e^{-2i\phi}}{r^2} {+}\frac{B e^{-i\phi}}{r}\Big] {-}  \du r e^{i\phi}  \Big).\nn
\eea
We have set $\lambda=q+r e^{i \phi}$ to arrive at the above formula. We numerically checked that the integral is independent of $r$. 
One can evaluate it analytically, if either $B=0$, or $C=0$, by expanding in powers of $r$, and retaining only terms which scale as $r^0$ (note $r^n\sim \rme^{i \phi n}$). 
The final result can be written as  a convolution of the two:
\begin{align}
\label{eq:VelDistSRKick}
P(\du) =& e^{wA-q\dot u}\int_0^{\du} \rmd \du_1 P_1(\du_1)P_2(\du-\du_1), \\
P_1(\du) = &  \delta(\du) + \sqrt{\frac{B w}{\du}} I_1\left(2 \sqrt{B w \du}\right) \\
P_2(\du) = & \delta(\du) + C u w \,
   {}_0F_2\left(\frac{3}{2},2\bigg|\frac{1}{4} C
   u^2 w\right).
\end{align}  
Note again that formally the $\delta$-function parts are an artifact of our expansion, which is not valid as $\lambda\to-\infty$ or $\du\to 0$. We expect them to be smeared out on a scale $e^{t/\tau}$, which goes to $0$ as $t \to -\infty$.
However, physically, these velocities are extremely small and unlikely to be observable. Thus the $\delta$-function term is physically sensible, and can be interpreted as the probability that all \textit{significant} avalanche activity has stopped.

\begin{figure}\includegraphics[width=\columnwidth]{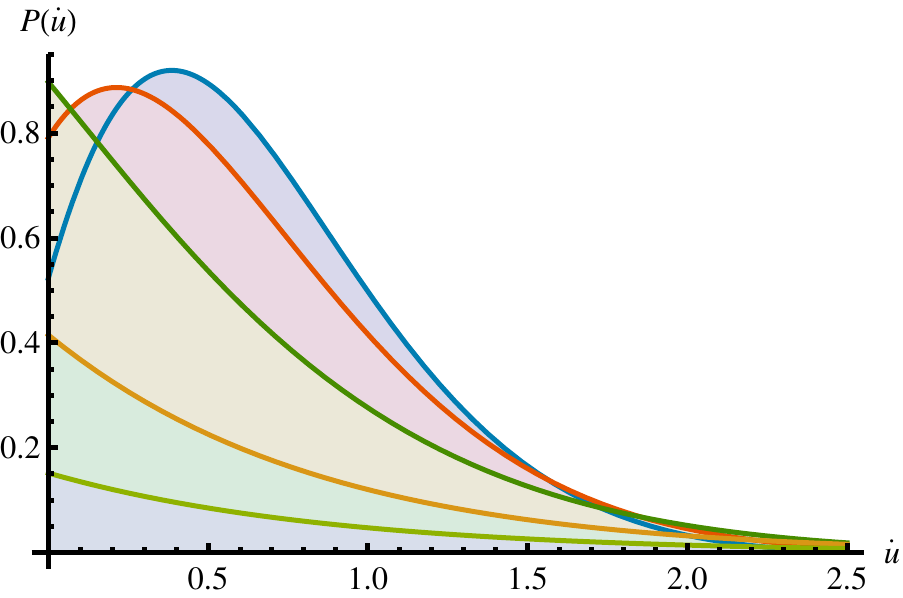}\caption{(Color online) Velocity distribution $P(\du_0)$ given by \eqref{eq:VelDistSRKick}  (without the $\delta$-part), after a kick at time $t<0$. The parameters were chosen to be $w=1,a=1, \tau=0.2$. The time of the kick varies from right to left as $t=-0.3,-0.4,-0.7,-1.5,-2.5$.} \label{fig:VelDistSRKick}\end{figure}

Numerically, the convolution can easily be computed. An example of the distributions for various times is shown in figure \ref{fig:VelDistSRKick}.
We see that for small times the distribution is peaked around the value $w$ imposed by the step in the force. Later on the typical value of the velocity approaches $0$, and the distribution becomes monotonous. Its area decreases since part of the probability is absorbed by the (smoothened) $\delta$-function near $\du=0$ ($\lambda =-\infty$), which we are unable to analyze here in more detail.

\section{Avalanche statistics at fixed size\label{sec:AvStatSize}}
In the previous section we saw, at least in the two limits $\eta/m^2 \ll \tau $ and $\eta/m^2 \gg \tau $, that avalanches following even an infinitesimal kick never completely stop. Computing observables conditioned to their duration of first return to $\du=0$, i.e.\ the sub-avalanche duration, requires introducing an  artificial   ``absorbing boundary'' at \(\du=0\) which will terminate the avalanche once $\du$ becomes zero\footnote{ The natural boundary at $\du=0$ would be reflecting, since if $\du_t$ becomes zero at some instant of time, it immediately restarts to positive velocities due to the decrease of the eddy current pressure in the next time step.}. This task is deferred to section \ref{sec:AvStatDur}. However, the mean velocity following a (finite or infinitesimal) kick still decreases, and the total avalanche size remains finite. We will now compute its distribution, and other observables conditioned on the total avalanche size.

\subsection{Avalanche sizes\label{sec:AvSizes}}
We define the size $S$ of a non-stationary avalanche following a kick of size $w_0$ at $t=0$ as $S=\int_{-\infty}^\infty \du(t) \rmd t$. The Laplace transform of the probability distribution of $S$ is given by Eq.~\eqref{eq:RetSolution},
\beq
\overline{e^{\lambda S}} = \overline{e^{\lambda \int_{-\infty}^\infty \du(t) \rmd t}} = e^{w_0 \tu_0}\ .
\eeq
Here $\tu_t$ is the solution of \eqref{eq:RetInstanton} with a time-independent source $\lambda(t) = \lambda$. This means that $\tu_t = \tu$ is also time-independent. Then, using $f(0)=1$ and $f(\infty) = 0$, the terms proportional to $a$ drop out from Eq.~\eqref{eq:RetInstanton} and we get
\beq
 - m^2 \tu + \sigma \tu^2 = -\lambda.
\eeq
Choosing the solution which tends to $0$ as $\lambda \to 0$, we get
\beq
\label{eq:SizeUtSol}
\tu = \frac{m^2-\sqrt{m^4-4 \lambda \sigma}}{2 \sigma}
\eeq
and 
\beq
\overline{e^{\lambda S}} = e^{w_0\textstyle\frac{m^2-\sqrt{m^4-4 \lambda \sigma}}{2 \sigma}}
\ .\eeq
Inverting the Laplace transform gives
\beq
\label{eq:SizeDist}
P(S) = \frac{w_0 }{2
 \sqrt{\pi\sigma} S^{\frac{3}{2}}}e^{-\textstyle\frac{(w_0-m^2 S)^2}{4 \sigma S}}.
\eeq
 with $\overline{S}=w_0$. Note that this extends to any finite kick of arbitrary shape replacing $w_0=\int_0^{\infty} \rmd t \dot w(t)$ \cite{LeDoussalWiese2012a}. This is precisely the  distribution obtained for the standard ABBM model and the mean-field theory of interfaces in \cite{LeDoussalWiese2009,DobrinevskiLeDoussalWiese2012,LeDoussalWiese2012a}. Of course, this can already be seen  from the fact that  the terms proportional to $a$ drop out from \eqref{eq:RetInstanton} when $\tu$ is time-independent. Note that this result is {\em independent} of the shape of the memory kernel $f$ in \eqref{eq:EOMU}.
This is a consequence of the monotonicity of the model, as discussed in the Introduction. In the limit of an infinitesimal kick, i.e.\ small $w_0$, one recovers the stationary avalanche-size density.

Universal corrections to the distribution \eqref{eq:SizeDist} are expected when one goes beyond the mean-field limit and considers $d$-dimensional elastic interfaces. Without retardation effects, the universal corrections at slow driving were obtained to one loop in an expansion around the critical dimension \cite{LeDoussalWiese2012a}. We expect them to remain unchanged by retardation effects, as seen in this section for the mean-field case.

\subsection{Avalanche shape at fixed size\label{sec:AvShapeFixedSize}}
The avalanche shape is usually obtained by computing the mean velocity as a function of time, in the ensemble of all avalanches of a fixed duration \cite{SethnaDahmenMyers2001,ColaioriZapperiDurin2004,ZapperiCastellanoColaioriDurin2005,PapanikolaouBohnSommerDurinZapperiSethna2011}. Here we shall instead consider the ensemble of all avalanches of a fixed size $S$. In a numerical simulation or in an experiment, the shape at fixed size is just as easily measurable as the shape at fixed duration. However, it is  easier to obtain theoretically with our methods, and it can be defined without a microscopic cutoff. We will thus compute the shape function defined via
\beq
\mathfrak{s}(t,S) := \int_0^\infty \rmd \du_t\, \du_t\, P(\du_t | S) = \frac{\int_0^\infty \rmd \du_t\, \du_t\, P(\du_t,S)}{P(S)}.
\eeq
$P(S)$ is the avalanche size distribution \eqref{eq:SizeDist}.

We follow the approach used in \cite{LeDoussalWiese2009,DobrinevskiLeDoussalWiese2012,LeDoussalWiese2012a} to obtain the avalanche shape in the standard ABBM model from the Martin-Siggia-Rose field theory. The driving $w_t$ performs a kick at $t=0$, i.e.\ we set $\dw_t = w_0 \delta(t)$.
We then consider the observable
\begin{align}
\label{eq:ShpDefH}
\hsh(t_0,\lambda):=\overline{\du_{t_0}\, e^{\lambda S}} = \partial_\mu \big|_{\mu=0} \overline{e^{\lambda S + \mu \du_{t_0}}}\ .
\end{align}
$\hsh$ is related to the shape function $\mathfrak{s}$ via a Laplace transform,
\beq
\hsh(t_0,\lambda) = \int_0^\infty \rmd S\, \mathfrak{s}(t,S)P(S)e^{\lambda S}.
\eeq
$\hsh$ as defined in \eqref{eq:ShpDefH} can be evaluated using \eqref{eq:RetSolution},
\beq
\label{eq:ShpSolH}
\hsh(t_0,\lambda) = \partial_\mu \big|_{\mu=0} e^{w_0 \tu_0(\mu)},
\eeq
where $\tu_0(\mu)$ is the solution of \eqref{eq:RetInstanton} with the source $\lambda_t = \lambda + \mu \delta(t-t_0)$.
To compute \eqref{eq:ShpSolH}, we need to solve the instanton equation \eqref{eq:RetInstanton} to first order in $\mu$. The solution for $\mu=0$ is  the constant $\tu(\lambda)$,  obtained previously in Eq.\ \eqref{eq:SizeUtSol} for the size distribution. The correction of order $\mu$, $\tu^{(1)}$, has to satisfy the linear (but still non-local) equation
\begin{align}
\label{eq:ShpLinInst}
& \partial_t \tu^{(1)}_t - (1 + a - 2 \tu) \tu^{(1)}_t - a \int^{\infty}_t \rmd s\,f'(s-t) \tu^{(1)}_s \nn\\& ~~~= -\mu\delta(t-t_0)\ .
\end{align}
We now restrict ourselves to the case of an exponentially decaying memory term, $ f(t)=e^{-t/\tau}$. Through the substitution $\tu^{(1)}_t = \mu(1-\tau\partial_t)g_t$, equation \eqref{eq:ShpLinInst} is transformed into a linear second-order ODE,
\beq
 (\partial_t - 1 - a + 2 \tu)(1-\tau\partial_t)g_t + a g_t = -\delta(t-t_0).
\eeq
The right-hand side yields the boundary conditions $g(t_0) = 0$, $g'(t_0) = - 1/\tau$. The resulting solution for $g_t$ is 
\beq
\nn
g_t = \frac{2r  e^{(t-t_0)/\tau} \left[e^{-\frac{2 a (t-t_0)}{r}}-e^{\frac{r (t-t_0)}{2 \tau }}\right]}{4 a \tau
   +r^2},
\eeq
where we defined $r$ via
\beq
\frac{(2+r)(r-2a\tau)}{2r\tau}=\sqrt{1-4\lambda}
\label{74}
.\eeq
The shape function \eqref{eq:ShpSolH} is then
\begin{align}
\hsh(t_0,\lambda) &= e^{w_0 \tu_0(\mu=0)} w_0 \tu^{(1)}(0) \nn\\
&= e^{\frac{w_0}{2}(1-\sqrt{1-4\lambda})} w_0\left[ g(0)-\tau g'(0)\right] \nn\\
&= e^{\frac{w_0}{2}(1-\sqrt{1-4\lambda})} w_0
\frac{4 a  e^{\frac{2 a t_0}{r}}+r^2 e^{-\frac{r}{2}t_0/\tau}}{4 a \tau+r^2} e^{-t_0/\tau}.
\end{align}
The shape at fixed size $S$ is finally obtained by inverting the Laplace transform. This is best done using the  coordinate $r$ introduced in Eq.~(\ref{74}):
\begin{align}
\nn
& \int \rmd\du_t\,\du_t\,P(\du_t,S) = - \int_{r_0 -i\infty}^{r_0 + i\infty} \frac{\rmd r}{2\pi i} \hsh \big(t,\lambda(r)\big) e^{-\lambda(r)S} \frac{\rmd \lambda}{\rmd r}\\
\nn
&= \int_{r_0 -i\infty}^{r_0 + i\infty} \frac{\rmd r}{2\pi i}\frac{(r-2 a\tau) (r+2) \left(4 a  \tau e^{\frac{2 a t}{r}} +r^2 e^{-\frac{t r}{2 \tau
   }}\right)}{8 \tau^2 
   r^3} \times \nn \\  \nn
\label{eq:ShpBeforeLT2}
&\quad ~~~~~~~~~\times \exp \left[\frac{S (r+2)^2(r-2a\tau)^2}{16 r^2\tau^2}-\frac{S}{4}-\frac{t}{\tau}\right] \\
&\quad ~~~~~~~~~\times w_0 \exp \left[ \frac{w_0}{2} \Big(1- \frac{(2+r)(r-2a\tau)}{2r\tau}\Big) \right].\!
\end{align}
$r_0$ fixes the location of the integration contour; it can be chosen arbitrarily, as long as $r_0 \neq 0$ \footnote{This is in order to avoid the singularity of the integrand at $r=0$. We checked that the result of the (numerical) integration of \eqref{eq:ShpBeforeLT2} is independent of the value and the sign of $r_0$.}.

Our final result for the avalanche shape at fixed size (following a kick of arbitrary strength $w_0$)  
is thus obtained by inserting this result into
\be 
\nn
\mathfrak{s}(t,S)  = 2 \sqrt{\pi} S^{3/2} e^{\textstyle\frac{(w_0-S)^2}{4 S}} \frac{1}{w_0} \int_0^\infty \rmd \du_t\, \du_t\, P(\du_t,S),
\ee 
where we have used (\ref{eq:SizeDist}). 
On this expression the limit of $w_0 \to 0$ is easy to take and provides the result for the stationary avalanches. 

One non-trivial check of this formula is that the resulting shape is properly normalized,
\begin{align*}
&\int_0^\infty \rmd t\,\mathfrak{s}(t,S) = \int_0^\infty \rmd t \int \rmd\du_t\,\du_t\,\frac{P(\du_t,S)}{P(S)} \\
&=  \int_{r_0 -i\infty}^{r_0+i\infty} \frac{\rmd r}{4 i \sqrt{\pi}}\frac{(r^2+4a\tau)}{r^2 \tau} \times \\
& \quad \times \exp \left[\frac{S (r+2)^2(r-2a\tau)^2}{16 r^2\tau^2}- \frac{S}{4}\right] \\
&\quad \times S^{\frac{3}{2}} e^{\textstyle\frac{(w_0-S)^2}{4 S}} \exp \left[ \frac{w_0}{2} (1- \frac{(2+r)(r-2a\tau)}{2r\tau}) \right] \\
&= \int_{-i\infty}^{i\infty} \frac{\rmd p}{4 i \sqrt{\pi} } S^{\frac{3}{2}} e^{\textstyle\frac{(w_0-S)^2}{4 S}} e^{\frac{S p^2}{16} + \frac{w_0}{2} (1- \frac{p}{2}) - \frac{S}{4}} = S,
\end{align*}
where in the second line we used the substitution $p = \frac{(r+2)(r-2a\tau)}{r\tau}$.

\begin{figure}
\includegraphics[width=\columnwidth]{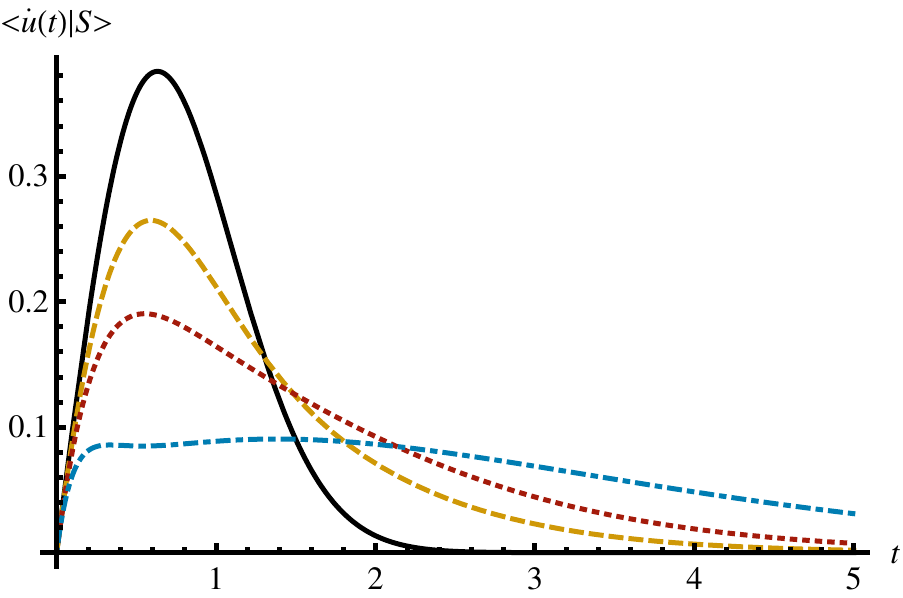}
\caption{(Color online). Avalanche shape at fixed size $S=0.8$, $\tau=0.5$ and $w_0=0$. Curves are, from top (black solid line) to bottom (blue dot-dashed line), $a=0$ (standard ABBM model, as given by Eq.~\eqref{eq:ShpABBM}), $a=1$, $a=2$, $a=5$.}
\label{fig:ShapeFixedSize}
\end{figure}

The integral \eqref{eq:ShpBeforeLT2} could be calculated in closed form in the case $a=0$ 
of the standard ABBM model. There one finds
\be 
\int \rmd\du_t\,\du_t\,P(\du_t,S) = \frac{(2 t + w_0) w_0 e^{-\frac{S^2 - 2 S w_0 + (2 t+w_0)^2}{4 S} } }{2 \sqrt{\pi } S^{3/2}}, \nn
\ee 
and the result for the shape,
\be 
\label{eq:ShpABBM}
\mathfrak{s}(t,S) =  (2 t + w_0) e^{-t(t+w_0)/S}.
\ee 
In the limit of an infinitesimal kick $w_0 \to 0$ we thus obtain the shape at fixed size for the standard ABBM model ($a=0$) 
for stationary avalanches as
\be 
\label{eq:ShpABBM2}
\mathfrak{s}(t,S) =  2 t  e^{-t^2/S}.
\ee 
For $a>0$ we could not find a closed expression, however the integral 
\eqref{eq:ShpBeforeLT2} is easily evaluated numerically. 
Some example curves are shown in figure \ref{fig:ShapeFixedSize} in the limit of small $w_0$. 
Observe that especially for large values of $a$, the additional time scale introduced 
by the eddy-current relaxation is clearly visible. 
Overall the shape stretches longer in time, and becomes non-monotonous, as  $a$ is increased.

\subsubsection{Tail of the shape function}

The behaviour of the avalanche shape for long times can also be understood analytically from equation \eqref{eq:ShpBeforeLT2}. 
For simplicity, we consider the case $w_0 =0$ in the following. For large $t$, the integral is dominated by its saddle-point. Since we have 
$e^{\frac{2 a t}{r}} \gg e^{-\frac{t r}{2 \tau}}$ for all times, the dominant exponential factor is
\beq
\nn
e^{H(r)} := \exp\! \left(\frac{S (r+2)^2(r-2a\tau)^2}{16 r^2\tau^2} - \frac{S}{4}- \frac{t}{\tau} + \frac{2at}{r}\right).
\eeq
Its maximum for large $t$ is obtained by solving $H'(r) = 0$:
\beq
\nn
r_m \approx 2 \sqrt[3]{\frac{2a t \tau ^2}{S}}+\frac{2}{3}
   (a \tau -1)  + \mathcal{O}(t)^{-1/3}.
\eeq
Determining the location of the saddle-point to higher order in $t$ is more complicated. The terms of $\mathcal{O}(t)^{-1/3}$ 
depend on whether the sub-exponential terms in \eqref{eq:ShpBeforeLT2} are included 
in the maximization procedure or not. However, to the order given here, $r_m$ is independent of such choices. 

\begin{figure}\includegraphics[width=\columnwidth]{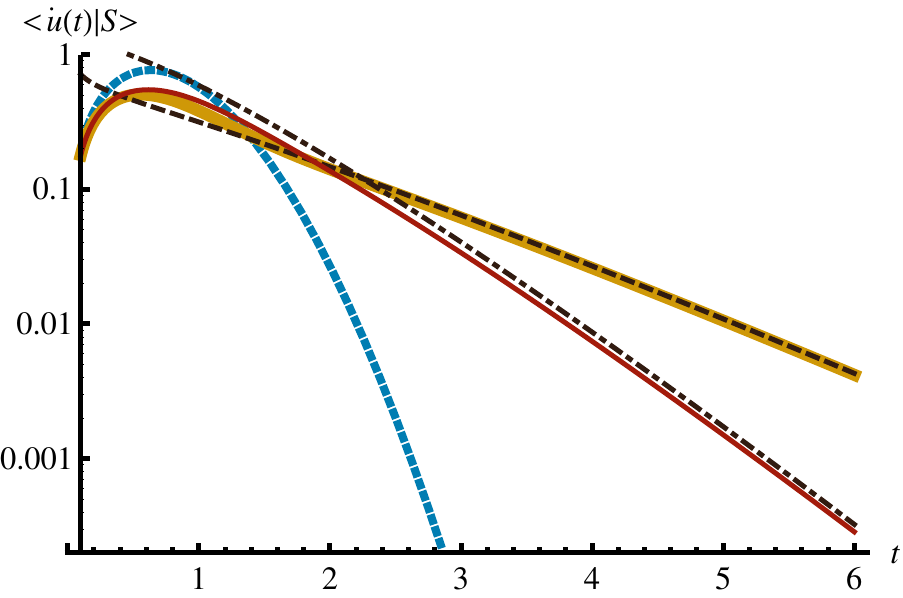}\caption{(Color online) Tail of the avalanche shape at fixed size $S=0.8$. Blue dotted line: Gaussian ABBM tail \eqref{eq:ShpABBM}. Red (thin) and yellow (thick) lines: numerical Laplace inversion of \eqref{eq:ShpBeforeLT2}, for $\tau=0.7$ and $\tau=0.4$, respectively. Black dot-dashed and dashed lines: Asymptotics \eqref{eq:ShpRetardedTail} for the corresponding values. $a=1$ in all cases. }\label{fig:ShapeTail}\end{figure}

Since the integral \eqref{eq:ShpBeforeLT2} does not depend on $r_0$ as discussed above, we can choose $r_0 = r_m$. Setting $r=u + iv$, we have $\partial_v^2 H(r) = -\partial_u^2 H(r)$ due to the Cauchy-Riemann equations.  Thus, we can approximate the integral \eqref{eq:ShpBeforeLT2} for large $t$ and fixed $S$ by

\begin{align}
\nn
\mathfrak{s}(t,S) \approx & \sqrt{\frac{2 \pi}{H''(r_m)}}\frac{(r_m-2 a\tau) (r_m+2) a }{4\pi \tau 
   r_m^3}\frac{e^{H(r_m)}}{P(S)} \\
\nn
\approx &
\exp\left[-\frac{t}{\tau
   }+\frac{3}{2} \left(\frac{at}{\tau}\right)^{2/3}\left(\frac{S}{2}\right)^{1/3}+ \right. \\
        \nn & +\left(\frac{at}{\tau}\right)^{1/3}\frac{1-a \tau}{\tau}\left(\frac{S}{2}\right)^{2/3} + \\
        \nn
        & \left. +\frac{S \left( a^2 \tau ^2-5 a \tau +1 \right)}{6 \tau ^2} + \mathcal{O}(t)^{-1/3}\right]\times \\
        & \times \left[\frac{S^{4/3}a^{2/3}}{\sqrt{3}(2t)^{1/3}\tau^{2/3}} + \mathcal{O}(t)^{-2/3}\right].
\label{eq:ShpRetardedTail}
\end{align}
Again, note that to this order both the exponent and the prefactor are independent of whether the sub-exponential terms are included in the maximization. In particular, the term of order $\mathcal{O}(t)^0$ in the exponent is independent of the term of order $\mathcal{O}(t)^{-1/3}$ in $r_m$. 
We thus see that the Gaussian tail of the shape in the standard ABBM model is replaced by an exponential tail, decaying on a time scale $ \tau$. This is confirmed by numerical Laplace inversion of Eq.~\eqref{eq:ShpBeforeLT2}, see figure \ref{fig:ShapeTail}. 
We also observe good agreement between the asymptotic expansion and the numerical result.

Using a similar method one can try to determine the tail of $\mathfrak{s}(t,S)$ for fixed $t$, at large $S$. 
In this limit, the maximum obtained by solving $H'(r) = 0$ is
\beq
\nn
r_m \approx 2 a \tau + \mathcal{O}(S)^{-1}.
\eeq
One finds
\begin{align*}
H(r_m) & =  - \frac{S}{4} + \mathcal{O}(S)^{-1}, \\
H''(r_m) & = \frac{(1+a\tau)^2}{8a^2 \tau^4} S + \mathcal{O}(S)^{0}.
\end{align*}
Noting that $P(S) \sim e^{-S/4}$, this means that the exponential factor in the saddle-point contribution to $\mathfrak{s}(t,S) \sim e^{H(r_m)}/P(S)$ vanishes to leading order. This indicates that $\mathfrak{s}(t,S)$ will scale as a power-law for fixed $t$ at large $S$. However, since the pre-exponential factors in \eqref{eq:ShpBeforeLT2} also vanish at $r_m = 2 a \tau$, obtaining a quantitative result requires a more controlled approximation.

\section{Sub-avalanche statistics\label{sec:AvStatDur}}

As we saw in section \ref{sec:VelKick}, an avalanche in the ABBM model with exponential retardation never strictly terminates, even after the driving has stopped.  It is thus interesting to explore the ``sub''-avalanches, or \textit{aftershocks}, inside an avalanche, and their durations $T_i$, \(i=1,...\). $T_i$ is defined  as the time it takes $\du$ to start from $\du=0$ at time \(t_{i-1}=\sum_{j=1}^{i-1}T_j\), go to positive values $\du >0$ and back to $\du=0$ at time \(t_{i}=\sum_{j=1}^{i}T_j\), without  touching $\du=0$ in the interval $]0t_{i-1},t_{i}[$.  In other words, it is the separation in time between successive passages of $\du$ at zero. The same question can be asked for avalanches at non-zero driving velocity $v>0$ in the standard ABBM model, which can also be seen as sub-avalanches of an infinite avalanche (there too $\du$ never vanishes on a finite time interval).\footnote{In the context of the standard ABBM models, these sub-avalanches are also called pulses \cite{WhiteDahmen2003}.} 
We will obtain detailed results in that case.

A convenient setting to study this problem is the Fokker-Planck approach, introducing an \textit{artificial} absorbing boundary at $\du=0$. It can be
implemented in the case of the simple exponential relaxation \eqref{tau} which reduces to two coupled Langevin equations for $\dot u$ and $h$ \eqref{eq:EOMExp}. We will
discover that, with such an absorbing boundary, equations \eqref{eq:RetSolution} and \eqref{eq:RetSolutionExp} for the generating functional of domain-wall velocities, as well as some details of the instanton method, need to be modified.

\subsection{Sub-avalanches in the standard ABBM model at finite driving velocity\label{sec:SubavABBM}}

In order to present our approach on a simple example, let us consider first the standard ABBM model with monotonous, but otherwise arbitrary driving $\dw(t) \geq 0$. 
The equation of motion \eqref{eq:EOMUd} with $a=0$ is, due to its Markovian nature,  completely characterized by the {\em propagator}
\beq
\mathcal{P}(\du_{\rm f};\tf | \du_{\rm i}; \ti) := E[\delta(\du_\tf-\du_{\rm f})|\du_\ti = \du_{\rm i}].
\eeq 
Here \(E\) means  ``expectation'', i.e.\ the average over the disorder.
As a function of the final velocity $\du_{\rm f}$, $P_t(\du) = \mathcal{P}(\du_{\rm f}=\du;\tf=t | \du_{\rm i}; \ti)$ satisfies the forward Fokker-Planck equation 
\begin{align}
\label{eq:FPEABBMForward}
        \partial_t P_t(\du)& = \partial_{\du}^2 \du P_t(\du) +\partial_{\du} (\du-\dw_t)P_t(\du) = \partial_{\du} J_t(\du),
\end{align}
where 
\beq
\label{eq:ABBMForwardCurrent}
J_t(\du) := \partial_{\du} \du P_t(\du) + (\du-\dw_t)P_t(\du)
\eeq
is (minus) the probability current. 
As a function of the initial condition $\du_{\rm i}$, $Q_t(\du) = \mathcal{P}(\du_{\rm f};\tf | \du_{\rm i}=\du; \ti=t)$ satisfies the backward Fokker-Planck equation
\begin{align}
\nn
        -\partial_t Q_t(\du)& = \du\partial_{\du}^2 Q_t(\du) - (\du-\dw_t)\partial_{\du} Q_t(\du)\\\label{eq:FPEABBMBackward}
        & = \partial_{\du}^2 \du Q_t(\du) + \partial_{\du}(\dw_t-2-\du) Q_t(\du) + Q_t(\du).
\end{align}
The propagator also satisfies the initial condition
\beq
\label{eq:PropABBMIC}
\mathcal{P}(\du_{\rm f};\ti | \du_{\rm i}; \ti)=\mathcal{P}(\du_{\rm f};\tf | \du_{\rm i}; \tf) = \delta(\du_{\rm f}-\du_{\rm i}).
\eeq
In order to obtain the solution of the forward or backward FPEs, one needs to complement them with a boundary condition at $\du=0$. We consider two cases:

(i) Propagator with \textit{reflecting boundary} $\mathcal{P}_{\text{refl}}$. This is the case we studied so far in this article, and in \cite{DobrinevskiLeDoussalWiese2012}. It is defined by a {\em vanishing probability current}: $J_t(\du=0^+)=0$, or
\beq
\label{eq:ReflABBMCurrent}
\lim_{\du \to 0} \partial_{\du} \du P_t(\du) + (\du-\dw_t)P_t(\du) = 0.
\eeq
Typically, this is satisfied by a power-law-like behaviour $P(\du)\sim \du^{-1+\dw_t}$ for $\du \to 0$ (see the examples in \cite{AlessandroBeatriceBertottiMontorsi1990,Colaiori2008,DobrinevskiLeDoussalWiese2012}).

(ii) Propagator with \textit{absorbing boundary} $\mathcal{P}_{\text{abs}}$. This is the relevant case  for  sub-avalanches. The problem is characterized  by a {\em vanishing propagator}, when starting from $\du=0^+$, i.e.\ $Q_t(\du=0^+)=0$. This implies (except for pathological cases) that the ``current'' for the backwards FPE vanishes, 
\beq
\label{eq:AbsABBMCurrent}
\lim_{\du \to 0} \partial_{\du} \du Q_t(\du) + (\dw_t-2-\du)Q_t(\du) =0.
\eeq
On the other hand, for the forward Fokker-Planck equation, $P_t(\du=0^+)= \mathcal{P}_{\text{refl}}(\du_{\rm f}=0^+,\tf=t|\du_{\rm i};\ti)$ will typically be a non-vanishing, non-trivial function of time, and the current $J_t$ will not vanish (as expected from physical intuition, since trajectories touching $\du=0$ are ``absorbed''). This is why treating the absorbing boundary using the forward Fokker-Planck equation is inconvenient; instead, the backward equation is natural here\footnote{Similarly, for the reflecting boundary, $Q_t(\du=0^+) = \mathcal{P}_{\text{refl}}(\du_{\rm f},\tf|\du_{\rm i}=0^+;\ti=t)$ will typically be a non-vanishing, non-trivial function of time. So, for the reflecting boundary, the backward equation is inconvenient and the forward equation is natural. For the absorbing boundary it is the other way around. This peculiar behaviour is due to the nature of the ABBM noise term, which vanishes for $\du=0$. For a standard Brownian motion, the absorbing boundary can be treated equally well using the forward or the backward Fokker-Planck equation: There we have $\mathcal{P}_{\rm abs}(\du_{\rm f}=0)=\mathcal{P}_{\rm abs}(\du_{\rm i}=0)=0.$}.

Let us now define Laplace transforms with respect to the final velocity  for $\hP$ and with respect to the initial velocity for $\hQ$:
\begin{align}
        \hP_{t}(\lambda|\du_{\rm i},\ti) & := \int_0^\infty\rmd \du_{\rm f}\,e^{\lambda\du_{\rm f}} \mathcal{P}(\du_{\rm f};\tf=t | \du_{\rm i}; \ti), \\
        \hQ_{t}(\lambda|\du_{\rm f},\tf) & := \int_0^\infty\rmd \du_{\rm i}\,e^{\lambda\du_{\rm i}} \mathcal{P}(\du_{\rm f};\tf | \du_{\rm i}; \ti = t). 
\end{align}
Laplace-transforming the forward Fokker-Planck equation \eqref{eq:FPEABBMForward}, with the reflecting boundary condition \eqref{eq:ReflABBMCurrent}, we obtain a first-order PDE for $\hP_t(\lambda)$
\beq
\label{eq:ReflABBMLT}
\partial_t \hP_t^{\text{refl}}(\lambda) = \left(\lambda^2- \lambda \right)\partial_\lambda \hP_t^{\text{refl}}(\lambda) + \dw_t \lambda \hP_t^{\text{refl}}(\lambda).
\eeq
Note that while boundary terms would arise in general, the reflecting boundary condition \eqref{eq:ReflABBMCurrent} ensures that they vanish.
Similarly, Laplace-transforming the backward Fokker-Planck equation \eqref{eq:FPEABBMBackward} with the absorbing boundary condition \eqref{eq:AbsABBMCurrent}, we obtain a first-order PDE for $\hQ_t(\lambda)$:
\begin{align}
\nn
-\partial_t \hQ_t^{\text{abs}}(\lambda) =& \left(\lambda^2+ \lambda \right)\partial_\lambda \hQ_t^{\text{abs}}(\lambda)  \\
\label{eq:AbsABBMLT}
&+ \left[(2-\dw_t)\lambda+1 \right] \hQ_t^{\text{abs}}(\lambda) .
\end{align}
 Again,  vanishing of boundary terms for the Laplace transformation is ensured by the absorbing boundary condition \eqref{eq:AbsABBMCurrent}.
On the other hand, the Laplace-transformed equations for $\hP_t^{\text{abs}}$ and $\hQ_t^{\text{refl}}$ are more complicated: There, the boundary terms at $\du=0$ do not vanish and are undetermined functions of time. Thus, in the following, we will always use the propagator in terms of the final condition $P$ or its Laplace-transform $\hP$ when discussing a reflecting boundary, and the propagator in terms of the initial condition $Q$ or its Laplace-transform $\hQ$ when discussing an absorbing boundary. 

Now, Eqs.\ \eqref{eq:ReflABBMLT} and \eqref{eq:AbsABBMLT} can be solved using the method of characteristics. In the
forward (reflecting boundary) case this method was shown to provide a general connection between the Fokker-Planck approach to the ABBM model and the dynamical path integral (instanton equation) approach  \cite{LeDoussalWiese2012a, LeDoussalPetkovicWiese2012}.  In the following we take a first step towards generalizing this to the case of an absorbing boundary. The solution of \eqref{eq:AbsABBMLT} is
\begin{align}
\label{eq:SolBackwardABBM}
&\hQ_\ti\big(\hu(\ti)\big) = e^{ \int_{\ti}^{\tf} \left[\left(2-\dw_t\right)\hu_t + 1 \right]\rmd t}\hQ_\tf\big(\hu(\tf)\big),
\end{align}
where $\hu$ satisfies the backward instanton equation \beq
\label{eq:AbsInstantonABBM}
-\partial_t \hu_t + \hu_t + \hu_t^2 = 0,
\eeq
with the boundary condition $\hu( \ti) = \lambda$.\footnote{Note that  this equation is causal -- solved by increasing the time -- in contrast to the forward instanton.}

The initial condition \eqref{eq:PropABBMIC} for the propagator   gives $\hQ_\tf (\lambda_f) = e^{\lambda_f \du_{\rm f}}$. Inserting this into \eqref{eq:SolBackwardABBM}, we obtain 
\begin{align}
\nn
&\int_0^\infty \rmd \du_{\rm i} \,e^{\lambda \du_{\rm i} }\mathcal{P}_{\text{abs}}(\du_{\rm f};\tf|\du_{\rm i};\ti) = \hQ_{\ti} (\lambda) \\
\label{eq:SolAbsPropABBM}
&= \exp\left\{\int_{\ti}^{\tf}\left[(2-\dw_{s})\hu_s + 1 \right]\rmd s + \du_{\rm f} \hu_\tf \right\}.
\end{align}
It is useful to recall that the fact that this solves the problem with an {\it absorbing boundary}
is an indirect consequence of our chain of arguments: It stems from the fact that we use 
the backward instanton equation (\ref{eq:AbsInstantonABBM}) which encodes the
solution (via the method of characteristics) of the Laplace-transformed backward FPE, which itself
contains {\it no boundary term} precisely in the case of an absorbing boundary.
For the case of a reflecting boundary, the analogous formula is
\begin{align}
\nn
&\int_0^\infty \rmd \du_{\rm f} \,e^{\lambda \du_{\rm f} }\mathcal{P}_{\text{refl}}(\du_{\rm f};\tf|\du_{\rm i};\ti) = \hP_{\tf} (\lambda) \\
&= \exp\left[\int_{\ti}^{\tf}\dw_s \tu(s)\, \rmd s + \du_{\rm i} \tu(\ti) \right].
\end{align}
As discussed in Ref.~\cite{LeDoussalPetkovicWiese2012}, this is equivalent to solving \eqref{eq:EOMUd} as above using the MSR field theory and the instanton equations. For details, see \cite{LeDoussalPetkovicWiese2012} section V C, and, in the present article, section \ref{sec:Characteristics}, where we discuss this in general for the ABBM model with retardation.

\subsubsection{Propagator at constant driving\label{sec:AbsPropABBM}}
Let us now apply \eqref{eq:SolAbsPropABBM} in order to determine the propagator of the standard ABBM model with an absorbing boundary, at a constant driving velocity $0<v<1$.

The solution of Eq.~\eqref{eq:AbsInstantonABBM} is
\beq
\label{eq:AbsInstSolABBM}
\hat u(t) = \frac{\lambda\,\theta({ t-\ti})}{e^{ \ti-t}(\lambda+1) - \lambda}
.\eeq
Inserting this into \eqref{eq:SolAbsPropABBM}, we obtain the Laplace transform (w.r.t.\ the initial condition) of the propagator with an absorbing boundary, at a constant driving velocity $v$,
\begin{align}
\nn
&\int\rmd \du_{\rm i} \,e^{\lambda \du_{\rm i} }\mathcal{P}_{\text{abs}}(\du_{\rm f};\tf|\du_{\rm i};\ti) =  \hQ_{\ti} (\lambda) \\
\nn
        & = \exp\left[ \int_{\ti}^{\tf} \left[\left(2-v\right)\hu(t) + 1 \right] \rmd t + \hu({ \tf}) \du_{\rm f}\right] \\
        \nn
        & = \exp\left[ (2-v)\int_{\ti}^{\tf} \hu(t)\rmd t + (\tf-\ti) + \hu({ \tf}) \du_{\rm f}\right] \\
        \nn
        & = \exp\left[\frac{\lambda e^{\tf-\ti } \du_{\rm f}}{1-\lambda \left(e^{\tf-\ti
   }-1\right)}+\tf-\ti \right] \times \\
        \label{eq:AbsPropABBMLT}
        & \quad \times \left[1-\lambda \left(e^{\tf-\ti
   }-1\right)\right]^{v-2}.
\end{align}
Inverting the Laplace transform from $\lambda$ to $\du_{\rm i}$ yields the propagator of the ABBM model with an absorbing boundary at $\du=0$,
              \begin{align}
\label{eq:AbsPropABBM}
&\mathcal{P}_{\text{abs}}(\du_{\rm f};\tf|\du_{\rm i};0) = \\
\nn
        & \exp \left(\frac{v}{2}\tf-\frac{\du_{\rm f} e^{\tf }+\du_{\rm i}}{e^{\tf
   }-1}\right)\frac{\left(\du_{\rm f} / \du_{\rm i}\right)^{\frac{v-1}{2}} }{2\sinh \frac{\tf}{2}}  I_{1-v}\left(\frac{\sqrt{\du_{\rm f}
   \du_{\rm i}}}{\sinh \frac{\tf}{2}}\right).
\end{align}
Here we set $\ti=0$ for simplicity since the result depends only on $t_{\rm f}-t_{\rm i}$. 

Our result is identical to  Eq.~(37) in the recent calculation \cite{LeblancAnghelutaDahmenGoldenfeld2013}, there obtained  using completely different methods (decomposition in eigenfunctions of the Fokker-Planck operator). The advantage of our approach is that it makes the connection to field theory clearer, and that it is easily generalizable to situations with a non-constant driving velocity (where the eigenfunction method is not applicable).

It is straightforward to check explicitly that \eqref{eq:AbsPropABBM} satisfies the backward FPE \eqref{eq:FPEABBMBackward}. Near $\du_{\rm i} = 0$ we have $\mathcal{P}_{\text{abs}}(\du_{\rm f};\tf|\du_{\rm i};0) \sim \du_{\rm i}^{1-v}$. For $0\leq v \leq 1$, it satisfies the absorbing boundary condition $\mathcal{P}_{\text{abs}}(\du_{\rm f};\tf|\du_{\rm i}=0;\ti)=0$ and \eqref{eq:AbsABBMCurrent}. On the other hand, near $\du_{\rm f} = 0$, $\mathcal{P}_{\text{abs}}$ is a non-vanishing constant, so the ``forward'' current \eqref{eq:ABBMForwardCurrent} does not vanish. For velocities $v>1$, our assumption on the absorbing boundary condition, $\mathcal{P}_{\text{abs}}(\du_{\rm f};\tf|\du_{\rm i}=0;\ti)=0$ and \eqref{eq:AbsABBMCurrent}, seems to break down. In fact, the absorbing boundary becomes equal to the reflecting one since $\du=0$ is unreachable\footnote{In fact, for $v>1$, one can still formally define avalanche-size distributions {\it conditioned} to $\dot u$  being close to 0 instead of precisely 0 as for $0\leq v <1$, 
see the appendix in \cite{LeDoussalWiese2009}.}. 

\subsubsection{Sub-avalanche durations in the standard ABBM model\label{sec:DurationABBM}}
Another simple example is the ``survival probability'', i.e.\ the probability never to touch the absorbing boundary $\du=0$ until time $\tf$, starting at some initial $\du_{\rm i}$ at time $\ti$:
\beq
\nn
 P_{\rm surv}(\tf |\du_{\rm i};\ti) := \int \rmd \du_{\rm f}\, \mathcal{P}_{\rm abs} (\du_{\rm f};\tf|\du_{\rm i};\ti)
\eeq
Its leading behaviour as $\du_{\rm i} \to 0$ corresponds to the probability of an avalanche duration $T > \tf-\ti$.

It can be obtained using \eqref{eq:SolBackwardABBM} as in the previous section, but with the initial condition 
\beq
\hQ_{\tf} (\lambda_f) = \int_0^\infty \rmd \du\,e^{\lambda_f \du} = -\frac{1}{\lambda_f},
\eeq
for $\lambda_f < 0$.

As above, inserting the solution \eqref{eq:AbsInstSolABBM} for $\hu(t)$, we obtain for the survival probability at time $\tf$ 
\begin{align}
\nn
&\int \rmd \du_{\rm i}\,e^{\lambda \du_{\rm i}} P_{\rm surv}(\tf|\du_{\rm i};\ti)= \nn\\ 
&=  \exp\left\{\int_{\ti}^{\tf}\left[(2-\dw_{s})\hu_s + 1 \right]\rmd s \right\}\left[-\frac{1}{\hu({\tf})}\right]  \nn\\
\nn
&=- \left[1+\lambda-\lambda e^{t_{\rm f}-t_{\rm i}}\right]^{v-2}\frac{(\lambda+1)
- \lambda e^{\tf-t_{\rm
i}} }{\lambda} \\
&=-\frac{\left[1+\lambda-\lambda e^{t_{\rm f}-t_{\rm i}}\right]^{v-1}}{\lambda}
\label{eq:ABBMPhatSurv}
\end{align}
This is, of course, identical to the integral of \eqref{eq:AbsPropABBMLT} over $\du_{\rm f}$. 
Inverting the Laplace transform, we obtain the survival probability at time $\tf>\ti$ when starting from $\du_{\rm i}$ at $t_i$. It is a function of $ \tf-\ti$ only.  For
$v<1$ it reads
\beq
 P_{\rm surv}(\tf-\ti |\dot u_{\rm i}) = 1- \frac{\Gamma\left(1-v,\frac{\du_{\rm i}}{e^{\tf-\ti}-1}\right)}{\Gamma(1-v)}.
\eeq
It increases from $0$ to $1$ as $\dot u$ increases from $0$ to $\infty$,
 while for $v \geq 1$ it is equal to unity for all $\dot u>0$, i.e.\ there are no zeroes of the velocity. Note that this result was obtained independently 
in \cite{MasoliverPerello2012} and \cite{LeblancAnghelutaDahmenGoldenfeld2013} by completely  different methods.
It can also be obtained by integrating the propagator \eqref{eq:AbsPropABBM} over $\du_{\rm f}$ from $0$ to $\infty$.

 The case $v<1$ is considered from now on. 
Taking a derivative w.r.t.\ the final time gives the probability density of first-passage times $T_0$ for $\du$ to become zero, given an initial velocity $\du_{\rm i}$,
\beq
\label{eq:ABBMPFirst}
P_{\rm first}(T_0|\dot u_{\rm i}) = \frac{ \du_{\rm i}^{1-v}}{\Gamma(1-v)(e^{T_0}-1)^{2-v}} e^{-\frac{\du_{\rm i}}{e^{T_0}-1} + T_0}.
\eeq
Since an avalanche always starts at $\du=0^+$, we can extract from the leading term in small $\du_{\rm i}$ a density of durations $T_0$,
\beq
v \rho_{\rm duration}(T_0) = 
e^{T_0} (1-v)
   \left(e^{T_0}-1\right)^{v-2}\frac{\sin (\pi 
   v)}{\pi }
\ .\eeq
In the limit $v\to 0$, density means units of $1/u \equiv 1/w$. Since $\rho_{\rm duration}(T_0)$ diverges like $1/T_0^{v-2}$, hence is not normalizable for $v<1$, we 
have chosen to normalized it as $\left<v T_0\right>_\rho=1$. Note that $\rho$ has a finite limit at $v=0^+$ 
which agrees with the avalanche-duration density obtained in
\cite{LeDoussalWiese2011}.

The probability density for an avalanche to continue a time  $T_0$  beyond an arbitrary chosen time $\ti$ is obtained by integrating over the distribution of $\du_0$ in the stationary state $P_{\rm stat}(\du_0) = \du_0^{-1+v} \frac{e^{-\du_0}}{\Gamma(v)}$, 
\begin{align}\nn
P_{\rm beyond}( T_0) &= \int_0^\infty \rmd \du_{\rm i}\,P_{\rm stat}(\du_{\rm i}) P_{\rm first}(T_0 | \dot u_{\rm i}) \\
& = \left(e^{T_0}-1\right)^{v-1} \frac{\sin (\pi  v)}{\pi }\ .
\end{align}
Note that this is a bona-fide probability distribution (normalized to unity).

Finally, we obtain again the density of avalanche durations by taking a derivative, 
\begin{align}
\label{eq:PofT}
v \rho(T) &= -\partial_{T_0}\big|_{T = T_0}P_{\rm beyond}(T ) \nn\\
&= \frac{e^T}{\left(e^T-1\right)^{2-v}} (1-v)\frac{\sin (\pi  v)}{\pi }\ .
\end{align}
This density can be interpreted as a probability $v \rho(T) \equiv P(T)$ in the following sense: Take a random time $t=0$. The velocity $\du_0$ is positive with probability 1. Thus, there is a first-passage time $-t_1$ at zero velocity on the left of $t=0$, and a first-passage time $+t_2$ at zero velocity on the right of $t=0$. The duration $T=t_2-(-t_1)=t_2+t_1$ is the distance between the two. It is thus normalized not by $\int_0^\infty P(T)\rmd T = 1$, but rather by
\beq
\nn
\int_0^\infty \rmd t_1 \int_0^\infty \rmd t_2\, P(T=t_2+t_1) =1.
\eeq
In terms of units, $P(T)$ is not a  probability density (which would give a dimensionless number when multiplied by a time interval $\rmd T$), but a ``double''-probability density which gives a dimensionless number when multiplied by \textit{two} time intervals $\rmd t_1$, $\rmd t_2$. In other words
$\int  T P(T) \rmd T=1$ since the probability that a randomly chosen time belongs to an avalanche is proportional to its duration $T$.

Note that for $v \to 0$ Eq.~\eqref{eq:PofT} reduces to the result known from \cite{Colaiori2008,LeDoussalWiese2011,DobrinevskiLeDoussalWiese2012}. The velocity-dependent power law for small $T$, $P(T) \propto T^{-2+v}$, was already predicted in \cite{Colaiori2008}. A similar result for the distribution of avalanche sizes at finite velocity in the standard ABBM model is discussed in appendix \ref{sec:AppABBMSizes}.

\subsection{Fokker-Planck equations and propagator including retardation}

Now let us go back to the more general ABBM model with retardation. 

The equations of motion \eqref{eq:EOMUdExp} are equivalent to a Fokker-Planck equation\footnote{\label{fn:Ito}To derive the forward Fokker-Planck equation (\ref{eq:FPEForward}), we set 
$$P_t(\dot u,h) = \left< \delta(\dot u-\dot u_{t}) \delta( h- h_{t})\right>\ .$$ 
Then apply It\^o calculus to $$ P_{t+\rmd t}(\dot u,h) = \left< \delta(\dot u-[\dot u_{t}+\rmd \dot u_{t}]) \delta( h- [h_{t}+\rmd h_{t}])\right>, $$ with  
$$
\rmd \du_{t} = \sqrt{\du_t} \rmd B_t + [ (\dw_{t}- \du_{t}) -a  \du_{t}
+ a  h_{t}]\rmd t\ ,$$ 
$$\tau\rmd h_{t} = [\du_{t} -h_{t}]\rmd t
$$ $$\left< \rmd B_t\rmd B_{t'} \right>= 2 \delta(t-t') \rmd t.$$
For the backward equation \eqref{eq:FPEBackward},  apply It\^o calculus to $$Q_{t+\rmd t}(\du_{t+\rmd t},h_{t+\rmd t}) = Q_{t}(\du_{t},h_{t})\ . $$ 
} generalizing \eqref{eq:FPEForward} for the joint probability distribution $P(\du,h)$,
\begin{align}
\nn
& \partial_t P_t(\du,h) \\ \nn
&= \partial_{\du}^2 \du P_t(\du,h)+\partial_{\du} \left(\du - \dw_t + a \du -a h\right)P_t(\du,h) \\
\label{eq:FPEForward}
 &~~~+\tau^{-1} \partial_h \left(h - \du\right)P_t(\du,h).
\end{align}
As discussed in \cite{LeDoussalPetkovicWiese2012}, this forward Fokker-Planck equation provides an alternative derivation for the generating function \eqref{eq:RetSolutionExp}. The instanton equations \eqref{eq:RetInstantonExpTu}, \eqref{eq:RetInstantonExpTh} are equivalent to the equations for the characteristics of the linear PDE \eqref{eq:FPEForward}, see \cite{LeDoussalPetkovicWiese2012} section V C for details. 
The transformation between the ``real space'' $\du,h$ and the ``Laplace space'' $\tu,\tih$ of the characteristics, i.e.~instantons is a very useful tool
whenever boundary terms are absent. This is the case for a zero probability current at $\du=0$, i.e.\ for a reflecting boundary condition.

To study the case of an absorbing boundary, as noted above for the pure ABBM model, it is useful to consider the flow of the probability density as a function of the initial condition $\du,h$ at time $\ti=t$, which satisfies the \textit{backward} Fokker-Planck equation${}^{\ref{fn:Ito}}$ \cite{Gardiner}
\begin{align}
\label{eq:FPEBackward}
&-\partial_t Q_t(\du,h) \nn\\
&= \du \partial_{\du}^2 Q_t(\du,h)- \left(\du - \dw_t + a \du -a h\right)\partial_{\du} Q_t(\du,h)\nn \\
& ~~~-\tau^{-1} \left(h - \du\right)\partial_h Q_t(\du,h) \\ 
 &= \partial_{\du}^2 \du Q_t(\du,h) - \partial_{\du} \left(\du + 2 - \dw_t + a \du -a h\right)Q_t(\du,h)\nn \\
\nn
&  ~~~-\tau^{-1} \partial_h\left(h - \du\right) Q_t(\du,h) + \left(1+a+\tau^{-1}\right)Q_t(\du,h)
\end{align}
Both \eqref{eq:FPEForward} and \eqref{eq:FPEBackward} are linear in the probability density $P$ or $Q$. Hence, they are completely characterized by the Green function or \textit{propagator} $\mathcal{P}(\du_{\rm f},h_{\rm f};t|\du_{\rm i},h_{\rm i},0)$, the probability to go from $\du_{\rm i},h_{\rm i}$ at $\ti$ to $\du_{\rm f}, h_{\rm f}$ at $\tf > \ti$. It satisfies \eqref{eq:FPEForward} as a function of $t=\tf, \du=\du_{\rm f},h=h_{\rm f}$ and  \eqref{eq:FPEBackward} as a function of $t=\tf, \du=\du_{\rm i},h=h_{\rm i}$; it  has the initial condition $\mathcal{P}(\du_{\rm f},h_{\rm f};\ti|\du_{\rm i},h_{\rm i},\ti) = \delta(\du_{\rm f}-\du_{\rm i})\delta(h_{\rm f}-h_{\rm i})$.

\subsection{Monotonicity, domains of definition, and boundaries}

It is important to note that $\dot u$ and $h$ satisfy together a monotonicity property: If $\dot w \geq 0$ and both $\dot u(t=0) \geq 0$ 
and $h(t=0) \geq 0$, then they remain so at all times. Although the quadrant $\dot u \geq 0$, $h \geq 0$ is the more physical one, 
as we see below it is  convenient to solve the FP equations in the half-space $\dot u \geq 0$ and then at the end restrict to the quadrant $h \geq 0$.
The reason for this is simple: No matter whether we impose an absorbing or a reflecting boundary at $\du=0$, one has a finite probability of reaching $\du_{\rm f} > 0, h_{\rm f} > 0$ starting from $\du_{\rm i} > 0, h_{\rm i} < 0$. So, as a function of the initial value $h_{\rm i}$, $\mathcal{P}(\du_{\rm f},h_{\rm f};\ti|\du_{\rm i},h_{\rm i},\ti)$ is smooth around $h_{\rm i}=0$, and has a finite value there. Its natural boundary is at $h_{\rm i} \to -\infty$, where $\mathcal{P} \to 0$. Thus, when considering the backward equation, we will work on the half-space $\dot u_{\rm i} \geq 0, h_{\rm i} \in \mathbb{R}$ in the following, in order to avoid undetermined boundary terms.

Due to the aforementioned monotonicity property, we know that the restriction to $h_{\rm i} > 0$ of the half-space propagator obtained in this way, will actually be equal to the propagator restricted to the physical quadrant from the beginning.

By taking linear functionals of the propagator, one obtains other observables. We give a few examples:

(i) Starting not from a fixed point, but from a distribution of initial values $P_{\rm i}(\du_{\rm i},h_{\rm i})$, the probability density of final values
\beq
 P_t(\du,h)=\int_0^\infty \rmd\du_{\rm i} \int_{-\infty}^\infty \rmd h_{\rm i}\,P_{\rm i}(\du_{\rm i},h_{\rm i})\mathcal{P}(\du,h;t|\du_{\rm i},h_{\rm i};\ti)
\eeq
 still satisfies the forward FPE \eqref{eq:FPEForward}. The initial condition is, naturally, $P_\ti = P_{\rm i}$.

(ii) One may be interested not in the probability density of the final point at $\du_{\rm f}, h_{\rm f}$ (given by the propagator), but the actual probability to land in a domain $D$, starting from an arbitrary initial condition $\du, h$. This is given by 
\begin{align}
&Q_t(\du,h)= \\
&\nn =\int_0^\infty \rmd\du_{\rm f} \int_{-\infty}^\infty  \rmd h_{\rm f}\,Q_i(\du_{\rm f},h_{\rm f})\mathcal{P}(\du_{\rm f},h_{\rm f};t|\du,h;\ti),
\end{align}
where $Q_i(\du_{\rm f},h_{\rm f}) = \mathbbm{1}_D$ is $1$ inside $D$ and $0$ outside. $Q_t(\du,h)$ satisfies the backward FPE \eqref{eq:FPEBackward}, with the initial condition $Q_\ti = Q_i = \mathbbm{1}_D$. This can be used to determine the distribution of avalanche durations, starting from an initial value $\du_{\rm i} > 0, h_{\rm i}$. Choosing $D$ to be the set $\{\du_{\rm f} > 0\}$, one obtains the probability to have any positive domain-wall velocity at $\tf$, i.e.\ the probability not to have touched the boundary $\du=0$ between $\ti$ and $\tf$

As mentioned above, our instanton solution \eqref{eq:RetSolutionExp} is, equivalent to the forward propagator $\mathcal{P}(\du_{\rm f},h_{\rm f};\tf|\du_{\rm i},h_{\rm i},\ti)$ from a given initial condition $\du_{\rm i}=0,h_{\rm i}=0$ to some final point $\du_{\rm f} > 0, h_{\rm f} > 0$, with a \textit{reflecting} boundary at $\du=0$ (the boundary at $h=0$ is unreachable when propagating forward). Imposing an \textit{absorbing} boundary at $\du=0$, as required for analyzing sub-avalanche durations, is less trivial. In contrast to the case of e.g.\ a standard Brownian motion, the probability current at the final point $\du_{\rm f}=0$ vanishes as soon as one sets $\mathcal{P}_{\text{abs}}(\du_{\rm f}=0,h_{\rm f};t|\du_{\rm i},h_{\rm i};\ti)=0$. The correct propagator with an absorbing boundary should thus have $\mathcal{P}_{\text{abs}}(\du_{\rm f}=0,h_{\rm f};t|\du_{\rm i},h_{\rm i};\ti)>0$, an undetermined function of time, as confirmed by the explicit calculation in section \ref{sec:AbsPropABBM} (see also the discussion in Appendix D of \cite{LeDoussalWiese2012a}). Hence obtaining it from the forward FPE \eqref{eq:FPEForward} is not easy. However, in terms of the initial condition, as motivated for the pure ABBM model above, we expect $\mathcal{P}_{\text{abs}}(\du_{\rm f},h_{\rm f};t|\du_{\rm i}=0,h_{\rm i},\ti)=0$. Then, the backward FPE \eqref{eq:FPEBackward} is easy to analyze using Laplace transforms, since the boundary term $Q(\du_{\rm i}=0,h_{\rm i})$ vanishes.

\subsection{Characteristics and instantons\label{sec:Characteristics}}

To solve the backward FPE, we define the Laplace transform $\hQ$ via
\beq
\label{eq:LTSurvProb}
\hQ_t(\lambda,\mu) := \int_0^\infty \rmd \du \int_{-\infty}^\infty \rmd h\, e^{\lambda \du + \mu h}Q_t(\du,h).
\eeq
Equation \eqref{eq:FPEBackward} then gives
\begin{align}
\nn
 -\partial_t \hQ_t(\lambda,\mu) =& \left[\lambda^2 + (1+a)\lambda - \tau^{-1} \mu\right] \partial_\lambda \hQ_t(\lambda,\mu)  \\
\nn
 &  \left[- a \lambda + \tau^{-1} \mu \right] \partial_\mu \hQ_t(\lambda,\mu) \\
\label{eq:FPEBackwardLT}
& + \left[\left(2-\dw_t\right)\lambda + \left(1+a+\tau^{-1}\right) \right] \hQ_t(\lambda,\mu).
\end{align}
Note that to obtain this equation (with vanishing boundary terms) we used $Q_t(\du=0,h)=0$, and the fact that the noise vanishes at $\du=0$ (which is a special property of the ABBM model). \footnote{Eq.~\eqref{eq:FPEBackwardLT} is valid for the LT (\ref{eq:LTSurvProb}) defined on the half-space $\dot u \geq 0,  h\in \mathbb{R}$. If one 
wishes to define the LT on the physical quadrant $\dot u \geq 0,h \geq 0,$ one cannot avoid a boundary term from $h=0$, since $Q_t(\du,h=0^+)$ does not vanish. Since there is no noise term in the evolution equation for $h$, the solution does not have to be continuous, and there is no contradiction to $Q_t(\du,h=0)=0$. One possibility to eliminate the boundary term is to add an extra diffusion term $\epsilon h \partial_h^2 Q_t$ on the r.h.s.\ of \eqref{eq:FPEBackward}, which leads to an additional term $\epsilon \mu^2 \partial_\mu \hat Q_t$ on the r.h.s.\ of \eqref{eq:FPEBackwardLT}, 
and $\epsilon \hat h^2$ on the l.h.s.\ of Eq.~\eqref{eq:RetInstAbsExpTh}. Then $Q_t$ vanishes at $h=0$, is continuous, and one can show that as $\epsilon \to 0$ it converges pointwise for any $h>0$ to the (discontinuous) restriction to the quadrant of the presently considered solution in  full space. We will however not need to further explore this method here, but can work with $h$ on the full real line.}

We now define the characteristics $\tu$, $\hat h$ via the \textit{backward} instanton equations 
\begin{align}
\label{eq:RetInstAbsExpTu}
&{ -}\partial_t \hat u(t) + (1 + a) \hat u(t) + \hat u(t)^2 - \hat h(t) = 0 \\
\label{eq:RetInstAbsExpTh}
&{ -}\tau\partial_t \hat h(t) +\hat h(t) - a \hat u(t)  =  0 \\
&{ -} \partial _t \hat q(t) =\big[(2-\dw_t)\hat u(t) + (1+a+\tau^{-1})\big]
\hat q(t)  
\end{align}
The boundary conditions are $ \hat u(\ti) = \lambda$, $ \hat h(\ti) = \mu/\tau$,  $\hat q(\ti) = \hat Q_{\ti}(\lambda,\mu)$. 
Note that the first two equations are identical to the standard instanton equations \eqref{eq:RetInstantonExpTu} and \eqref{eq:RetInstantonExpTh}.
The equation for $\hQ(\lambda,\mu)$ along a characteristic simplifies to
\begin{align}
& -\frac{\rmd}{\rmd t} \hQ_t\big(\hat u(t),\tau \hat h(t)\big) \nn \\
&=\big[(2-\dw_t)\hat u(t) + (1+a+\tau^{-1}) \big] \hQ_t\big(\hat u(t),\tau \hat h(t)\big). 
\end{align}
Its solution is
\bea
\label{eq:SolSurvProb}
\lefteqn{ \hQ_{ \ti}\big(\hat u({ \ti}),\tau\hat h({ \ti})\big)} & & \nn\\
\nn
&=& \exp\left\{\int_{\ti}^{\tf}\left[(2-\dw_{s})\hat u(s) + (1+a+ \tau^{-1}) \right]\rmd s\right\} \times \\
& & \times\hQ_{ \tf}\big(\hat u({ \tf}),\tau\hat h({ \tf})\big). 
\eea
The initial condition $\hQ_\tf$ depends on the observable we want to compute. For concreteness, let us think about the propagator $\mathcal{P}_{\text{abs}}$. There we have 
\begin{align}
\nn
Q_{ \tf}(\du,h) =& \delta(\du-\du_{\rm f})\delta(h-h_{\rm f}) \\
\label{eq:AbsBCProp}
 \Rightarrow \hQ_{ \tf}\left(\hat u(\ti),\tau\hat h(\ti)\right) =& e^{\du_{\rm f} \hu(\ti) + \tau h_{\rm f} \hh(\ti)}.
\end{align}
Then, the propagator $\mathcal{P}_{\text{abs}}$ is given by inverting the Laplace transform
\begin{align}
\nn
&\int_0^\infty \rmd \du_{\rm i} \int_{-\infty}^\infty \rmd h_{\rm i}\,e^{\lambda \du_{\rm i} +\mu h_{\rm i}}\mathcal{P}_{\text{abs}}(\du_{\rm f},h_{\rm f};\tf|\du_{\rm i},h_{\rm i};\ti) =  \\
\nn
&= \hQ_\ti (\lambda,\mu) \\
\label{eq:SolAbsProp}
& =\exp\left\{\int_{\ti}^{\tf}\left[(2-\dw_{s})\hat u(s) + (1+a+\tau^{-1}) \right]\rmd s\right. \\
\nn 
& \quad\quad \left. ~~~~+ \du_{\rm f} \hu({ \tf}) + \tau h_{\rm f} \hh({ \tf})\right\}.
\end{align}
To summarize, one can say the following: The usual instanton equations \eqref{eq:RetInstantonExpTu} and \eqref{eq:RetInstantonExpTh} encode the solution of our model with a reflecting boundary at $\du =0$. The backward instanton equations \eqref{eq:RetInstAbsExpTu} and \eqref{eq:RetInstAbsExpTh} encode the solution  with an absorbing boundary at $\du =0$, assuming the absorbing boundary indeed satisfies $\mathcal{P}_{\text{abs}}(\du_{\rm f},h_{\rm f};t|\du_{\rm i}=0^+,h_{\rm i},\ti)=0$.

 It would be interesting to extend this approach to a general retardation kernel, beyond the simple exponential.
In that case no local-in-time Fokker-Planck approach seems available. It is tempting to conjecture that
the correct generalization of the backward instanton equation (\ref{eq:RetInstAbsExpTu}) is
\begin{align}
\label{eq:RetInstAbs}
         - \partial_t \hu_t + (1+a)\hu_t +\hu_t^2 + a \int_{-\infty}^t \rmd s\, f'(t-s)\hu_s = -\lambda_t,
\end{align}
corresponding to \eqref{eq:RetInstanton} after mapping $\tu \to -\hu$. Note that \eqref{eq:RetInstAbs} reduces to \eqref{eq:RetInstAbsExpTu} and \eqref{eq:RetInstAbsExpTh} for the exponential kernel \eqref{tau}, by setting
\beq
\hh_t := \frac{a}{\tau}\int_{-\infty}^t e^{-(t-s)/\tau}\hu_s\, \rmd s,
\eeq
as would be required.

\subsection{Sub-avalanche durations with retardation}
In order to compute sub-avalanche durations, we need to eliminate the $h$ variable from \eqref{eq:SolAbsProp}. This is done by integrating  over $h_{\rm f}$ and fixing a value of $h_{\rm i}$.\footnote{Or convoluting later with a normalized distribution of $h_{\rm i}$. In any case,  this is not the same as setting $\mu=0$ in \eqref{eq:SolAbsProp}. The latter would be an integral over all $h_{\rm i}$, which in general diverges. For example, in the pure ABBM model the sub-avalanche duration is independent of $h_{\rm i}$.} In appendix \ref{sec:AppBackwardForward}, we motivate the following generalization of \eqref{eq:SolAbsProp} for this case,
\begin{align}
\nn
&\int_{-\infty}^{\infty}\rmd h_{\rm f}\int_0^\infty \rmd \du_{\rm i} \,e^{\lambda \du_{\rm i}}\mathcal{P}_{\text{abs}}(\du_{\rm f},h_{\rm f};\tf|\du_{\rm i},h_{\rm i};\ti) =  \\
\nn
&= \exp\left\{\int_{\ti}^{\tf}\left[(2-\dw_{s})\hu_s + (1+a+\tau^{-1}) \right]\rmd s  \right. \\
\nn
& \quad \quad \quad \left. + \du_{\rm f} \hu_\tf - \tau h_{\rm i} \hh_\ti\right\} \times \\
\label{eq:SolAbsPropInt2}
 & \quad \times \left[\partial_{\hh(\tf)}\big|_{\hh(\tf)=0}\hh(\ti)\right].
\end{align}
Now  $\hu$, $\hh$ are solutions of the backward instanton equations \eqref{eq:RetInstAbsExpTu}, \eqref{eq:RetInstAbsExpTh} with the boundary conditions
\beq
\label{eq:AbsBCFastRel}
\hh(\tf) = 0, \quad\quad\quad \hu(\ti) = \lambda.
\eeq
The effect of going from a fixed $h_{\rm f}$ to a fixed $h_{\rm i}$ is thus a change in boundary conditions for $\hh$, and the Jacobian factor in \eqref{eq:SolAbsPropInt2}. To understand its importance, let us see how \eqref{eq:SolAbsPropInt2} reduces to the pure ABBM solution \eqref{eq:SolAbsPropABBM} in the case of $a=0$. This is not trivial, for example the exponential factors in \eqref{eq:SolAbsProp}, \eqref{eq:SolAbsPropInt2} and \eqref{eq:SolAbsPropABBM} are quite different. 

\subsubsection{Recovering the pure ABBM model\label{sec:SubavRetToABBM}}
For $a=0$, we can solve Eq.~\eqref{eq:RetInstAbsExpTh} explicitly,
\beq
\label{eq:AbsInstSolHABBM}
\hh_{\rm f} = \hh_{\rm i} e^{(\tf-\ti)/\tau}.
\eeq
Thus, the Jacobian factor in \eqref{eq:SolAbsPropInt2} is 
\beq
\nn
\partial_{\hh(\tf)}\big|_{\hh(\tf)=0}\hh(\ti) = e^{-(\tf-\ti)/\tau}.
\eeq
Inserting this into \eqref{eq:SolAbsPropInt2} for $a=0$ gives
\begin{align}
\nn
&\int_{-\infty}^{\infty}\rmd h_{\rm f}\int_0^\infty \rmd \du_{\rm i} \,e^{\lambda \du_{\rm i}}\mathcal{P}_{\text{abs}}(\du_{\rm f},h_{\rm f};\tf|\du_{\rm i},h_{\rm i};\ti) =  \\
\label{eq:SolAbsPropInt3}
&= \exp\left\{\int_{\ti}^{\tf}\left[(2-\dw_{s})\hu_s + 1 \right]\rmd s + \du_{\rm f} \hu_\tf - \tau h_{\rm i} \hh_\ti\right\},
\end{align}
where now $\hh$, $\hu$ are solutions of \eqref{eq:RetInstAbsExpTu}, \eqref{eq:RetInstAbsExpTh} with $a=0$ and with the boundary conditions \eqref{eq:AbsBCFastRel}. Note that the Jacobian factor was necessary to cancel the term $(\tf-\ti)/\tau$  of the exponential. The solution \eqref{eq:AbsInstSolHABBM} for $\hh$ with the boundary condition $\hh_{\rm f} =0$ from \eqref{eq:AbsBCFastRel} implies that $\hh_t=0$ for all $t$. Hence, \eqref{eq:SolAbsPropInt3} reduces to the pure ABBM solution \eqref{eq:SolAbsPropABBM}, and the equation \eqref{eq:RetInstAbsExpTu} for $\hu$ reduces to the pure ABBM instanton equation \eqref{eq:AbsInstantonABBM}.

As expected, we obtain that in the limit $a=0$
\beq
\nn
\int_{-\infty}^{\infty} \rmd h_{\rm f} \mathcal{P}_{\text{abs}}(\du_{\rm f},h_{\rm f};\tf|\du_{\rm i},h_{\rm i};\ti) = \mathcal{P}_{\text{abs}}(\du_{\rm f};\tf|\du_{\rm i};\ti)
\eeq
is independent of $h_{\rm i}$ and reduces to the propagator of the standard ABBM model.

Now let us return to the more interesting case of the ABBM model with retardation. 
In the following section we will apply \eqref{eq:SolAbsPropInt2} to determine the correction to the duration distribution of the first sub-avalanche, for small $a$ (weak relaxation) and $\tau=1$. This is similar to the perturbation theory in $a$ performed in appendix \ref{sec:PertInA}. We also attempted to analyze the backward instanton solution in the physically interesting limits of fast and slow relaxation, as done in sections \ref{sec:QSLim} and \ref{sec:FastRel} for the forward (reflecting boundary) solution. However, we encountered technical difficulties and leave this for future work.

\subsubsection{Weak-relaxation limit}
Let us consider the solution of \eqref{eq:RetInstAbsExpTu}, \eqref{eq:RetInstAbsExpTh} with boundary condition
\beq
\label{eq:AbsBCWeakRel}
\hh(\tf) = \mu, \quad\quad\quad \hu(\ti) = \lambda.
\eeq
We will be interested in the limit of small $\mu$ (required for computing the Jacobian factor in \eqref{eq:SolAbsPropInt2}; the rest can be computed at $\mu=0$). To this end we expand the instanton to order $a, \mu$ as
\begin{align}
\nn
        \hu_t = & \hu_t^{(00)} + &&\mu \hu_t^{(10)}+a\hu_t^{(01)} + \mathcal{O}(a\mu,a^2, \mu^2) \\        \label{eq:AbsWeakRelExp}
        \hh_t = & && \mu \hh_t^{(10)} + a\hh_t^{(01)} + a\mu \hh_t^{(11)} + \mathcal{O}(a^2, \mu^2).
\end{align}
The correction of order $a$ to $P_{\rm surv}$, i.e.\ the cumulative distribution function for the avalanche durations, is obtained from \eqref{eq:SolAbsPropInt2} as
\begin{widetext}
\begin{align}
\nn
&       \hP_{\rm surv}(\lambda,h_{\rm i}) := \int_0^\infty \rmd \du_{\rm i}\,e^{\lambda \du_{\rm i}} P_{\rm surv}(\tf|\du_{\rm i},h_{\rm i};\ti) := \int_0^\infty \rmd \du_{\rm i}\,e^{\lambda \du_{\rm i}} \int_0^\infty \rmd \du_{\rm f} \int_{-\infty}^\infty \rmd h_{\rm f}  \mathcal{P}_{\rm abs}(\du_{\rm f},h_{\rm f};\tf|\du_{\rm i},h_{\rm i};\ti) = \nn\\ 
&
\nn
=\exp\left\{\int_{\ti}^{\tf}\left[(2-\dw_s)\hu_s + (1+a+\tau^{-1}) \right]\rmd s - \tau h_{\rm i} \hh_\ti \right\}\left[-\frac{1}{\hu({\tf})}\right]\left[\partial_{\hh(\tf)}\big|_{\hh(\tf)=0}\hh(\ti)\right]  \nn \\
\nn
& = \exp\left\{\int_{\ti}^{\tf}\left[(2-\dw_s)\left(\hu^{(00)}_s+a\hu^{(01)}_s\right) + (1+a+\tau^{-1}) \right]\rmd s - a \tau h_{\rm i} \hh_\ti^{(01)} \right\}\left[-\frac{1}{\hu^{(00)}(\tf)+a\hu^{(01)}(\tf)}\right]\left[\hh_\ti^{(10)}+a\hh_\ti^{(11)}\right]  \\
\label{eq:AbsWeakRelDur}
 & = \hP_{\rm surv}^{a=0}(\lambda) \left\{1+a\left[\int_{\ti}^{\tf}\left[(2-\dw_s)\hu^{(01)}_s+1 \right]\rmd s- \frac{\hu_\tf^{(01)}}{\hu_\tf^{(00)}} +  \frac{\hh_\ti^{(11)}}{\hh_\ti^{(10)}} \right] \right\} \exp\left[  -a\tau \hh_\ti^{(01)} h_{\rm i} \right].
\end{align}
\end{widetext}
The pure ABBM result $\hP_{\rm surv}^{a=0}(\lambda)$ is given by \eqref{eq:ABBMPhatSurv}. By time translation invariance, this expression only depends on $T:=\tf-\ti$. 
In appendix \ref{sec:AppPertInABackwards} we perform the perturbative calculation and determine explicitly the functions appearing in the expansion \eqref{eq:AbsWeakRelExp}, for the case $\tau=1$. In figure \ref{fig:BackwardInstantonTau1} we show the form of the resulting backward instantons, obtained from the perturbative expansion and a numerical solution. The case of general $\tau$ is more complicated and left for future research.

The survival probability $P_{\rm surv}$ is the probability of having a first-passage time to $\du=0$ of $T \geq \tf-\ti$.
The (normalized) probability distribution of first-passage times at $\du=0$, starting from $\du_{\rm i}$ is thus
\beq
P_{\rm first}(T|\du_{\rm i},h_{\rm i}) = \partial_\tf\big|_{\tf = \ti + T}P_{\rm surv}(\tf|\du_{\rm i},h_{\rm i};\ti).
\eeq
By time-translation invariance, this is independent of $\ti$. For the pure ABBM model, we computed $P_{\rm first}$ in \eqref{eq:ABBMPFirst}.
The (unnormalized) density of avalanche durations $\rho(T)$ at a fixed value of $h_{\rm i}$ is the leading order of $P_{\rm first}(\du_{\rm i},h_{\rm i})$ as $\du_{\rm i} \to 0$, in our case
\beq
\nn
\rho(T) := \lim_{\du_{\rm i} \to 0} \du_{\rm i}^{a h_{\rm i} + v-1} P_{\rm first}(T|\du_{\rm i},h_{\rm i}).
\eeq
We can then express $\rho(T)$ as the leading order of $\hP_{\rm surv}$, as $\lambda \to -\infty$:

\begin{widetext}
\begin{align}
\label{eq:AbsWeakRelCorrT}
&\int_T^{\infty}\rho(T')\rmd T'= \lim_{\lambda \to -\infty} \hP(\lambda,0) (-\lambda)^{-2+v+ah_{\rm i}} = \\
\nn
&\frac{e^{-a h_{\rm i} T}}{(e^{T}-1)^{1-v-ah_{\rm i}}}\left\{1+a    \frac{e^T \left[-2 (v-1)
   \text{Li}_2\left(1-e^T\right)-\left(T^2+1\right) v+T^2+T-1\right]+(2-v)T
   +v+1}{2 \left(e^T-1\right)} +\mathcal{O}(a)^2 \right\}.
\end{align}
This further simplifies for the case of a vanishing driving velocity, $v \to 0$. There we obtain
\begin{align*}
        \int_T^{\infty}\rho(T')\rmd T'& = \frac{1}{(e^{T}-1)^{1-a h_{\rm i}}}\left[1+a \frac{e^{T} \left(T^2+2
   \text{Li}_2\left(1-e^{T}\right)+T-1\right)+2 T+1}{2
   \left(e^{T}-1\right)} +\mathcal{O}(a)^2\right].
\end{align*}
\end{widetext}
On the other hand, expanding \eqref{eq:AbsWeakRelCorrT} for small $T$ we get
\begin{align}
\label{eq:AbsWeakRelSmallT}
\rho(T) =& \,T^{-2+v+a h_{\rm i}}\Big[a h_{\rm i} + v-1  \\
\nn
& \quad\quad \left.
+\frac{1}{2}(a h_{\rm i} + v)(1+a)\left(v-1-\frac{ah_{\rm i}}{1+a}\right) T \right. \\
\nn
& \quad\quad  + \mathcal{O}(T)^2\Big]
\end{align}
We observe that the power-law behavior near $T=0$, $\rho(T)\sim T^{-2+v}$, is not modified for the first sub-avalanche ($h_{\rm i}=0$), but is modified for later ones to
\beq
\label{eq:AbsWeakRelFinHI}
\rho(T)\sim T^{-2+v + a h_{\rm i}},
\eeq
 to leading order in $a$. This is natural, since for small avalanches $h$ remains essentially unchanged, and replaces in \eqref{eq:EOMUd} $v \to v + a h_{\rm i}$.
In order to obtain the sub-avalanche-duration distribution for stationary driving, one would need to average over all $h_{\rm i}$ taken from the stationary distribution $P(h_{\rm i}|\du_{\rm i}=0)$. Presumably, this would replace the correction $a h_{\rm i}$ to the exponent in \eqref{eq:AbsWeakRelFinHI} by a velocity-dependent correction. We leave the details  for further research. 

\begin{figure}\includegraphics[width=\columnwidth]{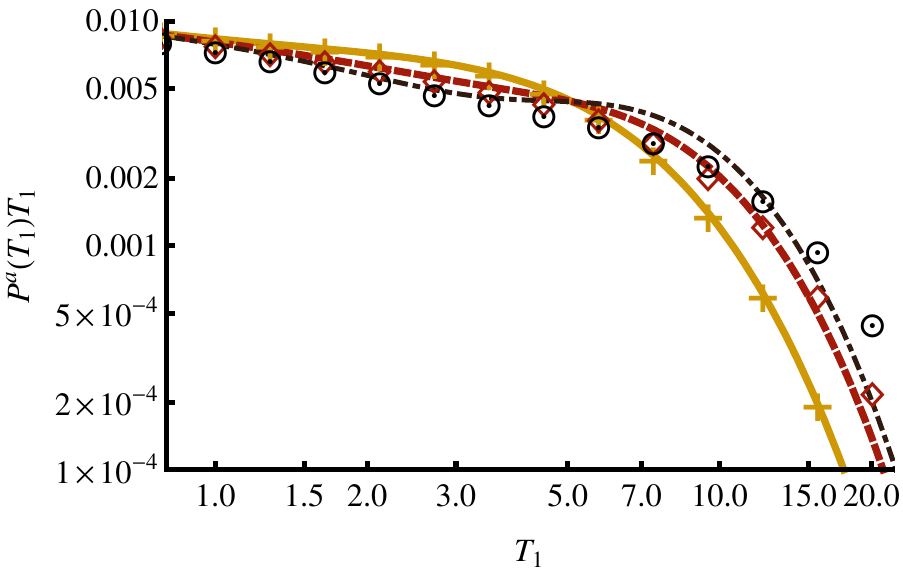}\caption{(Color online). Duration distribution of an avalanche starting at $\du=0,h=0$ and $\tau=1, v=0.6$. Crosses, diamonds and circles: Results from numerical integration of \eqref{eq:EOMUd} for $a=0$, $a=0.3$, $a=0.5$ with time step $\rmd t=10^{-5}$. Lines: Expansion in $a$,\eqref{eq:AbsWeakRelCorrT}. Yellow (thick) line: Pure ABBM model, $a=0$, Red (dashed) line: $a=0.3$, Black (dotted) line: $a=0.5$.}\label{fig:TFirstSub}\end{figure}

\begin{figure}\includegraphics[width=\columnwidth]{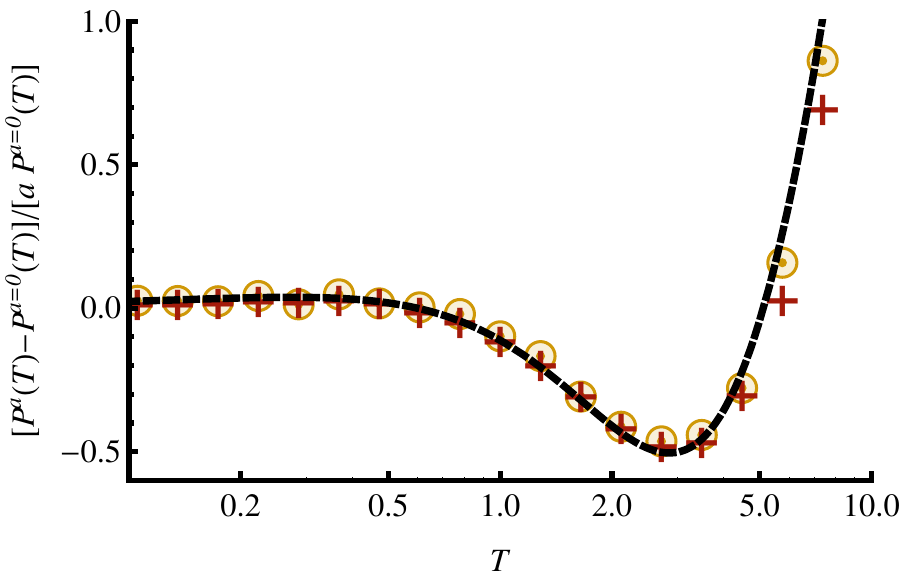}\caption{(Color online). Correction to the duration distribution of an avalanche starting at $\du=0 $,  $h=0$ due to retardation, $\left[P^{a}(T)-P^{a=0}(T)\right]/\left[a P^{a=0}(T)\right]$. $\tau=1, v=0.6$. (Yellow) circles, (red) squares: Results from numerical integration of \eqref{eq:EOMUd} for $a=0.1$, $a=0.3$ with time step $dt=10^{-4}$.  Dashed line: \eqref{eq:AbsWeakRelCorrT}.}\label{fig:PlotCorrPofTFirst}\end{figure}

\begin{figure}\includegraphics[width=\columnwidth]{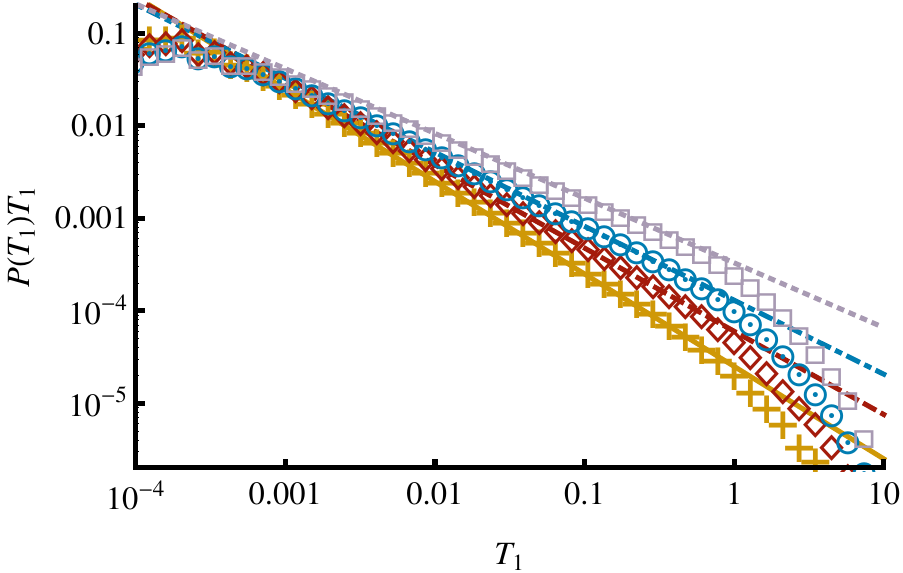}\caption{(Color online). Duration distribution of the first sub-avalanche, as a function of $h_{\rm i}$. $\tau=1$, $a=1$, $v=0$. (Yellow) crosses, (Red) diamonds, (Blue) circles and (grey) squares correspond to numerical simulations of \eqref{eq:EOMUd} for $h_{\rm i}=0$, $h_{\rm i}=0.1$, $h_{\rm i}=0.2$, $h_{\rm i}=0.3$ with time step $\rmd t=10^{-5}$. Solid, dashed, dot-dashed and dotted lines correspond to power laws $P(T_1)\sim T_1^{-2}, T_1^{-1.9}, T_1^{-1.8}, T_1^{-1.7}$.}\label{fig:PlotHIPofT1}\end{figure}

In figures \ref{fig:TFirstSub} and \ref{fig:PlotCorrPofTFirst}, we compare the result \eqref{eq:AbsWeakRelCorrT} to numerical simulations of the original model. One observes good agreement for small $a$ but larger deviations starting around $a \approx 0.5$.
In figure \ref{fig:PlotHIPofT1} we verify numerically the result \eqref{eq:AbsWeakRelFinHI} for the sub-avalanche-duration exponent as a function of $h_{\rm i}$. The agreement is very good,  even for $a=1$.


\subsubsection{Numerical results}
\begin{figure*}\includegraphics[width=0.95\columnwidth]{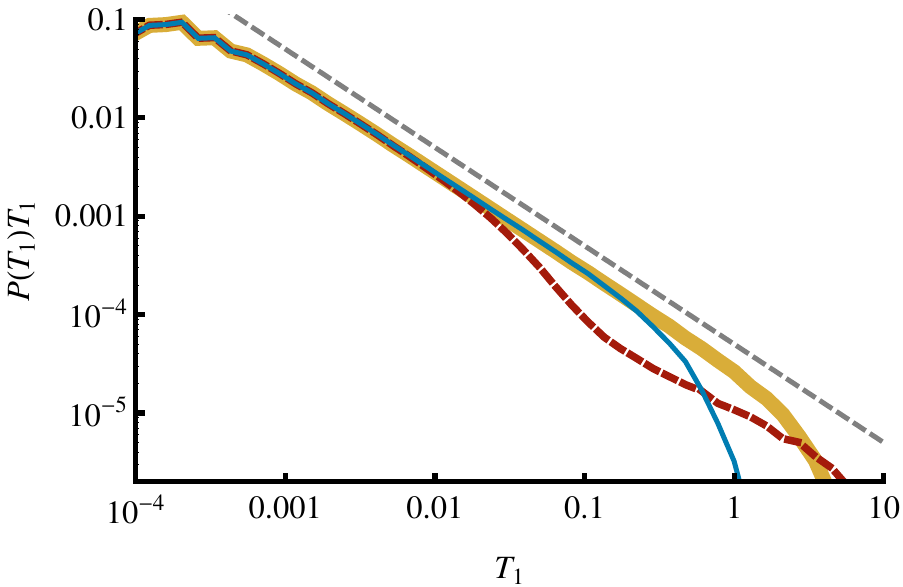}\includegraphics[width=0.95\columnwidth]{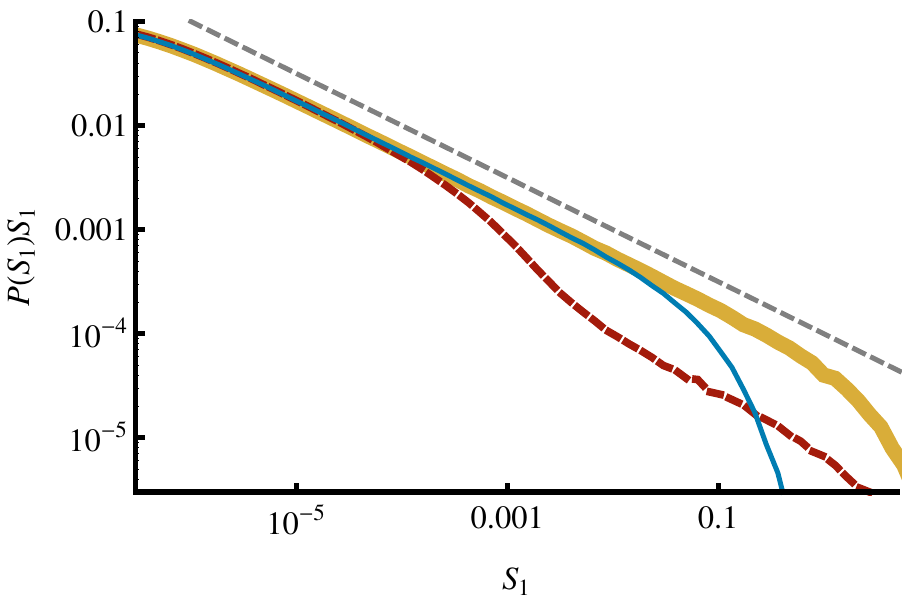}\caption{(Color online) First sub-avalanche sizes (right) and durations (left) from numerical simulation of the ABBM model with retardation, starting from 
$\du_{\rm i}=0, h_{\rm i}=0$, at $v=0$. Grey (dashed) lines: power laws $P(S)\sim S^{-3/2},P(T)\sim T^{-2}$. Yellow (thick) line: pure ABBM model $a=0$. Red (dashed) line: $a=60, \tau=0.05$. Blue (thin) line: $a=5, \tau=10$. }\label{fig:PlotDiv1}\end{figure*}

\begin{figure*}\includegraphics[width=0.95\columnwidth]{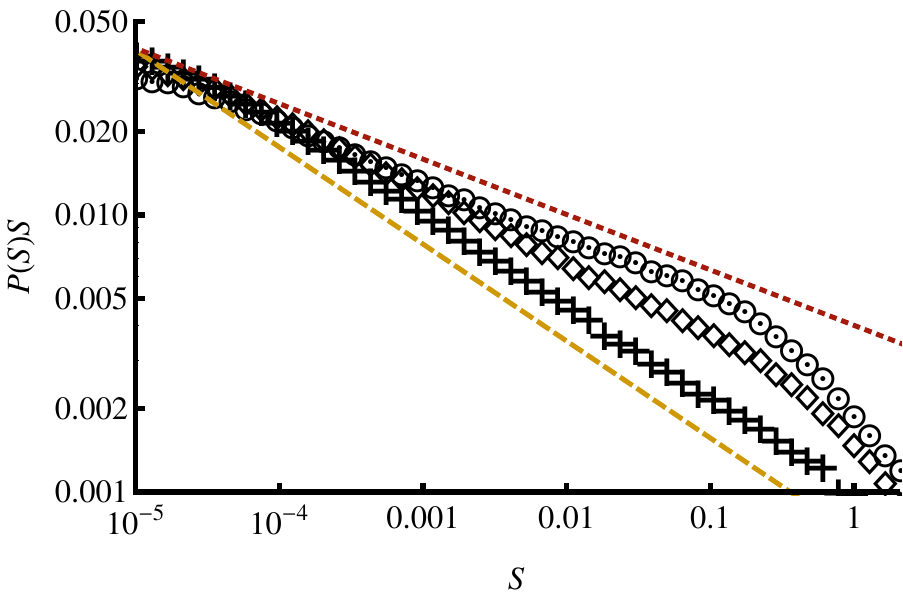}\includegraphics[width=0.95\columnwidth]{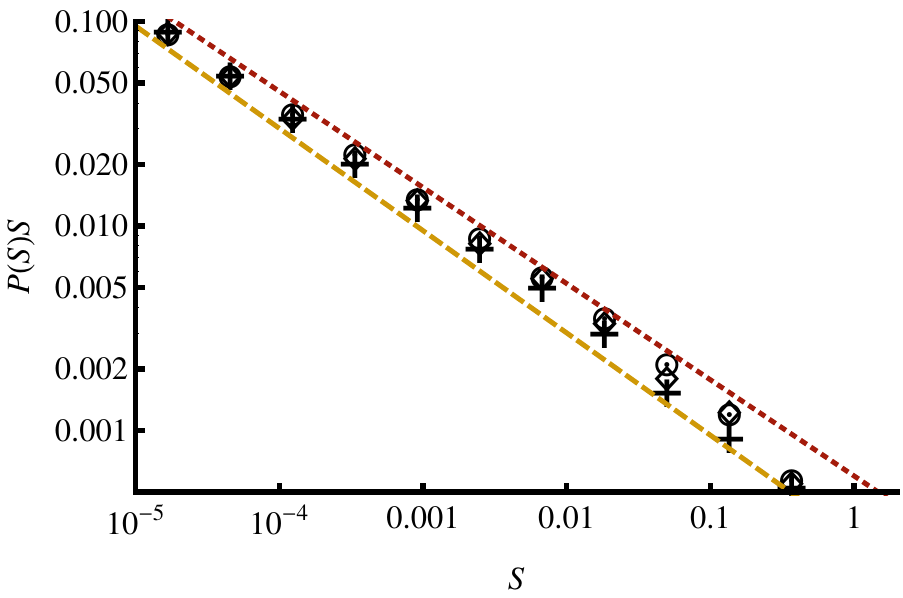}\caption{(Color online) Sub-avalanche sizes from numerical simulation of the ABBM model with retardation, in a steady state with $v=0.3$ (left) and $v=0.02$ (right). Yellow (dashed) lines: power laws $P(S)\sim S^{-(3-v)/2}$, left: $S^{-1.35}$, right: $S^{-1.49}$. Red (dotted) lines: power laws, left: $S^{-1.2}$, right: $S^{-1.47}$. Crosses: pure ABBM model $a=0$. Diamonds: $a=1$, Circles: $a=2$. In all cases, $\tau=1$.}\label{fig:PlotSubAvg1}\end{figure*}

Since the analytical results we obtained above for sub-avalanche sizes and durations are rather limited, we also give a few qualitative numerical results. In this section we neglect the difference between densities and probability distributions, and use $P$ instead of $\rho$ everywhere.
\begin{enumerate}
        \item Sub-avalanches at fixed initial $\du_{\rm i}$, $h_{\rm i}=0$ and $v=0$.  In \eqref{eq:AbsWeakRelSmallT}, we derived that the ABBM power-law $P(T)\propto T^{-2}$ for small $T$  (see \eqref{eq:PofT})  remains unchanged in this case, at least for $\tau=1$ and small $a$. In figure \ref{fig:PlotDiv1} we consider the sub-avalanche sizes and durations for general values of $a,\tau$. 
                We see that the mean-field zero-velocity power-laws $P(S)\sim S^{-3/2},P(T)\sim T^{-2}$ are clearly visible even when varying $a,\tau$. For large $a$ and small $\tau$, there is an interesting crossover in the shape of the distributions, showing a similar power law but with different amplitudes, depending on $a,\tau$. The case of large $\tau$ is equivalent to a modification of the mass, i.e.\ of the large-avalanche cutoff, as discussed in section \ref{sec:PhysicsQS}.
        \item Sub-avalanches at fixed $\du_{\rm i}$, and $h_{\rm i}$ taken from the stationary distribution at the driving velocity. We observe in figure \ref{fig:PlotSubAvg1} that this leads to a modification of the ABBM power-law exponent. However, the variation in the exponent due to retardation becomes smaller as the driving velocity decreases.  This is expected from \eqref{eq:AbsWeakRelFinHI}, since the typical value of $h_{\rm i}$ (and hence the correction to the exponent) decreases as $v \to 0$.
\end{enumerate}
From these numerical results we conjecture that sub-avalanche durations in the ABBM model with retardation,  at constant driving, satisfy
        \beq
        \nn
        P(T)\sim T^{-2+v(1+c_T)},\quad P(S)\sim S^{-3/2+v/2(1+c_S)}.
        \eeq
        Here $c_T, c_S$ depend on $a, \tau$, and vanish as $a\to 0$ and/or $\tau \to 0$.
        In other words, we conjecture that the zero-velocity exponents are unchanged, and only the driving-velocity-dependent part is modified. This is also consistent with our analytical results in sections \ref{sec:QSLim} and \ref{sec:FastRel} for the velocity distribution $P(\du)$. 
Verifying these conjectures in more detail, numerically or analytically, is an interesting task for the future.


\section{Conclusion and Outlook\label{sec:Conclusion}}

In this article we have analyzed in detail a general ABBM model with retardation. We showed that it satisfies the Middleton property (monotonicity of the dynamics). Using this, and the Brownian nature of the ABBM disorder, we have been able to reduce the calculation of the expectation value of a general observable in presence of monotonous driving to the problem of solving retarded non-linear  instanton equations. These equations can be implemented numerically for arbitrary retardation kernels, and are much simpler than the original model with quenched disorder. To obtain analytical results, we  focused on a model with exponential relaxation, which reduces to two coupled ``instanton" equations, local in time. We have derived explicit forms for a number of observables at stationary and non-stationary driving. We mostly focused on the two limits of fast relaxation (describing eddy current effects in magnetic Barkhausen noise) and slow relaxation (of interest for earthquake models).

In the limit of slow relaxation, the main physics is the splitting of a single avalanche of size $S$ of the standard
ABBM model, into a ``cluster of aftershocks", of the {\it  same} total size $S=\sum_\alpha S_\alpha$, 
but of much longer duration (strictly infinite for an exponential retardation kernel). This splitting is sharp in the limit of quasi-static driving and strong time-scale separation. 
Although we have been able to quantify some of these ideas, a more detailed analysis of the statistics of these sub-avalanches remains an important challenge for future work. In particular, if these ideas are to be extended to realistic earthquake dynamics, one needs to
investigate power-law retardation kernels motivated by the Omori law of decay of activity. Still, it is
of great interest to have found a tractable model with aftershocks, given that the standard ABBM model, which also models interfaces within 
mean-field theory (at the upper critical dimension) yields independent avalanches
following a Levy process \cite{LeDoussalWiese2008c,LeDoussalWiese2011b,LeDoussalWiese2012a}.\footnote{While this work was completed we learned of an independent
work by Jagla, Rosso \cite{JaglaRosso2013inpr}. These authors consider several variants of earthquake models, and one of them is similar to
the retarded model considered here. }

In the context of magnetic systems, the retardation describes the influence of eddy currents on the statistics of Barkhausen noise pulses. Experimentally, the time scale of eddy current relaxation is much shorter than that of the domain-wall motion, motivating our study of the fast-relaxation limit. In this limit, the effects of retardation are perturbative and vanish when the eddy-current relaxation-time tends to zero. We have obtained the leading corrections to the stationary velocity distribution, as well as the decay following a kick in the driving velocity. 

In both the slow-relaxation and the fast-relaxation limit, the influence of retardation turned out to be most pronounced in non-stationary observables. For example, we computed explicitly the tail of the avalanche shape at fixed size. While it is Gaussian in the standard ABBM model, in the ABBM model with retardation it decays exponentially. Furthermore, we showed that formally the avalanche activity following a kick in the driving velocity never stops (even if in the fast-relaxation limit, all significant avalanche activity still ceases), also in strong contrast to the pure ABBM model.

One important direction for future theoretical investigations is the role of the internal dimension of the interface. Here we reduce the description of the magnetic domain wall, which is a two-dimensional elastic interface in a three-dimensional medium, to a single degree of freedom (its center of mass). This mean-field description has been argued \cite{Colaiori2008} and recently shown in detail \cite{LeDoussalWiese2011,LeDoussalWiese2012a} to be accurate  for the center of mass of the interface above a critical internal dimension $d_{\rm c}$. For certain soft magnets, due to long-range elastic forces, one has $d_{\rm c}=2$ \cite{Colaiori2008}. Hence, some realistic domain walls have just the critical internal dimension. In that situation the internal degrees of freedom of the interface contribute only logarithmic corrections to the mean-field behavior as described in detail in \cite{LeDoussalWiese2012a}. Other ferromagnets are known to be in a non-mean-field universality class \cite{DurinZapperi2000}. There, correctly describing the details of Barkhausen noise requires combining the eddy current modifications discussed in this article with a treatment of the spatial degrees of freedom, for example using the functional renormalization group as in \cite{LeDoussalWiese2011, LeDoussalWiese2012a,DobrinevskiLeDoussalWiese2013inpr}. 

Another important avenue for further work is a detailed comparison of the analytical results discussed here with experiments on real magnetic domain walls. Considering the above results, non-stationary observables seem most promising. For example, the avalanche shape discussed in section \ref{sec:AvShapeFixedSize} shows a clear qualitative difference to the result of the standard ABBM model. As recognized in \cite{SethnaDahmenMyers2001,ZapperiCastellanoColaioriDurin2005} this is an inertial effect due to an effective negative domain-wall mass. However, inertia can be modelled in different ways -- for example, the retarded ABBM model considered here and in \cite{ZapperiCastellanoColaioriDurin2005}, the ABBM model with second-order dynamics considered in \cite{LeDoussalPetkovicWiese2012}, and the mean-field model with stress overshoots considered in \cite{SchwarzFisher2001}. We believe the avalanche shape allows identifying not just the existence and the sign of inertial effects, but their {\em  precise form}. To this end, more precise analytical and experimental results (which go beyond a single ``skewness'' quantity, and consider the entire shape) are necessary.

\acknowledgements
We thank Yanjiun Chen, Gianfranco Durin, Aleksandra Petkovic and Andrei Fedorenko for useful discussions. We are particularly grateful to Aleksandra Petkovic for bringing to our attention  Ref.~\cite{ZapperiCastellanoColaioriDurin2005}. We also thank Alberto Rosso and Eduardo Jagla for bringing to our attention their independent work
\cite{JaglaRosso2013inpr} and for discussions. This work was supported by ANR Grant No.~09-BLAN-0097-01/2, and by the CNRS through a doctoral fellowship for A.D.

\appendix
\section{Monotonicity of the domain-wall motion in the retarded ABBM model\label{sec:Monot}}
The model defined in \eqref{eq:EOMU} satisfies Middleton's theorem \cite{Middleton1992}: If the driving is monotonous, $w_t \geq w_s$ for $t \geq s$, then so is the motion of the domain wall\footnote{Note this is not true for models where domain-wall inertia is included by adding a second-order derivative in the ABBM equation of motion \cite{LeDoussalPetkovicWiese2012}. This makes the retarded ABBM model considered here quite special.}.

To prove this, note first that the velocity $\du_t$ is continuous. Hence, before any negative velocities can occur, there would be an instant $t > 0$ where $\du_t$ becomes zero for the first time. Then, using \eqref{eq:EOMU}, we get
\begin{align*}
\eta \partial_t \du_t &= \du_t F'(u_t) + m^2 [\dw(t) - \du(t)] - a \du(t) \\
&\quad - a \int_{-\infty}^t \rmd s f'(t-s) \du(s) \\
&= m^2 \dw(t) - a \int_{-\infty}^t \rmd s f'(t-s) \du(s) > 0.
\end{align*}
The first term is $\dw(t) \geq 0$ by monotonicity of the driving $w$. The second term is $\geq 0$ since $a>0$, $f'(t-s) \leq 0$ and $\du_s > 0 $ for all $s<t$. Similarly one checks that if $\ddot{u}_t=0$, then $\partial_t \ddot{u}_t \geq 0$. In other words, at any time where $\du_t = 0$, the first non-vanishing derivative of $\du_t$ is positive. Thus, the domain-wall motion is monotonous.

Note that the vanishing of $\dot u F'(u)$ for $\dot u=0$ is ensured if $F(u)$ is smooth. In the case of the ABBM Brownian landscape $F'(u)$ is a white noise.
One can then consider a version smoothed at short scale and take the continuum limit. Alternatively one sees on the formulation $\dot u F'(u) \leftrightarrow \sqrt{\dot u} \xi(t)$ 
in (\ref{noise1}) that this is not a problem.

\section{Position differences in the small-dissipation limit\label{sec:QSPositions}}
\begin{figure*}
\begin{minipage}[l]{0.48\textwidth}
\includegraphics[width=\textwidth]{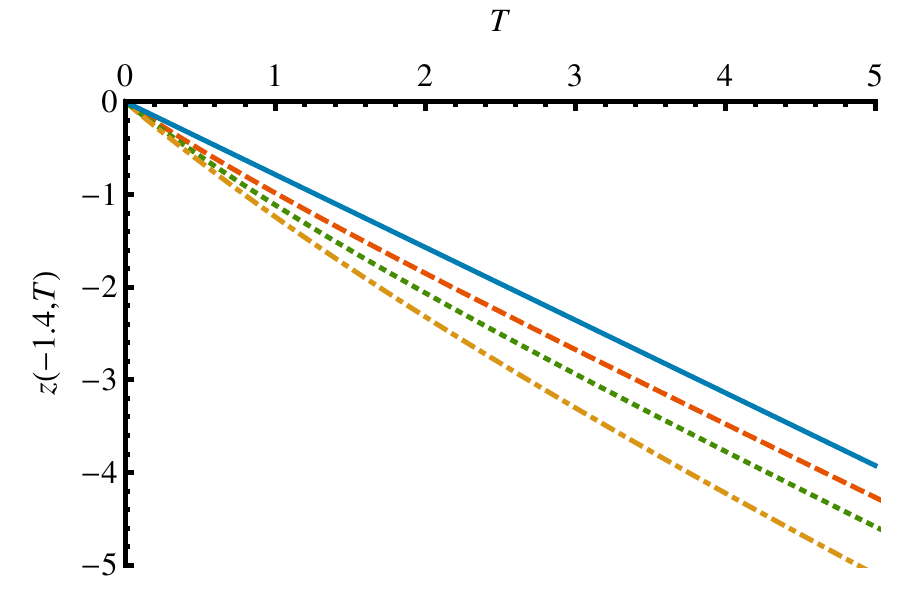}
{\small (a) $z$ at fixed $\lambda=-1.4$, as a function of $T$. Solid blue line: standard ABBM model \eqref{eq:QSPosGenFctABBM}. Red (dashed), green (dotted), yellow (dot-dashed) lines: ABBM model with retardation for $a=1, 2.3, 5.3$.\\ \mbox{~}}
\end{minipage}~
\begin{minipage}[l]{0.48\textwidth}
\includegraphics[width=\textwidth]{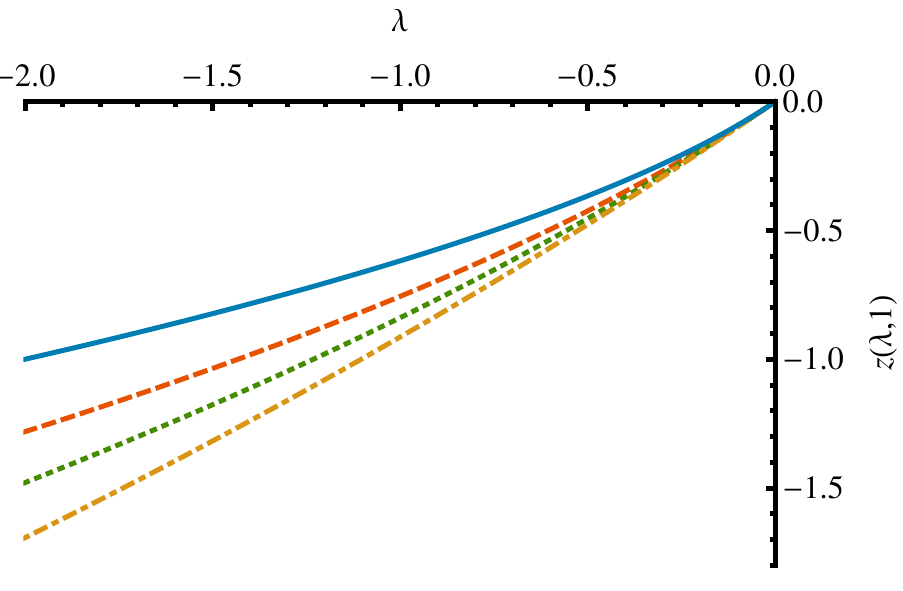}
{\small (b) $z$ at fixed $T=1$, as a function of $\lambda$. Dashed blue line: standard ABBM model \eqref{eq:QSPosGenFctABBM}. Solid blue line: standard ABBM model \eqref{eq:QSPosGenFctABBM}. Red (dashed), green (dotted), yellow (dot-dashed) lines: ABBM model with retardation for $a=1, 2.3, 5.3$.}
\end{minipage}
\caption{Generating function for position differences $z(\lambda,T)$ as defined in \eqref{eq:QSPosGenFct}.}
\label{fig:QSPosDiff}
\end{figure*}

As discussed in section \ref{sec:VelStat}, instantaneous velocities $\du_t$ are almost surely 0 in the small-dissipation limit $\eta=0$. However, the distribution of position differences $u_T-u_0$ has a finite limit. The generating function of position differences at constant driving velocity $v$ is given by \eqref{eq:RetSolutionExp}:
\beq
\label{eq:QSPosGenFct}
\overline{e^{\lambda(u_T-u_0)}} = \overline{e^{\int_0^T \rmd t\, \lambda\du_t}} = e^{v\,z(\lambda,T)},
\eeq
where $z(\lambda,T) = \int_t \tu_t$. $\tu_t$ is given by the instanton equations \eqref{eq:RetInstantonExpTu}, \eqref{eq:RetInstantonExpTh} with sources $\lambda(t) = \lambda \theta(t)\theta(T-t)$, $\mu(t) = 0$. In the dissipation-less limit $\eta=0$, \eqref{eq:RetInstantonExpTh} and \eqref{eq:RetInstantonExpTu} give an instanton equation similar to \eqref{eq:QSInstanton},
\beq
\label{eq:QSInstanton2}
(1+a-2\tu_t)\partial_t \tu_t = \tu_t-\tu_t^2 + \lambda'_t - \lambda_t.
\eeq
Here we thus work in the limit $\tau_m \ll \tau,\tau_v(=S_m/v)$, i.e.\ we set $\eta=0$. We express all times in units of $\tau$, i.e.\ our variable $t$ is the variable $s$ in section \ref{sec:QSEddyStat}. In other words, we set $\tau=1$, while space units remain such that $S_m=1$.
For $t<0$, the sources vanish and the solution is identical to that in section \ref{sec:QSEddyStat}. Thus, the integral over $\tu$ in that region has the closed form \eqref{eq:QSStatInt},
\begin{align}
\int_{-\infty}^0 \tu_t \rmd t = 2\tu_{0^-} +(1-a) \ln \left(1-\tu_{0^-}\right)\ .
\end{align}
Similarly, the integral over $\tu$ in the region $0<t<T$ is
\begin{align}
\int_0^T \tu_t \,\rmd t & = \int_{\tu_{0^+}}^{\tu_{T^-}} \frac{1+a-2\tu}{\tu-\tu^2 - \lambda}\tu\, \rmd \tu \\
\nn
& = \left[-\frac{1}{2} (a-1) \log (\lambda +(\tu-1) \tu)+\right. \\
\nn
& \quad \left. +\frac{(a+4 \lambda -1)
   \tanh ^{-1}\left(\frac{2 \tu-1}{\sqrt{1-4 \lambda }}\right)}{\sqrt{1-4
   \lambda }}+2 \tu\right]_{0^+}^{T^-}
\end{align}
The relationship between $\tu_{0^+}$ and $\tu_{T^-}$ is given implicitly by
\begin{align}
T &= \int_0^T \rmd s = \int_{\tu_{0^+}}^{\tu_{T^-}} \frac{1+a-2\tu}{\tu-\tu^2 - \lambda}\rmd \tu \nn\\
&= \left[\frac{2 a \tanh ^{-1}\left(\frac{2 \tu-1}{\sqrt{1-4 \lambda}}\right)}{\sqrt{1-4
   \lambda}}+\log (\lambda+(\tu-1) \tu)\right]_{0^+}^{T^-}.
\end{align}
The relationship between $\tu_{0^+}$ and $\tu_{0^-}$, as well as between $\tu_{T^-}$ and $\tu_{T^+} = 0$ is given by the matching conditions
\begin{align*}
\left[(1+a)\tu_{0^+} - \tu_{0^+}^2\right] &- \left[(1+a)\tu_{0^-} - \tu_{0^-}^2\right] = \lambda \\
\nn
 &- \left[(1+a)\tu_{T^-} - \tu_{T^-}^2\right]= -\lambda . 
\end{align*}
Altogether, this gives a complete (albeit implicit) solution for $\int_t \tu_t$ in terms of $T, \lambda$. This allows plotting the generating function \eqref{eq:QSPosGenFct}, figure \ref{fig:QSPosDiff}. It can also be compared to the result of the standard ABBM model,
\beq
\label{eq:QSPosGenFctABBM}
z_0(\lambda, T) = \frac{T}{2}\left(1-\sqrt{1-4\lambda}\right).
\eeq
While the result (\ref{eq:QSPosGenFctABBM}) holds  in the small-dissipation (equivalently large-time) limit $\tau_m \ll T$, a more general result was
obtained in \cite{LeDoussalWiese2012a} (Section II F) for the standard ABBM model in the case where $\tau_m$ and $T$ are comparable.

In figure \ref{fig:QSPosDiff}b  one observes that the slope $\partial_\lambda \big|_{\lambda=0} z(T,\lambda)=T$ is independent of the value of $a$. This is also seen from the explicit solution above:
The instanton equation \eqref{eq:QSInstanton2}, to linear order in $\lambda$ (equivalently to linear order in $\tu$) simplifies to
\beq
(1+a)\partial_t \tu_t = \tu_t + \lambda'_t - \lambda_t.
\eeq
Its solution is
\beq
\tu_t = \frac{1}{1+a}\int_{t}^{\infty} \rmd s\, e^{-(s-t)/(1+a)}\left(\lambda'_s-\lambda_s\right)
,\eeq
and its integral is
\beq
\int_{-\infty}^{\infty} \rmd t \tu_t = \int_{-\infty}^{\infty} \rmd s \left(\lambda'_s-\lambda_s\right) = \lambda T
.\eeq
Thus,  $z(T,\lambda) = \lambda T + \mathcal{O}(\lambda)^2$.
This means that the average displacement, $\overline{(u_T-u_0)} = v \partial_\lambda \big|_{\lambda=0} z(T,\lambda) = vT$. Of course, this is consistent with the fact that the mean velocity of the domain wall is fixed by the harmonic well $m^2(u_t-v t)$.

On the other hand, from figure \ref{fig:QSPosDiff} (b) one also sees that the curvature $\partial_\lambda^2 \big|_{\lambda=0} z(T,\lambda)$ decreases with increasing $a$. Thus, the fluctuations $\overline{(u_T-u_0)^2}^c$ are {\em  decreased} by retardation effects. This is in agreement with the intuition of eddy currents slowing the domain wall down when it is moving fast, and pushing it forward when it is moving slowly.

\section{Direct perturbation theory in $a$} \label{sec:PertInA}
Instead of looking at the slow-relaxation limit $\tau \gg \tau_m$  or fast-relaxation limit $\tau \ll \tau_m$ discussed in sections \ref{sec:QSLim} and \ref{sec:FastRel}, one may attempt to determine the corrections to the stationary velocity distribution for $a \ll 1$ through a direct perturbation expansion in $a$.
For this, we need to solve the instanton equations \eqref{eq:RetInstantonExpTu}, \eqref{eq:RetInstantonExpTh} for $\lambda(t)=\lambda \delta(t)$, $\mu(t)=0$. We  take $\tau$ fixed but $a \ll 1$ and work, as in most of the article, in the units where $\tau_m=1$, $S_m=1$.
We set
\beq
\label{eq:DirPertAnsatz}
\tu(t) = \tu_0(t) + a \tu_1(t) + \mathcal{O}(a)^2, \quad \tih(t) = a \tih_1(t) + \mathcal{O}(a)^2
\eeq
where $\tu_0(t)$ is the known instanton for the standard ABBM model \cite{LeDoussalWiese2011},
\beq
\nn
\tu_0(t) = \frac{\lambda}{(1 -\lambda)  e^{- t } +\lambda} \theta(-t).
\eeq
$\tih_1(t)$ for $t<0$ is then determined by \eqref{eq:RetInstantonExpTh},
\begin{align}
\nn
\tih_1(t) =& \frac{1}{\tau} \int^{0}_t \rmd s\,e^{-(s-t)}  \tu_0(s) \\
\nn
=& \, _2F_1\!\left(1,\tau^{-1} ;\tau^{-1} +1;e^{-t/\tau} \Big(1
   -\frac{1}{\lambda}\Big) \right)  \\
\nn
        &  -e^{t/\tau} \, _2F_1\!\left(1,\tau^{-1} ;\tau^{-1}
   +1;1-\frac{1 }{\lambda }\right)\ .
\end{align}
To obtain $\tu_1(t)$ for $t<0$, we  need to solve the linear equation
\beq
\tu_1'(t) - \tu_1(t) - \tu_0(t) +  2 \tu_1(t) \tu_0(t)  +  \tih_1(t) = 0
\ ,\eeq 
with $\tu_1(0)=0$. One simple case is $ \tau=1,$ where
\beq
 \tih_1(t) =  \frac{\lambda}{1-\lambda} e^t \ln\Big(\lambda + e^{-t} (1-\lambda)\Big) 
\eeq 
and 
\begin{align*}
\tu_1(t) =& \frac{\lambda  e^t}{2 (\lambda -1) [\lambda 
   (e^t-1)+1]^2} \times \nn\\ &\left[2 (\lambda -1)^2
   \text{Li}_2 \Big(\frac{\lambda }{\lambda -1}\Big) \right. \\
 &+(\lambda -1)\Big\{
   -\lambda -2 (\lambda -1) \text{Li}_2\Big(\frac{e^t \lambda
   }{\lambda -1}\Big)\nn\\ &+(\lambda -1) (t-2) t  \\
   & +3 (\lambda -1) \log\Big
   (\lambda  (e^t-1)+1\Big)  \nn\\ &+2 t \log\Big(\frac{e^t}{1-\lambda }\Big) \\
        &+2 \lambda  t \log\Big
   (e^{-t}(1-\lambda ) \Big)\Big\}\nn\\ &+\lambda ^2
   e^{2 t} \log(\lambda -(\lambda -1) e^{-t})\\
   & -\lambda 
   (\lambda -1) e^t \Big(4 \log (\lambda -(\lambda -1)
   e^{-t})-1\Big)\bigg]. 
\end{align*}
This solution has the asymptotics
\beq
\nn
\lim_{\lambda \to -\infty} \tu_1 (t) = \frac{e^t \left(2 t-4 e^t+e^{2 t}+3\right)}{2 \left(e^t-1\right)^2} \ln (-\lambda).
\eeq
We see that $\tu_1(t) \to -\infty$ as $\lambda \to -\infty$, for any fixed $t<0$. Consistent with the discussion in sections \ref{sec:VelStat} and \ref{sec:VelKick}, we recover the result that there is no $\delta(\du)$ contribution,  $p_{\du=0}=0$ following a kick, and avalanches never end. 

Instead of the stationary velocity distribution, one can also consider the stationary distribution of eddy-current pressure as in section \ref{sec:QSEddyStat}.
In contrast to \eqref{eq:DirPertAnsatz}, the contribution $\tih_0(t)$ of order $\mathcal{O}(a)^0$ to $\tih$ does not vanish. 
To order $\mathcal{O}(a)^0$ the expressions \eqref{eq:RetInstantonExpTu} and \eqref{eq:RetInstantonExpTh} in dimensionless units reduce to
\begin{align}
\nn
        \partial_t \tu_0 -\tu_0 + \tu_0^2 + \tih &= 0 \\
\nn
        \tau \partial_t \tih_0 - \tih_0 &= -\mu \delta(t).
\end{align}
Thus
\beq
\nn
        \tih_0(t) = \frac{\mu}{\tau}\, e^{t/\tau}  \theta(-t),
\eeq
and the equation for $\tu_0$ becomes
\beq
\label{eq:DirPertStatH}
\partial_t \tu_0 -\tu_0 + \tu_0^2 = -\frac{\mu}{\tau}\, e^{t/\tau} \theta(-t).
\eeq
It is  the instanton equation for the standard ABBM model ($a=0$), but with a time-dependent source $\lambda_t = \frac{\mu}{\tau}\, e^{t/\tau}\theta(-t)$. This is natural, stating that to $\mathcal{O}(a)^0$, the distribution of $\mu h(t=0)$ is  the distribution of the observable $\int_t \lambda_t \du_t = \frac{\mu}{\tau} \int_{-\infty}^0 \rmd t\, e^{t/\tau}\du_t$ in the pure ABBM model.

The solution of \eqref{eq:DirPertStatH} is $\tu_0(t) = \frac{\psi'(t)}{\psi(t)}$, where
\begin{alignat*}{3}
0 & = & & \tau \psi''(t) - \tau \psi'(t) + \mu e^{t/\tau} \psi(t) \\
\Rightarrow  \psi(t) & = && \left[c_1\, J_{-\tau}\left(2e^{t/(2\tau)}\sqrt{\mu\tau}\right) \right. \\
& && \left. +c_2\, J_{\tau}\left(2e^{t/(2\tau)}\sqrt{\mu\tau}\right)\right]e^{t/2}.
\end{alignat*}
Fixing $c_1/c_2$ using the boundary condition $\tu_0(t=0)=0$, we obtain
\begin{align}
\nn
        \int\rmd t\, \tu_0(t) & = \ln \frac{\psi(0)}{\psi(-\infty)} = -\ln \,_0 F_1(\tau,-\mu \tau).
\end{align}
Thus, the generating function of the stationary distribution of $h$ is
\beq
\label{eq:DirPertGenH}
\overline{e^{\mu h}} = \left[\,_0 F_1(\tau,-\mu \tau)\right]^{-v}\left[1+\mathcal{O}(a)\right].
\eeq
We can now make contact with the result of section \ref{sec:QSEddyStat}. Defining $\mu := \tau \mu_r$, $v_r := v \tau$ and using that 
\begin{align*}
\lim_{\tau\to\infty} \frac{1}{\tau}\ln \,_0 F_1(\tau,-\mu_r \tau^2) =& -1+\sqrt{1-4\mu_r}  \\
& -\ln \frac{1}{2}(1+\sqrt{1-4\mu_r}),
\end{align*}
Eq.~\eqref{eq:DirPertGenH} in the limit $\tau\to\infty$ reduces to the $a=0$ limit of \eqref{eq:QSStatLT}. 
Its Laplace inverse is  given by \eqref{eq:QSStatH} for $a=0$.
However, computing the Laplace inverse of \eqref{eq:DirPertGenH} for general $\tau, v$ seems to be doable only numerically. Likewise, we did not manage to obtain simple expressions for $\tu_1, \tih_1$ at the next-to-leading order, $\mathcal{O}(a)$. 

In the limit $v \to 0^+$ one defines the density $\rho(h)= \partial_v P(h)|_{v=0^+}$ and one finds
\be 
\label{K1}
h \rho(h) = - {\rm LT}^{-1}_{s \to h} \frac{\partial_s [ _0 F_1(\tau,s \tau)] }{_0 F_1(\tau,s \tau)},
\ee  
where we defined $s:=-\mu;$   for the following suppose $s>0$. 
For half-integer values of $\tau$ (\ref{K1}) can be expressed in terms of elementary functions,
for instance for $\tau=1/2$:
\begin{align*}
h \rho(h)& = {\rm LT}^{-1}_{s \to h} \frac{\tanh(\sqrt{2 s})}{\sqrt{2 s}}  = \sum_{k=1}^{\infty}{\rm LT}^{-1}_{s \to h} \frac{1}{\frac{\pi^2 (2k-1)^2}{8}+s} \\
& =  \sum_{k=1}^{\infty}e^{-\pi^2 (2k-1)^2 h /8} = \frac{1}{2}\theta_2\left(0;e^{-h \pi^2/2}\right),
\end{align*}
where $\theta_2$ is the Jacobi theta function. For small $h$ it diverges as $\rho(h) \simeq \frac{1}{\sqrt{2 \pi}} h^{-3/2}$, similar to the result (\ref{aga}) 
found in the limit of large $\tau$ and fixed (not necessarily small)  $a$.

\section{Perturbation theory in $a$ for the backward equation\label{sec:AppPertInABackwards}}
\begin{figure}
\includegraphics[width=\columnwidth]{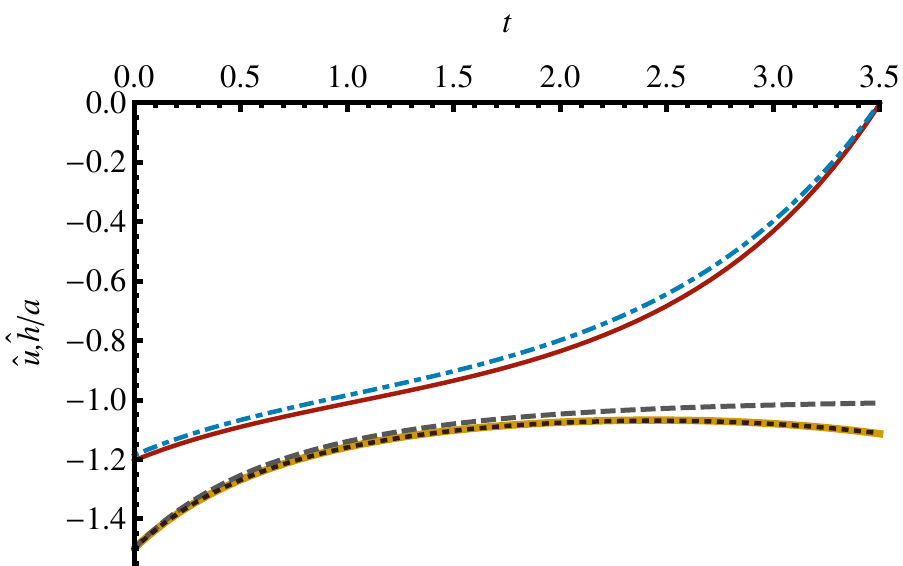}\caption{(Color online) Numerical solution and perturbative expansion of \eqref{eq:RetInstAbsExpTu}, \eqref{eq:RetInstAbsExpTh} for $a=0.2$, $\tau=1$. Boundary conditions are \eqref{eq:AbsBCFastRel} with $\ti=0$, $\tf=3.5$, $\mu=0$, $\lambda=-1.5$. Thick (yellow) line: $\hu_t$ from numerical solution. Thin (red) line: $\hh_t /a$ from numerical solution. Grey (dashed) line: pure ABBM instanton $\hu_t^{(00)}$, given by \eqref{eq:AbsInstSolABBM}. Blue (dot-dashed) line: $\hh_t^{(01)}$ given by \eqref{eq:HH01}. Black (dotted) line: $\hu_t^{(00)}+a\hu_t^{(01)}$ given by \eqref{eq:AbsInstSolABBM}, \eqref{eq:HU01}.}\label{fig:BackwardInstantonTau1}\end{figure}

In this section, we compute the 6 functions $\hu_t^{(00)}, \hu_t^{(10)}, \hu_t^{(01)}$ and $\hh_t^{(10)}, \hh_t^{(01)}, \hh_t^{(11)}$ used in the perturbative expansion  \eqref{eq:AbsWeakRelExp} for  sub-avalanche durations. Expanding \eqref{eq:RetInstAbsExpTu}, \eqref{eq:RetInstAbsExpTh} in $a$ and $\mu$, and inserting the ansatz \eqref{eq:AbsWeakRelExp}, we obtain the following set of equations
\begin{align}
\label{eq:PertInAEqU00}
        -\partial_t \hu_t^{(00)} + \hu_t^{(00)} + \left(\hu_t^{(00)}\right)^2 & = -\lambda \delta(t-\ti) \\\label{eq:PertInAEqH10}
        -\tau\partial_t \hh_t^{(10)} + \hh_t^{(10)} & = \delta(t-\tf) \\        \label{eq:PertInAEqU10}
\left[-\partial_t+1 + 2 \hu_t^{(00)}\right]\hu_t^{(10)} - \hh_t^{(10)} &= 0 \end{align}
\begin{align}
        \label{eq:PertInAEqH01}
        -\tau\partial_t \hh_t^{(01)} + \hh_t^{(01)} - \hu_t^{(00)} & = 0 \\        \label{eq:PertInAEqU01}
\left[-\partial_t+1 + 2 \hu_t^{(00)}\right]\hu_t^{(01)} + \hu_t^{(00)} - \hh_t^{(01)} &= 0 \\
        \label{eq:PertInAEqH11}
        -\tau\partial_t \hh_t^{(11)} + \hh_t^{(11)} - \hu_t^{(10)} & = 0
\end{align}
Some numerical solutions, and their comparison to this perturbative expansion, can be seen in figure \ref{fig:BackwardInstantonTau1}.

\eqref{eq:PertInAEqU00} shows that $\hu_t^{(00)}$ is given by the pure ABBM solution \eqref{eq:AbsInstSolABBM}. Similarly, \eqref{eq:PertInAEqH10} shows that $\hh_t^{(10)}$ is also given by the expression from the pure ABBM model \eqref{eq:AbsInstSolHABBM},
\beq
\hh_t^{(10)} = e^{-(\tf-t)/\tau}.
\eeq
The term $\hu_t^{(10)}$ can still be computed at order $a^0$, by solving \eqref{eq:PertInAEqU10} with the boundary condition $\hu_\ti^{(10)} = 0$. The solution reads
\begin{widetext}
\begin{align}
\nn
\hu_t^{(10)} &= -\int_\ti^t \rmd s_1 \, \exp\left\{ \int_{s_1}^t\rmd s_2 \, \left[1+2\hu_{s_2}^{(00)}\right] \right\} e^{-(\tf-s_1)/\tau} \\
 & =  \frac{\tau  e^{-\frac{\tf}{\tau }}}{\left(\tau ^2-1\right) \left[\lambda
   e^t-(\lambda+1) e^{\ti}\right]^2} \left\{-\lambda^2 (\tau -1)e^{t
   \left(\frac{1}{\tau }+2\right)}+2 (\lambda+1) \lambda \left(\tau ^2-1\right)
   e^{\frac{t}{\tau }+t+\ti} \right. \nn\\
        & \quad \left. -\left[2 \lambda (\lambda+1) \tau ^2+2 \lambda \tau +\tau
   +1\right] e^{t+\frac{\ti}{\tau }+\ti}+(\lambda+1)^2 (\tau +1)
   e^{\frac{t}{\tau }+2 \ti}\right\}.
\end{align}
On the other hand, $\hh_t^{(01)}$ is obtained by solving        \eqref{eq:PertInAEqH01} 
with the boundary condition $\hh_\tf^{(01)} = 0$. The solution reads
\begin{align}
\nn
\hh_t^{(01)} &= -\frac{1}{\tau}\int_\tf^t \rmd s_1 \, e^{(t-s_1)/\tau } \frac{\lambda}{e^{\ti-s_1}(\lambda+1)-\lambda} \\
 & =  \frac{\lambda  e^{\frac{t}{\tau }-\ti}}{(\lambda +1) (\tau -1)} \left[e^{\frac{(\tau -1)
   \tf}{\tau }} \, _2F_1\left(1,\frac{\tau -1}{\tau
   };2-\frac{1}{\tau };\frac{e^{\tf-\ti} \lambda }{\lambda
   +1}\right)-e^{\frac{t (\tau -1)}{\tau }} \, _2F_1\left(1,\frac{\tau
   -1}{\tau };2-\frac{1}{\tau };\frac{e^{t-\ti} \lambda }{\lambda
   +1}\right)\right]
\end{align}
\end{widetext}
Obtaining the higher-order terms for arbitrary $\tau > 0$ is now complicated, due to the appearance of hypergeometric functions. However, the latter simplify significantly in the limit $\tau \to 1$. From now on, we will consider this limit only. We then have
\begin{align}
\nn
\hu_t^{(10)} = &\frac{e^{t-\tf}}{2
   \left[\lambda  e^t-(\lambda +1) e^{\ti}\right]^2} \left\{-\lambda^2 e^{2 t}+4 (\lambda
   +1) \lambda  e^{t+\ti} \right. \\
\nn
        &\left.+e^{2 \ti} \lambda  \left[-3 \lambda -2
   (\lambda +2) t+2 (\lambda +2) \ti-4\right] \right. \\
                & \left. -2e^{2 \ti} (t-\ti)\right\} \\
\label{eq:HH01}
\hh_t^{(01)} =& \frac{\lambda e^{t-\ti}}{\lambda+1} \left[\tf-t + \log \frac{(\lambda+1) e^{\ti}-\lambda
   e^t}{(\lambda+1) e^{\ti}-\lambda
   e^{\tf}}\right]
\end{align}
To obtain $\hu_t^{(01)}$, we need to solve \eqref{eq:PertInAEqU01} with the boundary condition $\hu_\ti^{(01)} = 0$. The solution reads
\begin{align}
\nn
\hu_t^{(01)} =& \int_\ti^t \rmd s_1 \, \exp\left\{ \int_{s_1}^t\rmd s_2 \, \left[1+2\hu_{s_2}^{(00)}\right] \right\} \times \\
\label{eq:HU01}
& \quad\quad\quad \times\left(\hu_{s_1}^{(00)}-\hh_{s_1}^{(01)}\right) 
                                                                                                                                \end{align}
This integral can be evaluated in closed form and gives a lengthy expression in terms of logarithms and dilogarithms. 
However, the expression \eqref{eq:AbsWeakRelDur} for the avalanche duration only depends on $\int_{\ti}^{\tf}\rmd t\, \hu_t^{(01)}$, and on $\hu_\tf^{(01)}$. These two terms depend only on $T:=\tf-\ti$, and are simpler:
\begin{widetext}
\begin{align}
\nn
&\hu_\tf^{(01)} = \int_\ti^\tf \rmd s_1 \, \exp\left\{ \int_{s_1}^\tf\rmd s_2 \, \left[1+2\hu_{s_2}^{(00)}\right] \right\} \left(\hu_{s_1}^{(00)}-\hh_{s_1}^{(01)}\right) \\
 & =  \frac{e^{T} \lambda  }{2 (\lambda +1) \left(-e^{T} \lambda +\lambda
   +1\right)^2}\left[-T^2 (\lambda +1)^2-T
   \left(\lambda ^2-2\right)-\left(e^{T}-1\right) \lambda 
   (\lambda +1)+(2 T-3) (\lambda +1)^2 \log
   \left(1-\frac{e^{T} \lambda }{\lambda +1}\right) \right. \nn\\
& \quad \left. +\lambda  (3
   \lambda +4) \log \left(-e^{T} \lambda +\lambda +1\right)+2
   (\lambda +1)^2 \text{Li}_2\left(\frac{e^{T} \lambda }{\lambda
   +1}\right)+3 (\lambda +1)^2 \log \left(\frac{1}{\lambda +1}\right)-2
   (\lambda +1)^2 \text{Li}_2\left(\frac{\lambda }{\lambda
   +1}\right)\right] \\
        \nn
        & \int_{\ti}^{\tf}\rmd t\, \hu_t^{(01)} = \frac{1}{2 (\lambda +1) \left[\left(e^{T}-1\right)
   \lambda -1\right]}  \left\{\lambda\left[T \lambda +e^{T} (T
   (T \lambda +T-2)+\lambda +1)+2 T-\lambda
   -1\right] \right. \\
        \nn
        & \quad \left. +\left[2 e^{T} \lambda  ((\lambda +1) \log (\lambda
   )-(\lambda +1) \log (\lambda +1)+1)+1\right] \log
   \left[-e^{T} \lambda +\lambda +1\right] \right. \\
        & \quad \left. -2 e^{T}
   \lambda  (\lambda +1) \left[\text{Li}_2\left(\frac{1}{\lambda
   +1}\right)-\text{Li}_2\left(1-\frac{e^{T} \lambda }{\lambda
   +1}\right)\right]\right\}
\end{align}
To obtain $\hh_t^{(11)}$,  we need to solve \eqref{eq:PertInAEqH11} 
with the boundary condition $\hh_\tf^{(11)} = 0$. Note that the formula \eqref{eq:AbsWeakRelDur} for the avalanche duration only contains $\hh_\ti^{(11)}$. The result for this value again only depends on $T:=\tf-\ti$: 
\begin{align}
\nn
\hh_\ti^{(11)} &= \frac{1}{2} e^{-T} \left[\frac{T^2 (\lambda +1)^2
   \left(\left(e^{T}-1\right) \lambda -1\right)+T \lambda 
   \left(e^{T} \left(\lambda ^2-2\right)-(\lambda +1) (3 \lambda
   +4)\right)+\left(e^{T}-1\right) \lambda  (2 \lambda +3)
   (\lambda +1)}{(\lambda +1)^2 \left(-e^{T} \lambda +\lambda
   +1\right)} \right. \\
        & \quad \left. -\frac{(2 \lambda +3) \log \left(-e^{T} \lambda
   +\lambda +1\right)}{(\lambda +1)^2}+2 T \log
   \left(1-\frac{e^{T} \lambda }{\lambda +1}\right)+2
   \text{Li}_2\left(\frac{e^{T} \lambda }{\lambda +1}\right)-2
   \text{Li}_2\left(\frac{\lambda }{\lambda +1}\right)\right].
\end{align}
Finally, $\hh_\ti^{(01)}$ is obtained from \eqref{eq:HH01} as
\beq
\hh_\ti^{(01)} = \frac{\lambda}{1+\lambda}\left[T-\log\left(1+\lambda-\lambda e^T\right)\right]
\eeq
In figure \ref{fig:BackwardInstantonTau1}, we show that these perturbative expressions compare well to a direct numerical solution of the backward-instanton equations.

In order to apply \eqref{eq:AbsWeakRelDur} we need to compute the leading behaviour as $\lambda \to -\infty$. We get
\begin{align}
\lim_{\lambda\to-\infty} \hh_\ti^{(11)} =& \frac{T}{e^{T}-1} -\frac{1}{2} e^{-T} \left[2
   \text{Li}_2\left(1-e^{T}\right)+(T+1)
   (T+2)\right]\\\lim_{\lambda\to-\infty} \int_{\ti}^{\tf}\rmd t\, \hu_t^{(01)} = & \frac{e^{T} T^2+2 e^{T}
   \text{Li}_2\left(1-e^{T}\right)+T+e^{T}-1}{2
   \left(e^{T}-1\right)}\\
\lim_{\lambda\to-\infty} \hu_\tf^{(01)} = & -\frac{e^{T}}{2 \left(e^{T}-1\right)^2}\left[T^2+2
   \text{Li}_2\left(1-e^{T}\right)+T+e^{T}-1\right] \\
        \lim_{\lambda\to-\infty} (-\lambda)^{a h_{\rm i}}\exp\left(-a h_{\rm i}\hh_\ti^{(01)}\right) = & \left(1-e^{-T}\right)^{-a h_{\rm i}}
\end{align}
\end{widetext}
Inserting these results into \eqref{eq:AbsWeakRelDur}, we obtain \eqref{eq:AbsWeakRelCorrT}.

\section{Avalanche sizes at finite driving velocity $v$ in the standard ABBM model\label{sec:AppABBMSizes}}
Similar to the computation of avalanche durations in the standard ABBM model at finite driving velocity in section \ref{sec:DurationABBM}, 
we can also compute the avalanche sizes. The discussion follows closely Ref.\ \cite{LeDoussalWiese2009} where some explicit
expressions were obtained in the small-$m$ limit (at fixed $x=v/v_m$).
Let us now consider the stochastic process $\du(u)$, i.e.\ the domain-wall velocity as a function of its position.
Its first-passage distribution   was derived in \cite{LeDoussalWiese2009} (equivalent to formula (E30) there\footnote{${\sf v}$ and ${\sf v'}$ must be replaced by $v$ in the first argument of the hypergeometric function there.}) and later in \cite{MasoliverPerello2012}:
\begin{align}
&\int_0^\infty \rmd u\, e^{\lambda u}P_{\du}(u|\du_0) = e^{-\frac{1}{2}(1-\sqrt{1-4\lambda})(\du_0-\du)} \times \\
&~~~~~~~~~~~~~~~~~~~~~~~\times
\frac{U\left[\frac{v}{2}\left(1-\frac{1}{\sqrt{1-4\lambda}}\right), v, \sqrt{1-4\lambda}\du_0\right]}{U\left[\frac{v}{2}\left(1-\frac{1}{\sqrt{1-4\lambda}}\right), v, \sqrt{1-4\lambda}\du\right]}. \nn
\end{align}
As $\du\to 0$, this has a finite limit for $v < 1$ (the first-passage distribution at $\du=0$),
\begin{align*}
&\int_0^\infty \rmd u\, e^{\lambda u}P_0(u|\du_0) = e^{-\frac{\du_0}{2}(1-\sqrt{1-4\lambda})}\times \\
&~~~~~~~~~~~~~~~~~~~~~~~~~\times
\frac{U\left[\frac{v}{2}\left(1-\frac{1}{\sqrt{1-4\lambda}}\right), v, \sqrt{1-4\lambda}\du_0\right]}{\Gamma(1-v)\big/ \Gamma\left[1-\frac{v}{2}\left(1+\frac{1}{\sqrt{1-4\lambda}}\right)\right]}
\end{align*}
Integrating this over the stationary distribution for $\du_0(u)$, $P(\du_0) = \frac{1}{\Gamma(v+1)}\du_0^{v}e^{-\du_0}$, we get the Laplace-transformed cumulative distribution of avalanche sizes in the form
\bea
&& \int\limits_0^\infty e^{\lambda u} \frac{P(S\geq u)}{\int_0^\infty dS' S' P(S')}  \rmd u= -\frac{(1-4\lambda)^{\frac{1-v}{2}}}{\lambda v B\left[1-v,\frac{v}{2}\left(1-\frac{1}{\sqrt{1-4\lambda}}\right)\right]} \nn \\
&& \label{dist1}
\eea
Here, $B(x,y)$ is the usual Beta function. (\ref{dist1}) is valid since the probability that $u=0$ belongs to an avalanche of size $S$ is
$S P(S)/\int_0^\infty \rmd S' S' P(S')$ and the conditional distribution of $u$ is then $P(u|S)=\theta(S-u)/S$. Putting this together
produces (\ref{dist1}). One can check that taking the large-$\lambda$ limit and Laplace inverting one recovers
exactly the formula below (E.28) in \cite{LeDoussalWiese2009}, valid in the small-$m$ limit (at fixed $x=v/v_m$).

The Laplace transform of the avalanche density $\rho(S) = P(S)/[\int_0^\infty \rmd S' S' P(S')]$ is 
obtained by integration by part of (\ref{dist1}),\beq
\label{eq:PofS}
\int_0^\infty \rmd S (e^{\lambda S} -1) \rho(S) = -  \frac{(1-4\lambda)^{\frac{1-v}{2}}}{v B\left[1-v,\frac{v}{2}\left(1-\frac{1}{\sqrt{1-4\lambda}}\right)\right]}.
\eeq
A nontrivial check of \eqref{eq:PofS} is that it satisfies the normalization condition 
$\int \rmd S \,S \rho(S) =1$ automatically.

Using standard relations for the Beta function \cite{AbramowitzStegun1965}, one can rewrite \eqref{eq:PofS} as
\begin{align}
\nn
\int_0^\infty dS (e^{\lambda S} -1) \rho(S) = &\frac{\sin \pi v}{2\pi} (1-4\lambda)^{-\frac{v}{2}}\left(1+\sqrt{1-4\lambda}\right) \\
\label{eq:PofS2}
& \times B\left[v,-\frac{v}{2}\left(1+\frac{1}{\sqrt{1-4\lambda}}\right)\right].
\end{align}
        Using the substitution $r := 1+ \sqrt{1-4\lambda}$, we can write the inverse Laplace transform in the compact form
\begin{align*}
\rho(S)=&\int_{r_0-i\infty}^{r_0+i\infty} \frac{\rmd r}{2\pi i} \frac{\sin (\pi  v)}{4 \pi }\times \\
& \quad \times r e^{\frac{1}{4} (r-2) r S} (r-1)^{1-v} 
   B\left(v,\frac{r v}{2-2 r}\right),
\end{align*}
where $r_0 > 1$. 

For the case $v=0$, \eqref{eq:PofS} reduces to the known expression \cite{LeDoussalWiese2011b,LeDoussalWiese2012a}
\beq
\nn
 \int_0^\infty \rmd S\, (e^{\lambda S} -1) \rho(S) =  \frac{1}{2}\left(1-\sqrt{1-4\lambda}\right),
\eeq
which leads to the standard ABBM avalanche-size density $\rho(S)=\frac{1}{2 \sqrt{\pi} S^{3/2}} e^{-S/4}$. 

For any $0<v<1$, we can obtain the behaviour of $\rho(S)$ at small $S$ from the $\lambda \to -\infty$ limit of \eqref{eq:PofS2}. In this limit, the Beta function tends to a constant and we obtain
\begin{align*}
&\int_0^\infty \rmd S\, (e^{\lambda S} -1) \rho(S) =\\
&= \frac{\sin \pi v}{2\pi}(-4\lambda)^{\frac{1-v}{2}} \left[1+\mathcal{O}(|\lambda|)^{-1/2}\right]B\left(v,-\frac{v}{2}\right)
\end{align*}
Inverting the Laplace transform, we see that near $S=0$
\beq
\rho(S) = S^{-(3-v)/2}\left[\frac{(v-1)\Gamma(-v/2)\sin(\pi v)}{4\pi^{3/2}} + \mathcal{O}(S)^{1/2}\right].
\eeq
This is in agreement with previous results \cite{DurinZapperi2006b,Colaiori2008,LeDoussalWiese2009}, and extends them by giving the prefactor of the power law.

\section{Fixing initial conditions instead of final conditions\label{sec:AppBackwardForward}}
In this section, we discuss how to transform \eqref{eq:SolAbsProp}, a formula with a fixed value of $h_{\rm f}$ and a Laplace transform taken with respect to $h_{\rm i}$,  into \eqref{eq:SolAbsPropInt2}, a formula with a fixed value of $h_{\rm i}$ and a Laplace transform taken with respect to $h_{\rm f}$.

We start by integrating \eqref{eq:SolAbsProp} over $h_{\rm f}$,
\begin{widetext}
\begin{align}
\nn
&\int_{-\infty}^{\infty}\rmd h_{\rm f}\int_0^\infty \rmd \du_{\rm i} \int_{-\infty}^\infty \rmd h_{\rm i}\,e^{\lambda \du_{\rm i} +\mu h_{\rm i}}\mathcal{P}_{\text{abs}}(\du_{\rm f},h_{\rm f};\tf|\du_{\rm i},h_{\rm i};\ti) =  \\
\label{eq:SolAbsPropInt1}
&
= \exp\left\{\int_{\ti}^{\tf}\left[(2-\dw_{s})\hu_s + (1+a+\tau^{-1}) \right]\rmd s + \du_{\rm f} \hu_\tf\right\}(2\pi)\delta\left[i\tau \hh(\tf)\right].
\end{align}
We now invert the Laplace transform from $\mu = \tau \hh(\ti)$ to $h_{\rm i}$ using a complex integral,
\begin{align}
\nn
&\int_{-\infty}^{\infty}\rmd h_{\rm f}\int_0^\infty \rmd \du_{\rm i} \,e^{\lambda \du_{\rm i}}\mathcal{P}_{\text{abs}}(\du_{\rm f},h_{\rm f};\tf|\du_{\rm i},h_{\rm i};\ti) =  \\
\nn
& = \int_{-i\infty}^{i\infty} \rmd \mu \, \exp\left\{\int_{\ti}^{\tf}\left[(2-\dw_{s})\hu_s + (1+a+\tau^{-1}) \right]\rmd s + \du_{\rm f} \hu_\tf\right\} \delta(\hh_\tf) \exp\left(-\tau \hh_\ti h_{\rm i}\right) \\
\nn
&= \exp\left\{\int_{\ti}^{\tf}\left[(2-\dw_{s})\hu_s + (1+a+\tau^{-1}) \right]\rmd s + \du_{\rm f} \hu_\tf - \tau h_{\rm i} \hh_\ti\right\} \left[\partial_{\hh(\tf)}\big|_{\hh(\tf)=0}\hh(\ti)\right].
\end{align}
\end{widetext}
This  now is Eq.~\eqref{eq:SolAbsPropInt2}.
Due to the $\delta$-function, the only value of $\hh(\ti)$ that contributes is the one which leads to $\hh(\tf)=0$. So, the effect of going from a fixed $h_{\rm f}$ to a fixed $h_{\rm i}$ in the propagator is a change in the boundary conditions for the pair of backward instanton equations \eqref{eq:RetInstAbsExpTu}, \eqref{eq:RetInstAbsExpTh}. When integrating over all $h_{\rm f}$, we have to impose the boundary conditions \eqref{eq:AbsBCFastRel}
\beq
\nn
\hh(\tf) = 0, \quad\quad\quad \hu(\ti) = \lambda.
\eeq
In addition, we get the ``Jacobian'' factor in \eqref{eq:SolAbsPropInt2}. In the pure ABBM case $a=0$, as discussed in section \ref{sec:SubavRetToABBM}, it just cancels the $(\tf-\ti)/\tau$ factor in the exponential. For $a>0$ it is more complicated.

\section{Some exact relations for the propagator}
Finding general solutions to the forward instanton equations \eqref{eq:RetInstantonExpTu}, \eqref{eq:RetInstantonExpTh} and the backward instanton equations \eqref{eq:RetInstAbsExpTu}, \eqref{eq:RetInstAbsExpTh} is difficult. However, simple particular solutions exist, where the instanton is constant in time. These imply exact relations on particular observables in the ABBM model with retardation, which we  discuss below.

\subsection{Forward instanton}
A particular solution of \eqref{eq:RetInstantonExpTu}, \eqref{eq:RetInstantonExpTh} is
\begin{align}
\nn
\lambda_t & = \delta(t-\tf), & \mu_t & = a\tau \delta(t-\tf), \\  
\tu_t & = \theta(\tf-t), & \tih_t & = a \theta(\tf-t).
\end{align}
To see the significance of this solution, consider starting at fixed $\du_{\rm i},h_{\rm i}$ at $t=0$, and driving with a constant velocity $\dw_t=v$ for $0<t<t_{\rm f}$. Using \eqref{eq:RetSolutionExp}, and accounting for the initial condition as in \cite{DobrinevskiLeDoussalWiese2012}, Eq.~(4), we get
\begin{align}
\nn
\overline{e^{\du(\tf) + a\tau h(\tf)}} =& e^{\du_{\rm i}\tu_0 + \tau h_{\rm i}\tih_0 + \int_{0}^{\tf}\rmd t\, \dw_t\tu_t} \\
=& e^{\du_{\rm i}+a \tau h_{\rm i}+v\tf}.
\end{align}
This implies the following exact relation on the propagator of the ABBM model with retardation at constant driving velocity $v$,
\begin{align}\nn
&\int_0^\infty \rmd \du_{\rm f} \int_{0}^\infty \rmd h_{\rm f}\,e^{\du_{\rm f} + a\tau h_{\rm f}} \mathcal{P}(\du_{\rm f},h_{\rm f};\tf|\du_{\rm i},h_{\rm i};0) \\
& = e^{\du_{\rm i} + a\tau h_{\rm i} + v\tf}.
\end{align}
For the case $a=0$ (pure ABBM model), this relation can be checked explicitly using the formula for the ABBM propagator, Eq.~(19) in \cite{DobrinevskiLeDoussalWiese2012}. It generalizes similarly to the case of arbitrary driving $\dw$.

\subsection{Backward instanton}
Now let us apply the same idea to the backward instanton, used in section \ref{sec:AvStatDur}, in order to perform calculations with an absorbing boundary at $\du=0$. 
A particular solution of \eqref{eq:RetInstAbsExpTu}, \eqref{eq:RetInstAbsExpTh} is
\begin{align}
\nn
  \lambda_t & = -\delta(t-\tf), & \mu_t & = -a\tau \delta(t-\tf), \\
   \hu_t & = -\theta(\tf-t), & \hh_t & = -a \theta(\tf-t).
\end{align}
To see the significance of this solution, consider the density to arrive at $\du_{\rm f},h_{\rm f}$ at $t=\tf$, while driving with a constant velocity $\dw_t=v$ for $0<t<\tf$, as a function of the initial condition at $t=0$. 
 Using \eqref{eq:SolSurvProb}, we get
\begin{align}
\nn
\overline{e^{-\du(0) - a\tau h(0)}} =& e^{\du_{\rm f}\hu_{\tf} + \tau h_{\rm f}\hh_{\tf} + \int_{0}^{\tf}\rmd t\, (2-\dw_t)\hu_t + (1+a+\tau^{-1})} \\
=& e^{ - \du_{\rm f} - a \tau h_{\rm f}+(v-1+a+\tau^{-1})\tf}
\end{align}
This implies the following exact relation on the propagator of the ABBM model with retardation at constant driving velocity $v$, with an absorbing boundary at $\du=0$:
\begin{align}\nn
&\int_0^\infty \rmd \du_{\rm i} \int_{-\infty}^\infty \rmd h_{\rm i}\,e^{-\du_{\rm i} - a h_{\rm i}} \mathcal{P}_{\text{abs}}(\du_{\rm f},h_{\rm f},\tf|\du_{\rm i},h_{\rm i},0) \\
&= e^{-\du_{\rm i} - a h_{\rm i} + (v-1+a)\tf}.
\end{align}
Again, for the pure ABBM model $a=0$ this relation can easily be checked using the expression \eqref{eq:AbsPropABBM} for the propagator with an absorbing boundary at $\du=0$.

\bibliography{ABBM}

\tableofcontents

\end{document}